\documentclass[11pt,a4paper]{article}
\usepackage{jheppub}
\usepackage{bm}
\usepackage{feynmf}
\usepackage{mathrsfs}
\usepackage{slashed}
\usepackage{mathtools}
\usepackage[dvipsnames]{xcolor}
\usepackage{physics}
\usepackage{hyperref}
\usepackage{graphicx, animate}
\usepackage{tikz}
\usepackage{listings}
\usepackage{orcidlink}
\usepackage{float}

\usepackage[utf8]{inputenc}
\usepackage{soul}

\allowdisplaybreaks
 
\newcommand\sumint[1]{\int\kern-1.5em\sum\nolimits_{#1}}

\newcommand\del{\partial}

\newcommand{\beq}{\begin{equation}\begin{aligned}}
\newcommand{\eeq}{\end{aligned}\end{equation}}
\newcommand{\av}[1]{\ensuremath{\langle#1\rangle}}
\newcommand{\deff}{\ensuremath{\vcentcolon =}}
\newcommand{\pl}{\text{\tiny Pl}}
\renewcommand{\pv}{\text{\tiny PV}}
\newcommand{\cl}{\text{\tiny cl}}
\newcommand{\qu}{\text{\tiny qu}}
\newcommand{\tinytext}[1]{\text{\tiny #1}}

\renewcommand{\max}{\text{\tiny max}}
\newcommand{\cut}{\text{\tiny cut}}
\newcommand{\AH}{\text{\tiny AH}}

\newcommand{\causal}{\text{\tiny causal}}

\newcommand{\ud}{\mathrm{d}}

\def\p{\partial}

\graphicspath{ {./main_images/} }

\title{Quantum correlations in a gravitational collapse simulation with \texttt{SpheriCo.jl}}

\author[a]{Benjamin Berczi\orcidlink{0000-0001-9390-6124},}
\author[b]{Magdalena Eriksson \orcidlink{0000-0001-7193-620X},}
\author[a,c]{Thanasis Giannakopoulos \orcidlink{0000-0002-3055-9652},}
\author[a,c]{Paul M. Saffin \orcidlink{0000-0002-4290-3377}}
\date{~}

\affiliation[a]{School of Physics and Astronomy, University of Nottingham, University Park, Nottingham NG7
2RD, United Kingdom}
\affiliation[b]{AEC, Institute for Theoretical Physics, University of Bern,
Sidlerstrasse 5, CH-3012 Bern, Switzerland}
\affiliation[c]{Nottingham Centre of Gravity, University of Nottingham,
University Park, Nottingham, NG7 2RD, United Kingdom}

\emailAdd{benji.berczi@gmail.com}
\emailAdd{magdalena.eriksson@unibe.ch}
\emailAdd{pmzag1@exmail.nottingham.ac.uk}
\emailAdd{paul.saffin@nottingham.ac.uk}

\abstract{
    We report on work using a newly developed code, \texttt{SpheriCo.jl}, that computes the gravitational collapse of a spherical scalar field, where the scalar can be either a classical field, or a quantum field operator. By utilising summation-by-parts methods for the numerical derivatives we are able to simulate the collapse longer than was possible previously due to enhanced numerical stability. We present a suite of tests for the code that tests its accuracy and stability, both for the classical and quantum fields. We are able to observe critical behavior of gravitational collapse for the classical setup, in agreement with expected results. The code is also used to compute two-point correlation functions, with results that hint at a non-trivial correlation across the horizon of Hawking quanta.
}

\begin{document} 
\maketitle
\flushbottom

\section{Introduction} \label{Sec:intro}

The study of the classical problem of gravitational collapse of a scalar field with spherical symmetry has now reached a good level of maturity, with analytic progress  \cite{Christodoulou:1986zr,Christodoulou:1986du,Christodoulou:1987vv,Christodoulou:1987vu} as well as numerical \cite{alcubierre2008introduction}. In particular, the phenomenon of critical collapse was identified in \cite{Choptuik:1992jv}, along with the echos that are observed as the collapse continues \cite{Choptuik:1992jv,Garfinkle:1994jb,Hamade:1995ce}.
Understanding the same system when the classical scalar is promoted to a quantum field is yet to be fully understood, but some progress has been made in this direction \cite{Russo:1992ax,Strominger:1993tt,Piran:1993tq,Lowe:1992ed} in two spacetime dimensions. In this semiclassical system we retain a classical geometry, but its dynamics are sourced by the expectation value of the quantum stress-energy tensor. One issue that we face when doing this is the divergence of expectation values of quantities such as the stress-energy tensor, meaning that we need to find suitable regularization techniques that do not spoil the co-ordinate invariance of general relativity. One such approach is a point-splitting procedure \cite{Levi:2015eea,Levi:2016esr,Levi:2016quh,Levi:2016exv}. Another approach to regularization was taken in \cite{Berczi:2020nqy,Berczi:2021hdh}, where Pauli-Villar ghost fields were introduced, whose contribution to the expectation value of the stress-energy tensor is designed to cancel the divergences. Further study was made in \cite{Guenther:2020kro} of the effect of radial modes, with a report on the extension that includes angular modes presented in \cite{Hoelbling:2021axl}.

Simulating quantum fields is inherently numerically expensive, as each mode of the quantum field operator must be dynamically evolved and there are as many modes as there are Fourier modes. The restriction to spherical symmetry allows for a large reduction in numerical effort, as the angular part of the calculation may, to some extent, be performed analytically. The cost that one pays is the appearance of spherical co-ordinate singularities, such as $\frac{1}{r}$, in the equations of motion, which must be treated with care in numerical simulations. Here we take the approach of redefining our dynamical variables, along with a summation-by-parts representation of certain operators, following the approach of ~\cite{Gundlach:2010et}. With this improvement we then follow \cite{Berczi:2020nqy,Berczi:2021hdh} in that we use a Pauli-Villars (PV) regularization scheme. We further extend the techniques of \cite{Berczi:2020nqy,Berczi:2021hdh} by analytically computing the relevant divergent counterterms for the Planck mass and the cosmological constant within this PV regularization scheme.

The numerical approach we use then allows us to simulate the collapsing system for longer times than was possible previously for the semiclassical setup, since the equations near the origin are better approximated. By simulating quantum fields in dynamical backgrounds we may also examine behavior that is not present in the classical system, such as looking for Hawking radiation \cite{Hawking:1975vcx} and correlations in the scalar field between points that are separated across the horizon \cite{Schutzhold:2010ig,Fontana:2023zqz}, which are interpreted as correlations between pairs of Hawking quanta. These correlations have also been examined in the case of analogue horizons \cite{Carusotto:2008ep}, with correlations in fluctuations developing alongside the appearance of the analogue horizon. In the black hole setting, initial calculations \cite{Balbinot:2021bnp} suggest that such correlations may not appear, owing to the singularity swallowing up the Hawking particle that enters the black hole. In this paper we compute such correlations along a spacelike hypersurface in a dynamical, asymptotically flat spacetime, where an apparent horizon forms. For this, we have developed \texttt{SpheriCo.jl} \cite{SpheriCo.jl}, an open-access code written in the \texttt{Julia} programming language \cite{Julia_lang}. Even though we detect non-trivial correlation between points inside and outside the horizon in certain setups, we believe that a more systematic study is necessary to gain deeper understanding of this behavior and its relation to the various parameters of the problem, such as the size of the apparent horizon.

The structure of the paper is as follows. In section \ref{Sec:classical_setup} we introduce the equations of motion, together with the initial and boundary data for the classical setup. In section \ref{Sec:semiclassical_setup} we explain the semiclassical approach of promoting the classical scalar field to a quantum scalar field operator and the subsequent equations of motion we evolve and we describe the regularization scheme, as well as the initial and boundary data. Section \ref{Sec:numerical_implementation} is devoted to the numerical implementation followed in \texttt{SpheriCo.jl} to solve the classical and semiclassical setups. For the semiclassical case, the implementation includes both scenarios with and without backreaction on the background geometry. In section \ref{Sec:classical_tests} we present various convergence and physical tests to validate the performance of our code on the classical setup and similarly for the semiclassical one in section \ref{Sec:semiclassical_tests}. Our main new result, the quantum correlators, are in section \ref{Sec:quantum_correlators}. Finally, we conclude in section \ref{Sec:outlook}. Throughout the text, we use the notation $\dot{f} := \p_t f(t,r)$ and $f' := \p_r f(t,r)$, but we also use explicit partial derivative notation when it enhances readability. In our simulations we use units where $c = \hbar = 1$. We typically take $G_N$ to be $1$ or $1/8 \pi$, and we explicitly state our choice in each case.

\section{The classical setup} \label{Sec:classical_setup}

For both the classical and semi-classical simulations we consider a spherically symmetric four-dimensional spacetime, with line element
\beq
    \dd s^2 = -\alpha^2(t,r)\dd t^2+A(t,r)\dd r^2+ r^2B(t,r)\left[\dd\theta^2+\sin^2\theta\dd\varphi^2\right],
    \label{eq:lineelement}
\eeq
where $\alpha(t,r)$ is the lapse function and $A(t,r)$ and $B(t,r)$ are scalar functions. In the classical case, the Einstein field equations are given by
\beq
    G_{ab} + \Lambda g_{ab} = \frac{1}{M_\pl^2}{T}_{ab},
    \label{eq:classicalEinsteinEqs}
\eeq
where $G_{ab}\deff R_{ab}-\frac{1}{2}Rg_{ab}$ is the Einstein tensor, $T_{ab}$ is the stress-energy tensor, $\Lambda$ the cosmological constant, $M_\pl=(8\pi G_N)^{-1/2}$ is the reduced Planck mass and  $a,b,\dotso=0,\dots,3$ run over spacetime dimensions. 

Numerical evaluations of the Einstein equations in spherically symmetric spacetimes are often plagued by co-ordinate singularities at the origin, i.e. $1/r$ terms appearing in the evolution equations. One way to circumvent this issue is by introducing convenient new variables in order to rewrite the (second-order) Einstein equations as first-order differential equations. Rather than evolving the system in terms of the variables $\alpha, \ A$ and $B$, we follow \cite{Alcubierre:2004gn} and recast it in terms of the variables
\begin{equation}
    \begin{aligned}
        K_A &\deff -\frac{1}{2\alpha}\frac{\dot{A}}{A},\\
        K_B &\deff -\frac{1}{2\alpha}\frac{\dot{B}}{B},
    \end{aligned}
    \qquad 
    \begin{aligned}
        D_A &\deff\frac{A'}{A}, \\
        D_B &\deff\frac{B'}{B},
    \end{aligned}
    \qquad 
    \begin{aligned}
        D_\alpha &\deff \frac{\alpha'}{\alpha},\\
        \lambda &\deff\frac{1}{r}\left(1-\frac{A}{B}\right),
    \end{aligned}
    \label{eq:FirstOrderGravVariables}
\end{equation}
where $\dot{A}\deff\partial_t A$ and $A'\deff \partial_r A$. In these co-ordinates, local flatness of the metric at $r=0$ corresponds to $A(t,0)=B(t,0)$, for which $\lambda=0$, where $\lambda$ is an auxiliary variable introduced to deal with the co-ordinate singularity at $r=0$. In addition, the variables $K_A$ and $D_A$ are replaced with $K$ and $\tilde{U}$ respectively, where
\beq
    K \deff K_A+2K_B, \qquad \tilde{U}\deff D_A-2D_B-\frac{4B\lambda}{A}.
    \label{eq:KandUtildeDefs}
\eeq
In terms of these variables, the non-vanishing components of the Einstein tensor are
\begin{subequations}
  \begin{align}
    G^t{}_t &= \frac{1}{A}\left[ D_B'+\frac{1}{r}\left( \lambda + D_B + \tilde{U} -\frac{4\lambda B}{A} \right)-D_B\left( \frac{D_B}{4}+\frac{\tilde{U}}{2}+\frac{2\lambda B}{A}\right) \right]\nonumber \\
    &\quad -K_B(2K-3K_B), \\
    G^t{}_r &= \frac{2}{\alpha}\left[ -K_B' + \left( \frac{1}{r}+\frac{D_B}{2} \right)(K-3K_B) \right], \\
    G^r{}_r &= \frac{2}{\alpha}\left[\dot{K}_B-\frac{3\alpha K_B^2}{2} + \frac{\alpha}{2r^2A}\left(1-\frac{A}{B}\right)+\frac{\alpha D_B}{2rA}+\frac{\alpha D_B^2}{8A}+\frac{\alpha D_\alpha}{rA}+\frac{\alpha D_BD_\alpha}{2A} \right], \\
    G^\theta{}_\theta &= \frac{1}{\alpha}(\dot{K}-\dot{K}_B)-K^2+3K_B(K-K_B)+\frac{1}{2A}(D_B'+2D_\alpha')\nonumber\\
    &\quad -\frac{1}{4A}\left[ (D_B+2D_\alpha)\left(\tilde{U}+D_B+\frac{4B\lambda}{A}\right)-4D_\alpha^2\right]\nonumber\\
    &\quad +\frac{1}{2rA}\left(2D_\alpha-\tilde{U}-\frac{4B\lambda}{A}\right).
    \end{align}
\end{subequations}
Here we note that the $G^t{}_t$ and $G^t{}_r$ components contain no time derivatives of the evolving variables and are therefore constraint equations. They are referred to as the Hamiltonian and momentum constraint equations, respectively and can be expressed as
\begin{align}
    H &\deff D_B'+\frac{1}{r}\left(\lambda+D_B-\tilde{U}-\frac{4\lambda B}{A}\right) - D_B\left(\frac{D_B}{4}+\frac{\tilde{U}}{2}+\frac{2\lambda B}{A}\right)\nonumber\\
    &\quad- AK_B(2K-3K_B)-AG^t{}_t=0, \label{eq:HamiltonianConstraintGeneral}\\
    P &\deff K_B'-\left(\frac{1}{r}+\frac{D_B}{2}\right)(K-K_B)+\frac{\alpha}{2}G^t{}_r =0,  \label{eq:MomentumConstrainGeneral}
\end{align}
where the terms $G^t{}_t$ and $G^t{}_r$ should be understood to be replaced by the components of the stress-energy tensor and cosmological constant term via \eqref{eq:classicalEinsteinEqs}.

During a numerical evolution, the constraint equations \eqref{eq:HamiltonianConstraintGeneral} and \eqref{eq:MomentumConstrainGeneral} are violated to some degree, which drives the numerical solution away from the physical one and can potentially lead to code instabilities. Having a scheme that forces these constraints to be damped during the evolution can be of great benefit. A famous example is the first successful simulation of black hole binaries \cite{Pretorius:2005gq}, where the constraint damping scheme of \cite{Gundlach:2005eh} was utilized. Here, we attempt to minimize the numerical violation of \eqref{eq:HamiltonianConstraintGeneral} and \eqref{eq:MomentumConstrainGeneral} by combining the constraint damping scheme of \cite{Bernuzzi:2009ex} with the regular system of \cite{Alcubierre:2004gn}. The former is based on the $Z_4$ formulation of \cite{Bona:2003fj, Bona:2004yp} in which a 4-vector $\textbf{Z}$, is added to the Einstein equations:
\beq
    G_{ab} + \Lambda g_{ab} + \nabla_aZ_b + \nabla_b Z_a = \frac{1}{M_\pl^2}{T}_{ab} + \kappa_1\left[n_aZ_b+n_bZ_a-(1+\kappa_2)g_{ab}n_cZ^c\right],
    \label{eq:classicalEinsteinEqs_damping}
\eeq
where $\kappa_1$ and $\kappa_2$ are parameters that control the constraint damping. The components $Z^a$ are associated with the Hamiltonian and momentum constraints after a $3+1$ decomposition of \eqref{eq:classicalEinsteinEqs_damping}. $n^a$ are the components of the time-like normal-vector, $\mathbf{n}$, orthogonal to the spatial hypersurface $\gamma_{ij}$ in a $3+1$ (ADM) decomposition of the metric, 
\beq
    \dd s^2=-\alpha^2\dd t^2+\gamma_{ij}\dd x^i\dd x^j,
\eeq
where the hypersurface is defined as $\gamma_{ab}\deff g_{ab}+n_an_b$ with $n_a$ satisfying $n^a\gamma_{ai}=0$ and $n^a n_a=-1$.

As a final step to define the classical system, we must fix the redundant gauge dependence of the metric \eqref{eq:lineelement}. The gauge dependence is due to the freedom to choose $\alpha$, which determines how spacetime is sliced into hypersurfaces of constant time. In order to close the system of the Einstein equations, we impose the Bona-Masso slicing condition \cite{Bona:1994dr} via the evolution equation 
\beq
    \dot{\alpha} = -2\alpha^2f(\alpha)[K-2\Theta],
    \label{eq:alphaChoice}
\eeq
where we choose $f(\alpha)=2/\alpha$, often referred to as the ``$1+\log$" condition \cite{Anninos:1995am}. Here $K\deff\gamma^{ij}(-\gamma^a{}_i\gamma^b{}_j\nabla_a n_b)= K_A+2K_B$ is the trace of the extrinsic curvature and $\Theta\deff-n^aZ_a$ is the time-like projection of $Z_a$, which is added to \eqref{eq:alphaChoice} in order to ensure that the system is strongly hyperbolic \cite{Bernuzzi:2009ex}.

\subsection{The classical equations of motion}
\label{Subsec:classical_eoms}

We consider a massless spherically symmetric scalar field, $\Phi(t,r)$, with equation of motion $\Box \Phi=0$. In order to also recast the matter sector of the Einstein equations into a first-order differential form, we introduce the variables
\beq
    \Pi \deff \frac{\sqrt{A}B}{\alpha}\dot{\Phi}, \qquad \Psi\deff \Phi'.
\eeq
The stress-energy tensor is $T_{ab}=\partial_a\Phi\partial_b\Phi-\frac{1}{2}g_{ab}g^{cd}\partial_c\Phi\partial_d\Phi$, where the components in the $3+1$ formulation read 
\begin{subequations}
  \begin{align}
    \rho &\deff n^a n^b T_{ab} = \frac{1}{2A} \left( \frac{\Pi^2}{B^2} + \Psi^2 \right), \\
    j_A  &\deff -n^a T_{ar} = - \frac{\Pi \Psi}{B \sqrt{A}}, \\
    S_A  &\deff \gamma^{rr} T_{rr} = \frac{1}{2A} \left( \frac{\Pi^2}{B^2} + \Psi^2 \right), \\
    S_B  &\deff \gamma^{\theta \theta} T_{\theta \theta} = \frac{1}{2A} \left( \frac{\Pi^2}{B^2} - \Psi^2 \right),
  \end{align}
  \label{eq:stress_tensor_projections}%
\end{subequations}
and $\rho\deff \rho_\text{\tiny ADM}$ is the ADM energy density and we assume that the classical field $\Phi$ is spherically symmetric, i.e. $\partial_\theta\Phi=\partial_\varphi\Phi=0$.
In terms of these variables, the equations of motion for the matter quantities read
\begin{subequations}
  \begin{align}
    \dot{\Phi} &= \frac{\alpha \Pi}{\sqrt{A} B}, \label{eq:Phi_eom}
    \\
    \dot{\Pi} &= \frac{ \alpha B}{\sqrt{A}}\left[ \Psi' + \frac{2 \Psi}{r} + \Psi \left( D_\alpha - \frac{\tilde{U}}{2} - \frac{2 B \lambda}{A} \right) \right],\label{eq:Pi_eom}
    \\
    \dot{\Psi} &= \frac{\alpha}{\sqrt{A} B} \left[ \Pi' + \Pi \left(D_\alpha -\tilde{U} - 3 D_B - \frac{4 B \lambda}{A} \right) \right],
      \label{eq:Psi_eom}
      \end{align}
      \label{eq:classicalEoMs_matter}%
\end{subequations}
and for the gravitational quantities,
\begin{subequations}
  \begin{align}
    \dot{A} &= -2 \alpha A \left( K - 2 K_B \right), \label{eq:A_eom}
    \\
    \dot{B} &= -2  \alpha B K_B, \label{eq:B_eom}
    \\
    \dot{D}_B &= -  2 \alpha \left( K_B' + D_\alpha K_B \right), \label{eq:DB_eom}
    \\
    \dot{\tilde{U}} &= - 2 \alpha \left[ K'  + 2D_B \left(3 K_B - K\right)
      \right.  \nonumber \\
    & \quad \left.
      + D_\alpha
      \left(K - 4 K_B \right) + \frac{4 B \left(-3 K_B + K\right) \lambda}{A}
      +\frac{2 j_A}{M_p^2}\right], \label{eq:Utld_eom}
    \\
    \dot{K}
    & = \frac{\alpha}{A}
      \left(
      \frac{2 B D_\alpha  \lambda }{A}
      - D_\alpha'
      - D_\alpha^2
      - \frac{2 D_\alpha}{r}
      + \frac{D_\alpha \tilde{U} }{2}
      - 2 Z_r'
      \right)
      + 3 \kappa_1\left(1 + \kappa_2 \right) \alpha \Theta
      \nonumber
    \\
    & \quad
      - \alpha \left( \frac{4 Z_r}{r B}
      + 4 K_B K
      - 6 K_B^2
      - K^2
      \right)
      + \frac{\alpha}{M_\pl^2}
      \left( \frac{S_A}{2}
      + S_B 
      - \Lambda
      + \frac{\rho}{2}
      \right)
      ,
      \label{eq:K_eom}
    \\
    \dot{K}_B
    & = 
      \frac{\alpha}{A}
      \left(
      \frac{B D_B \lambda}{A}
      + \frac{2 B \lambda}{r A}
      - \frac{D_\alpha }{r}
      - \frac{D_B' }{2}
      - \frac{D_\alpha D_B }{2}
      - \frac{D_B}{r}
      + \frac{ D_B \tilde{U} }{4}
      - \frac{\lambda}{r}
      + \frac{\tilde{U} }{2 r}
      \right)
      \nonumber \\
    & \quad
      + \alpha
      \left[
      \frac{2 Z_r}{r B}
      + K_B K
      + \frac{1}{2 M_\pl^2}
      \left(
      S_A
      - 2 \Lambda
      - \rho
      \right)
      - \kappa_1 \left(1 + \kappa_2\right) \Theta
      \right]
      ,
      \label{eq:KB_eom}
    \\
    \dot{\lambda}
    & =
      \frac{\alpha}{B}
      \left(
      3 A D_B K_B
      - A D_B K
      + \frac{A j_A}{M_\pl^2}
      + 2 A K_B'
      \right)
      ,
      \label{eq:lambda_eom}
    \\
    \dot{\alpha}
    & =
      \alpha \left( 4 \Theta - 2 K \right)
      ,
      \label{eq:alpha_eom}
    \\
    \dot{D}_\alpha
    & =
      4 \Theta' - 2 K'
      ,
      \label{eq:Dalpha_eom}
    \\
    \dot{\Theta}
    & =
      \alpha
      \left[
      - \frac{H}{A}
      + \frac{Z_r'}{A}
      + \frac{2 Z_r}{r B}
      - \kappa_1 \left(2 + \kappa_2\right) \Theta
      \right]
      ,
      \label{eq:Theta_eom}
    \\
    \dot{Z}_r
    & =
      - 2 \alpha P
      + \alpha \left( \Theta'
      - \kappa_1 Z_r
      \right)
      ,
      \label{eq:Zr_eom}
  \end{align}
  \label{eq:classicalEoMs_grav}%
\end{subequations}
where $Z_r\deff \gamma^a{}_rZ_a$ is the space-like (radial) projection of $Z_a$. In addition, the Hamiltonian and momentum constraints read
\begin{subequations}
  \begin{align}
    H
    & =
      - \frac{2 B D_B \lambda}{A}
      - \frac{4 B \lambda}{r A}
      - 2 A K_B K
      + 3 A K_B^2
      + \frac{A \rho}{M_\pl^2}
      \nonumber \\
    &   \quad + D_B'
     - \frac{1}{2} D_B \tilde{U}
      - \frac{1}{4} D_B^2
      + \frac{D_B}{r}
      + \frac{\lambda}{r}
      - \frac{\tilde{U}}{r}
      \label{eq:our_Ham}
      , \\
    P
    & = -\left(\frac{1}{2} D_B + \frac{1}{r} \right)
      \left( K -3 K_B \right)
      + \frac{j_A}{2 M_\pl^2}
      + K_B'
      \label{eq:our_mom}
      ,
  \end{align}
  \label{eq:our_Ham_mom_constraints}%
\end{subequations}
which are equated to \eqref{eq:HamiltonianConstraintGeneral} and \eqref{eq:MomentumConstrainGeneral}, respectively. When simulating the classical system, we will set the cosmological constant to vanish, i.e. $\Lambda=0$. However, it will be kept non-zero for the semi-classical setup with backreaction described in section \ref{Sec:semiclassical_setup}, where it will act as a counterterm to absorb a divergent contribution from the expectation value of the quantised stress-energy tensor. In summary, the variables of the system can be collected in the state vector
\beq
    \mathbf{u} = \left(\Phi,\Pi,\Psi,A,B,D_B,\tilde{U},K,K_B,\lambda,\alpha,D_\alpha,\Theta,Z_r\right)^T,
    \label{eq:state_vector}
\eeq
consisting of matter and gravitational quantities, complemented with $\Theta$ and $Z_r$, which control the violation of the Hamiltonian and momentum constraints, respectively. Here, we have dropped the explicit dependence of the variables on the $t,r$ co-ordinates, for convenience.

\subsection{Classical initial data and boundary conditions}
\label{Subsec:classical_ID_BC}
To solve the evolution equations in \eqref{eq:classicalEoMs_matter} and \eqref{eq:classicalEoMs_grav} we must also provide initial data and boundary conditions. We look for a solution in a domain of interest $\mathcal{D}$ by specifying a time and radius interval for our simulation,
\beq
    \mathcal{D} = \left\{ \left(t,r\right) \in \mathbb{R}^2 : 0 \geq
  t \geq t_\text{\tiny max} , 0 \geq r \geq r_\text{\tiny max} \right\}.
\eeq
The elements of the state vector $\mathbf{u}$ are then provided with initial data at $(t=0,r)$ and boundary conditions at $(t,r=r_\text{\tiny max})$.

For the matter sector, we provide as initial data for the scalar field $\Phi(0,r)$ a Gaussian profile of amplitude $a$, centred symmetrically at a radial distance $b$ from the origin $r=0$, with a width $c$, i.e.,
\beq    
    \Phi(0,r) = a \left[e^{-\frac{(r-b)^2}{c^2}} + e^{-\frac{(r+b)^2}{c^2}}\right].
  \label{eq:Phi_ID}
\eeq
The two Gaussian functions centred at $r=\pm b$ are used to ensure symmetry of $\Phi(0,r)$ around the origin $r=0$. The remaining scalar variables are initiated as 
\beq
    \Psi(0,r)=\Phi'(0,r), \quad \Pi(0,r)=0.
    \label{eq:Psi_Pi_ID}
\eeq
The initial data parameters $a, \, b$ and $c$ for the classical scalar field profile determine whether a black hole forms in our simulations. Specifically, we interpret black hole formation as the dynamic emergence of an apparent horizon. The apparent horizon is located at the outermost radius where the expansion of null rays $\theta_\text{\tiny exp} = \frac{1}{\sqrt{A}} \left( \frac{2}{r} + \frac{B'}{B} \right)- 2 K_B $ vanishes. The choices of $a, \, b$ and $c$  that lead to black hole formation are referred to as \textit{supercritical data}, while those that do not result in black hole formation are called \textit{subcritical data}.

On the gravitational side, the geometric quantities are initialised as 
\beq
    \alpha(0,r) &= B(0,r) = 1, \\
    D_\alpha(0,r) &= D_B(0,r)=K(0,r)=K_B(0,r)=0,
    \label{eq:initialCondsGeoQuants}
\eeq
for which the initial data for $K,\ K_B$ and $\Pi$ trivially satisfy the momentum constraint \eqref{eq:our_mom}. To satisfy the Hamiltonian constraint \eqref{eq:our_Ham}, we start by setting $A(0,0)=1$. With the choices \eqref{eq:initialCondsGeoQuants}, the non-zero variables in \eqref{eq:our_Ham} simplify to
\beq
    \tilde{U}(0,r) = \frac{1}{A(0,r)}\left[A'(0,r)-4\lambda(0,r)\right],
    \quad 
    \lambda(0,r) = \frac{1}{r}\left[1-A(0,r)\right],
    \quad 
    \rho(0,r)=\frac{\Psi^2(0,r)}{2A(0,r)},
    \nonumber
\eeq
where we have used the definitions in \eqref{eq:FirstOrderGravVariables}. With these initial values, the Hamiltonian constraint takes the form
\beq
    A'(0,r) = A(0,r)\left[ \frac{1-A(0,r)}{r}+\frac{\Psi^2(0,r)}{2M_\pl^2}r \right].
    \label{eq:ACond_initialHamiltionan}
\eeq
The initial value for $A(0,r)$ is then obtained by integrating \eqref{eq:ACond_initialHamiltionan} over the entire radial domain from $r=0$ to $r_{\max}$.  Finally, the variables related to the constraint damping are initialised as
\beq
    \Theta(0,r)=Z_r(0,r)=0.
\eeq

We understand the spatial boundary at $r=r_\text{\tiny max}$ as a limitation of our numerical setup rather than a physical boundary. Consequently, we choose to prescribe outgoing boundary conditions at this value. This can be achieved by using the hyperbolicity analysis in appendix \ref{app:classical_hyp} and setting the ingoing characteristic variables of the system to zero. By solving for $K,\, K_B$ and $D_B$, we obtain 
\beq
    D_B &= \sqrt{A}\Theta+Z_r, \\
    K_B &= \frac{\sqrt{A}\Theta+Z_r}{2\sqrt{A}}, \\
    K   &= -\frac{1}{\sqrt{2A}(\alpha-2)}\left[4\sqrt{2A}\Theta+2D_\alpha\sqrt{a}-4\sqrt{A\alpha}\Theta+4Z_r\sqrt{\alpha}-2\sqrt{2}Z_r\alpha-D_\alpha\alpha^{3/2}\right].
    \nonumber
\eeq
where the argument $(t,r_{\max})$ of all ingoing quantities is suppressed.
For the scalar field, we set radiative (Sommerfeld) boundary conditions \cite{alcubierre2008introduction} by setting
\begin{align*}
  \dot{\mathcal{F}}(t,r_\max) = - \left[ v_\text{\tiny out} \mathcal{F}'(t,r) + \frac{\mathcal{F}(t,r)}{r}\right]_{r=r_\max}, \quad \mathcal{F}=\{\Phi,\Psi,\Pi\},
\end{align*}
for $t \in[0,t_{\max}]$ and $v_\text{\tiny out}=\alpha/\sqrt{A}$ is the speed of the relevant outgoing characteristic variable defined in \eqref{eq:Jordan_form_Ar}.

\section{The semiclassical setup} \label{Sec:semiclassical_setup}

In the semiclassical approximation, the matter fields are promoted to quantum operators $\Phi\to\hat\Phi$, while the gravitational degrees of freedom are treated classically, being sourced by the expectation value of the stress tensor. This approximation holds on large spacetime length scales where quantum gravitational effects are negligible compared to those of the matter sector.\footnote{Specifically, it is valid when the characteristic length scales of the spacetime are much larger than the Planck length, such as when the Ricci scalar satisfies $R\ll 1/l_\pl^2$. For spacetimes describing collapsing objects like stars or black holes, this condition is met when the mass of the object is much larger than the Planck mass $M_\pl$.}
In addition, during the gravitational collapse of a scalar field to a black hole, the curvature increase is expected to be below what is necessary to render quantum-gravity effects important \cite{Cardoso:2022eqx}.

In the semiclassical framework the Einstein field equations take the form
\beq
    G_{ab} = \frac{1}{M_\pl^2}\left[\av{\chi|\hat{T}_{ab}|\chi} -\Lambda g_{ab}\right],
    \label{eq:semiclassicalEinsteinEqs}
\eeq
where $\hat{T}_{ab}\deff \partial_a\hat{\Phi}\partial_b\hat{\Phi}-\frac{1}{2}g_{ab}g^{cd}\partial_c\hat{\Phi}\partial_d\hat{\Phi}$ and $\ket{\chi}$ is the quantum state.
As we work in a spherically symmetric geometry, it is convenient to expand the quantum field in a basis of spherical harmonics $Y^m_l(\theta,\phi)$ as
\beq
    \hat{\Phi}(t,r,\theta,\varphi) = \sum_{l,m}\int\dd k\left[\hat{a}_{klm}\tilde{u}_{kl}(t,r)Y^m_l(\theta, \varphi) + \hat{a}_{klm}^\dagger\tilde{u}^*_{kl}(t,r)Y^{m*}_l(\theta, \varphi)\right],
    \label{eq:mode_expansion}
\eeq
in terms of a set of mode functions $\tilde{u}_{kl}(t,r)$. We choose the quantum state to be a coherent state $\ket{\chi}$~\cite{Berczi:2020nqy, Berczi:2021hdh}, which has the property that the expectation value of the quantum field operator satisfies the classical field equations and so we call it the classical field and denote it $\phi$, i.e. $\phi(t,r)\deff \av{\chi|\hat{\Phi}(t,r)|\chi}$. In this case, $\ket{\chi}$ is an eigenstate of the annihilation operator $\hat{a}_{klm}$, such that $\hat{a}_{klm}\ket{\chi}=\mathcal{E}\ket{\chi}$, where $\mathcal{E}$ represents the eigenvalue. The annihilation operator is defined to annihilate the vacuum state $\ket{0}$, i.e. $a_{klm}\ket{0}=0$. In a dynamic spacetime, the vacuum state is ambiguous and we choose to define it by the Minkowski mode functions of the asymptotic past. In practice, this amounts to defining the initial quantum data using the Minkowski spacetime solutions, as described in section \ref{Subsec:quantum_ID_BC}.

\subsection{Quantum equations of motion}
\label{Subsec:quantum_eoms}

Analogous to the classical setup, we formulate the equations of motion in a first-order form, in terms of the first temporal and spatial derivatives of the field $\hat{\Phi}$. In addition, it proves convenient to work with the rescaled mode functions
\beq
    u_{kl}(t,r) \deff \frac{\tilde{u}_{kl}(t,r)}{r^l}.
    \label{eq:u_kl_def}
\eeq
We define the conjugate momentum and spatial derivative variables as
\beq
    \pi_{kl}(t,r) \deff \frac{\sqrt{A} B}{\alpha} \dot{u}_{kl}(t,r),
    \quad
    \psi_{kl}(t,r) \deff u_{kl}'(t,r),
  \label{eq:pi_psi_kl_def}
\eeq
so that the equations of motion take the form
\begin{subequations}
    \begin{align}
    \dot{u}_{kl}
    & =
      \frac{\alpha}{\sqrt{A}B} \pi_{kl},
      \label{eq:u_quantum_evol}
    \\
    \dot{\psi}_{kl}
    & =
      \frac{\alpha}{\sqrt{A}B}
      \left[
      \left( D_\alpha - \frac{D_A}{2} - D_B \right) \pi_{kl}
      + \pi_{kl}' \right],
      \label{eq:psi_quantum_evol}
    \\
    \dot{\pi}_{kl}
    &=
      \frac{\alpha B}{\sqrt{A}}
      \left[
      \left( D_\alpha + D_B - \frac{D_A}{2} \right)
      \left( \frac{l}{r} u_{kl} + \psi_{kl} \right) 
      + \frac{2(l+1)}{r}\psi_{kl} + \psi_{kl}' \right]
      \nonumber
    \\
    &
      \quad + \lambda\frac{l(l+1)}{r}\frac{\alpha B}{\sqrt{A}} u_{kl}
      - \mu^2\alpha \sqrt{A} B u_{kl},
      \label{eq:pi_quantum_evol}
    \end{align}
    \label{eq:quantum_evol_eqs}%
\end{subequations}
where the argument $(t,r)$ has been suppressed and we have included a mass, $\mu$, for the quantum field. We do this because we will soon need to introduce Pauli-Villar ghost fields in order to regularize the stress-energy tensor and these ghost fields will be massive. We also define 
\beq
    D_A \deff \tilde{U} + 2 D_B + \frac{4 \lambda B}{A}.
\eeq

\subsection{Regularized Einstein equations}
\label{Subsec:RegSEtensor}

With a quantised matter sector, the scalar field quantities entering the stress-energy tensor in \eqref{eq:stress_tensor_projections} get replaced with quantum expectation values. Concretely, this entails the substitutions
\beq
    \Phi^2 &\rightarrow \av{\chi|\hat{\Phi}^2|\chi}, \\
    \Pi^2 &\rightarrow \av{\chi|\hat{\Pi}^2|\chi}\deff \frac{AB^2}{\alpha^2}\av{\chi|(\p_t\hat{\Phi})^2|\chi}, \\ 
    \Pi\Psi &\rightarrow \av{\chi|\hat{\Pi}\hat{\Psi}|\chi}\deff \frac{\sqrt{A}B}{\alpha}\av{\chi|\p_t \hat{\Phi}\p_r\hat{\Phi}|\chi},
\eeq
and so on. 

The expectation values of the stress-energy tensor contain UV divergences that are directly related to the local curvature of spacetime. These divergences can therefore be regularized using purely geometrical quantities, so that the regularized stress-energy tensor can be expressed as 
\beq
    \av{\chi|\hat{T}_{ab}|\chi}_\text{reg} = \av{\chi|\hat{T}_{ab}|\chi}-\Lambda g_{ab}-\delta M_\pl^2 G_{ab}-\mathcal{O}(g_{ab}R^2,R_{ac}R^c{}_b,R_{acdf}R^{fdc}{}_b),
    \label{eq:Treg}
\eeq
for some regularization scheme. Since the lattice breaks diffeomorphism invariance, implementing the continuum curvature-invariant counterterms of \eqref{eq:Treg} becomes challenging in practice. One way to address this issue is through Pauli-Villars regularization\footnote{Recently, new point-splitting techniques for numerical simulations in this context have been proposed in \cite{Levi:2015eea,Levi:2016esr}.}, which introduces fictitious heavy fields to cancel out the UV divergences while conserving covariance. By incorporating Pauli-Villars fields alongside the scalar field $\hat{\Phi}$ in the simulation, the regularization occurs at the integrand level in momentum space, ensuring that all quantities computed on the lattice are already regularized, meaning that it is not the lattice that provides the regularization.

When introducing Pauli-Villars fields with masses $\mathcal{O}(M_\pv)$, much larger than any physical mass scales or inverse length scales, the Pauli-Villars mass $M_\pv$ effectively works as a UV cutoff. On dimensional grounds we then expect that the counterterms in \eqref{eq:Treg} to depend on the Pauli-Villars mass as
\beq
    \Lambda\sim M_\pv^4, \quad \delta M_\pl^2\sim M_\pv^2,
    \label{eq:MPVpowerdepCounterterms}
\eeq
while the coefficients of the higher-order curvature invariants in \eqref{eq:Treg} should display at most a logarithmic dependence of $M_\pv$. In order to ease our computational efforts, we may set the renormalisation condition, for the range of Pauli-Villars masses we consider, to be that the coefficients of these higher-order geometrical terms vanish. Then we need only determine the counterterms necessary to cancel the quartic and quadratic divergences in \eqref{eq:MPVpowerdepCounterterms}.

While a single Pauli-Villars field is sufficient to cancel the original divergences in the stress-energy tensor components caused by the physical field $\hat{\Phi}$, the added Pauli-Villars field introduces new divergences proportional to the Pauli-Villars mass $M_\pv$. As described in detail in \cite{Berczi:2021hdh}, it turns out that adding five Pauli-Villars ghost fields is sufficient to fully regularize the system. In order to incorporate the Pauli-Villars ghost fields, the matter Lagrangian can be extended to
\beq
    \mathcal{L}_\text{\tiny matter} =  - \sum_{n=0}^5(-1)^n\left[\frac{1}{2}\p_\mu \hat{\Phi}_n\p^\mu \hat{\Phi}_n+\tfrac{1}{2}\mu_n^2\hat{\Phi}_n\right],
    \label{eq:PVextendedLagrangian}
\eeq
where the index choice $n=0$ corresponds to the physical field $\hat{\Phi}_0\deff \hat{\Phi}$, while $n=1,\dots,5$ denote the ghost fields. The mode expansion of the extended scalar sector is analogous to the single-field case in \eqref{eq:mode_expansion}, i.e., 
\beq
    \hat{\Phi}_n(t,r) = \sum_{l,m}\int\dd k\left[\hat{a}_{klm;n}\tilde{u}_{kl;n}(t,r)Y^m_l(\theta, \varphi) + \hat{a}_{klm;n}^\dagger\tilde{u}^*_{kl;n}(t,r)Y^{m*}_l(\theta, \varphi)\right].
    \label{eq:extended_mode_expansion}
\eeq
The mass of the physical field is $\mu_0= 0$ and the masses of the Pauli-Villars fields can be set to 
\beq
    \mu_1=\mu_3=M_\pv, \quad \mu_2=\mu_4=\sqrt{3}M_\pv, \quad \mu_5=2 M_\pv,
    \label{eq:PVmasses}
\eeq
so as to regulate the theory. 

Due to diffeomorphism invariance of the theory and the Pauli-Villar regulator, any stress tensor counterterm must be expressible as metric/curvature tensors, or their covariant derivatives. The coefficients of these terms may be discovered by computing the stress tensor on different backgrounds. For example, putting the quantum field on Minkowski spacetime allows us to compute the coefficient of the $g_{ab}$ counterterm. However, this geometry will not allow us to compute the $G_{ab}$ counterterm, because $G_{ab}$ vanishes for Minkowski. One way to compute the $G_{ab}$ counterterm is to put the quantum field on a cosmological FRW geometry and if we consider a slowly expanding geometry then we may perform a WKB analysis to compute the coefficient. This calculation is performed in Appdendix \ref{app:counterterms} and we find the general form of the expectation value of the stress-energy tensor to be
\beq
    \av{\chi|\hat{T}_{ab}|\chi} = \text{finite} - \ln(\frac{3^9}{12^{16}})\frac{M_\pv^4}{8(2\pi)^2}g_{ab}+\ln(\frac{2^4}{3^3})\frac{M_\pv^2}{12(2\pi)^2}G_{ab}+\mathcal{O}(g_{ab}R^2,\dots).
    \label{eq:TregExplicit}
\eeq
This then sets the counterterms in \eqref{eq:MPVpowerdepCounterterms} to
\beq
    \Lambda = \ln(\frac{3^9}{12^{16}})\frac{M_\pv^4}{8(2\pi)^2}, \quad 
    \delta M_\pl^2=-\ln(\frac{2^4}{3^3})\frac{M_\pv^2}{12(2\pi)^2}.
    \label{eq:CountertermsEq}
\eeq
With these counterterms, the effective cosmological constant is set to recover Minkowski spacetime when the matter sector corresponds to vacuum (i.e. when $\phi\deff \av{\hat{\Phi}}=0$).
In addition, the Planck mass counterterm $\delta M_\pl^2$ allows us to define an effective Planck mass
\beq
     M_{\pl,\text{\tiny eff}}^2\deff M_{\pl}^2-\ln(\frac{2^4}{3^3})\frac{M_\pv^2}{12(2\pi)^2},
\eeq
which will be used in the simulation when backreaction is considered.

With the extended matter sector of \eqref{eq:PVextendedLagrangian}, the quantum expectation values entering the stress-energy tensor quantities are
\begin{subequations}
    \begin{align}
        \av{\chi|\hat{\Phi}^2|\chi} &= \phi^2+\frac{\hbar c^2}{4\pi}\sum_{n=0}^5\sum_l\int \dd k\,(-1)^n (2l+1)|\tilde{u}_{kl;n}|^2, \label{eq:twopointcorrelator}\\
        \av{\chi|(\p_t\hat{\Phi})^2|\chi} &= (\p_t\phi)^2+\frac{\hbar c^2}{4\pi}\sum_{n=0}^5\sum_l\int \dd k\, (-1)^n (2l+1)|\p_t\tilde{u}_{kl;n}|^2, \\
        \av{\chi|(\p_r\hat{\Phi})^2|\chi} &= (\p_r\phi)^2+\frac{\hbar c^2}{4\pi}\sum_{n=0}^5\sum_l\int \dd k\, (-1)^n (2l+1)|\p_r\tilde{u}_{kl;n}|^2, \\
        \av{\chi|\p_t\hat{\Phi} \p_r\hat{\Phi}|\chi} &= \p_t\phi\p_r\phi+\frac{\hbar c^2}{4\pi}\sum_{n=0}^5\sum_l\int \dd k\, (-1)^n (2l+1)\tfrac{1}{2}\left[\p_t\tilde{u}_{kl;n}\p_r\tilde{u}_{kl;n}^* +
        \p_r\tilde{u}_{kl;n}\p_t\tilde{u}_{kl;n}^*\right], \\
        \av{\chi|(\p_\theta\hat{\Phi})^2|\chi} &= (\p_\theta\phi)^2+\frac{\hbar c^2}{4\pi}\sum_{n=0}^5\sum_l\int \dd k\, (-1)^n \tfrac{1}{2}l(l+1)(2l+1)|\tilde{u}_{kl;n}|^2,
    \end{align}
    \label{eq:bilinears}%
\end{subequations}
where we have defined $\hat{\mathcal{O}}\deff\sum_{n=0}^5\hat{\mathcal{O}}_n$ and the display of $\hbar$ emphasizes the quantum contributions. The temporal and spatial derivatives of $\tilde{u}_{kl;n}$ appearing in \eqref{eq:bilinears} are related to the rescaled variable $u_{kl;n}$ of \eqref{eq:u_kl_def} as
\beq
    \p_t\tilde{u}_{kl;n} = r^l\frac{\alpha}{\sqrt{A}B}\pi_{kl;n}, \quad \p_r\tilde{u}_{kl;n} = lr^{l-1}u_{kl;n}+r^l\psi_{kl;n}.
    \label{eq:u_tilde_redef}
\eeq
By using the above expectation values in the expressions for the projections $\rho, j_A, S_A$ and $S_B$ of the stress-energy tensor in \eqref{eq:stress_tensor_projections}, the new components of the semi-classical stress-energy tensor separates into classical and quantum contributions as
\begin{subequations}
\begin{align}
    \rho &= \rho_\cl + \frac{\hbar c^2}{4\pi} \rho_\qu, \\
    j_A  &= j_{A,\cl} + \frac{\hbar c^2}{4\pi} j_{A,\qu}, \\
    S_A  &= S_{A,\cl} + \frac{\hbar c^2}{4\pi} S_{A,\qu}, \\
    S_B  &= S_{B,\cl} + \frac{\hbar c^2}{4\pi} S_{B,\qu}.
\end{align}
\end{subequations}
The classical components are given as in \eqref{eq:stress_tensor_projections} and the quantum components are defined as
\begin{subequations}
\begin{align}
    \rho_\qu  &\deff \sum_{n=0}^5\sum_l\int\dd k\, (-1)^n\bigg[\frac{2l+1}{2\alpha^2}|\p_t\tilde{u}_{kl;n}|^2+\frac{2l+1}{2A}|\p_r\tilde{u}_{kl;n}|^2 \nonumber\\
    &\quad+\frac{l(l+1)(2l+1)}{2B}|r^{l-1}\tilde{u}_{kl;n}|^2+\frac{\mu_n^2}{2}|\tilde{u}_{kl;n}|^2\bigg],
    \label{eq:rho_quantum}
    \\
    j_{A,\qu} &\deff -\sum_{n=0}^5\sum_l\int\dd k\, (-1)^n \frac{2l+1}{2\alpha}\left[ \p_t\tilde{u}_{kl;n}\p_r\tilde{u}_{kl;n} + \p_r\tilde{u}_{kl;a}\p_t\tilde{u}_{kl;n} \right], 
    \label{eq:jA_quantum}
    \\
    S_{A,\qu} &\deff  \sum_{n=0}^5\sum_l\int\dd k\, (-1)^n \bigg[\frac{2l+1}{2\alpha^2}|\p_t\tilde{u}_{kl;n}|^2+\frac{2l+1}{2A}|\p_r\tilde{u}_{kl;n}|^2 \nonumber\\
    &\quad-\frac{l(l+1)(2l+1)}{2B}|r^{l-1}\tilde{u}_{kl;n}|^2-\frac{\mu_n^2}{2}|\tilde{u}_{kl;n}|^2\bigg], 
    \label{eq:SA_quantum}
    \\
    S_{B,\qu} &\deff  \sum_{n=0}^5\sum_l\int\dd k\, (-1)^n \bigg[\frac{2l+1}{2\alpha^2}|\p_t\tilde{u}_{kl;n}|^2+\frac{2l+1}{2A}|\p_r\tilde{u}_{kl;n}|^2-\frac{\mu_n^2}{2}|\tilde{u}_{kl;n}|^2\bigg].
    \label{eq:SB_quantum}
\end{align}
\label{eq:stress_tensor_projection_quantum}
\end{subequations}

\subsection{Quantum initial data and boundary conditions}
\label{Subsec:quantum_ID_BC}

The quantum evolution equations must be specified with appropriate initial conditions. Minkowski space initial data provides a natural starting point, as it reflects the asymptotically flat spacetime typical of realistic gravitational collapse scenarios. With this choice, the mode solution to the equation of motion $\Box \hat{\Phi}_n=\mu^2_n\hat\Phi$ is given by
\beq
    u^{\text{\tiny Mink}}_{kl;n}(t,r) = \frac{k}{\sqrt{\pi \omega_n}}  \frac{j_l(k r)}{r^l}e^{-i \omega_n t},
\eeq
where $\omega_n=\sqrt{k^2+\mu_n^2}$ and $j_l(kr)$ are the spherical Bessel functions of the first kind. The initial conditions for the scalar and Pauli-Villars fields at $t=0$ are then given by
\begin{subequations}
\begin{align}
  u_{kl;n}(0,r)
  & = \frac{k}{\sqrt{\pi \omega_n}} \frac{j_l(k r)}{r^l},
    \label{eq:u_klmu_ID}
  \\
  \psi_{kl;n}(0,r)
  & = \frac{k}{\sqrt{\pi \omega_n}}
    \left[ \frac{\p_r j_l(kr)}{r^l} - \frac{l j_l(kr)}{r^{l+1}} \right],
    \label{eq:psi_klmu_ID}
  \\
  \pi_{kl;n}(0,r)
  & = - i \omega_n \frac{k}{\sqrt{\pi \omega_n}} \frac{j_l(kr)}{r^l}.
    \label{eq:pi_klmu_ID}
\end{align}
\label{eq:quantum_ID}%
\end{subequations}
The boundary conditions for the quantum modes at $r_\max$ are the same as those for the classical scalar fields, namely, radiative.

\section{Numerical
  implementation} \label{Sec:numerical_implementation}

To simulate the system described by \eqref{eq:classicalEoMs_matter}, \eqref{eq:classicalEoMs_grav} and \eqref{eq:quantum_evol_eqs}, we discretise it on a uniform grid for the radial co-ordinate $r\in [0,r_\tinytext{max}]$. In order to take derivatives at $r=0$ using finite-difference operators to second-order accuracy, as described in section \ref{Subsec:FD_operators}, we include two ghost points on the radial grid. Given a maximal radial co-ordinate $r_\tinytext{max}$ and number of radial points $N_r$, the grid spacing is $h = r_\tinytext{max}/(N_r-3)$, where the two ghost points are located at the left of the origin $r=0$, such that the radial grid is $\{-2 \, h, -h, 0, h, \dots, (N_r-3) \, h\}$.

When simulating fields of a given wavelength $\tilde{\lambda}$, we must require that the radial grid spacing is much smaller than the smallest wavelength, i.e. $h\ll \tilde{\lambda}_\text{\tiny min}\deff2\pi/ k_\text{\tiny max}$, in order for the lattice to resolve it. For the discrete time step $\dd t$ we require that $(\dd t)^{-1} \gg \omega_\text{\tiny max}\deff \sqrt{k_\text{\tiny max}^2+(\mu_5)^2}$, where $\mu_5=2M_\pv$ is the mass of the heaviest Pauli-Villars ghost field, in order to have enough resolution in time to resolve the fastest oscillations in the simulation. For the Pauli-Villars masses, we need $M_\pv\gg2\pi/\tilde{\lambda}$, so that the ghost fields do not influence the dynamics of the physical field. In practise, to calculate $\tilde{\lambda}$ for the physical field, we consider twice the radial distance between the maximum of $\Phi(0,r)$ and the location where it has dropped to $5\%$ of the maximum.

\subsection{Finite difference
  operators} \label{Subsec:FD_operators}

We denote as~$f_i = f(r_i)$ the discrete approximation of a
continuous function $f(r)$ at the point $r=r_i$ of our radial grid. At the beginning of every timestep, we populate the ghost points via
\begin{align*}
  f(-h) = f(h)
  \,,
  \quad
  f(-2h) = f(2h)
  \,,
\end{align*}
for the even variables and with
\begin{align*}
  f(-h) = -f(h)
  \,,
  \quad
  f(-2h) = -f(2h)
  \,,
\end{align*}
for the odd ones. We use the centred finite difference (FD) operator
\begin{align}
  D_r f_i \deff \frac{f_{i+1} - f_{i-1}}{2 h}
  \,,
  \label{eq:standard_FD_bulk}
\end{align}
to approximate~$\p_r f(r_i)$ with error~$\mathcal{O}(h^2)$ for the grid
points~$\{0, h, \dots, (N_r-4) \, h \}$. At the last grid
point~$r_{\max} = (N_r-3) \,  h$, we use the backward finite difference operator
\begin{align}
  D_r f_{N_r-3} \deff \frac{f_{N_r-3} - f_{N_r-4}}{h}
  \,,
  \label{eq:standard_FD_rmax}
\end{align}
that is accurate to first order. We refer to~$D_r$ as the
\textit{standard} FD operator and use it to replace all the first
order radial derivatives that appear on the right-hand-side (RHS) of
our evolution equations.

Both in the classical \eqref{eq:classicalEoMs_matter}--\eqref{eq:classicalEoMs_grav} and
quantum~\eqref{eq:quantum_evol_eqs} evolution systems there are
terms~$ \sim g/r$, with~$g$ a regular odd function. Even though these
terms are regular, they can lead to code instabilities that
develop in the region around~$r=0$. This problem is especially difficult
in the quantum system as the mode number~$l$ increases. For very
large~$l$ the differential equation actually becomes stiff, as
the~$l/r$ terms in \eqref{eq:pi_quantum_evol} dominate over the derivative
terms in size near~$r=0$. Stiff systems can be evolved by using
appropriate time integrators such as implicit-explicit Runge-Kutta (RK)
schemes~\cite{IMEX}. Here, we take a different approach to mitigate
this problem by using the second-order accurate summation-by-parts
(SBP) operators of~\cite{Gundlach:2010et}. We justify
this choice in section \ref{Subsec:1/r_terms}. Denoting these second-order SBP 
operators as~$\tilde{D}_r(l)$, they approximate a radial derivative of an odd function $g(r)$ that is regular
at~$r=0$ according to
\begin{align*}
  \tilde{D}_r(l) \, g_i =  g'(r_i) + \frac{l \, g(r_i)}{r_i} + \mathcal{O}(h^2)
  \,,
\end{align*}
where $l$ is a positive integer. The operation~$\tilde{D}_r(l) \, g_i$ is calculated via
\begin{align*}
  \tilde{D}_r(l) \,  g_i \deff \frac{\textrm{w}_{i+1} \, g_{i+1}
  - \textrm{w}_{i-1} \, g_{i-1}}{\textrm{w}_i \, 2  h}
  \,,
\end{align*}
for the grid points~$\{h, \dots , (N_r-4) \, h \}$, where $\textrm{w}_i$ are yet-to-be determined weight functions. Since $g$ vanishes on the grid
point~$r_0 = 0$, we can use the l'Hôpital rule, such that ~$\tilde{D}_r(l) \, g_0 = (1+l) \,  g'(r_0)$, or
\begin{align*}
  \tilde{D}_r(l) \, g_0 \deff (1+l) \frac{g_1 - g_{-1}}{2 h}
  \,.
\end{align*}
Finally, at the last gridpoint~$r_{\max} = (N_r-3) \, h$ we apply the
backward SBP operator
\begin{align*}
  \tilde{D}_r(l) \, g_{N_r-3} \deff \frac{\textrm{w}_{N_r-3} \, g_{N_r-3}
  - \textrm{w}_{N_r-4} \, g_{N_r-4}}{\textrm{w}_{N_r-3} \, h}
  \,,
\end{align*}
which is first-order accurate. To complete this calculation, we need
the weights~$\textrm{w}_i$. To obtain them we first
define~$\textrm{w}_i \deff i^l \,\bar{\textrm{w}}_i$. Then, we set
\begin{align*}
  \bar{\textrm{w}}_0 = \frac{l!}{2^l}
  \,, \quad
  \bar{\textrm{w}}_1 = (1+l) \bar{\textrm{w}}_0
  \,,
\end{align*}
and solve the recursive relation
\begin{align*}
  \bar{\textrm{w}}_i = \frac{2 (l+1)}{i} \left(1 - \frac{1}{i} \right)^l \bar{\textrm{w}}_{i-1}
  +  \left(1 - \frac{2}{i} \right)^{l+1} \bar{\textrm{w}}_{i-2}
  \,,
\end{align*}
for~$2 \leq i \leq N_r-3$. Note that these weights are different for
different values of~$l$. The interested reader can find more details
on this and higher-order accurate SBP operators
in~\cite{Gundlach:2010et}.

\subsection{Approximating~$\sim 1/r$
  terms} \label{Subsec:1/r_terms}

The SBP operators of~\cite{Gundlach:2010et} were developed for the
wave equation on a flat spacetime (of arbitrary dimensions), which
after a spherical harmonic decomposition and a first-order reduction provides the system
\begin{align}
  \dot{\psi} = \pi' \,, \quad
  \dot{\pi} = \psi' + p \frac{\psi}{r}
  \,,
  \label{eq:flat_wave_sys}
\end{align}
where~$p$ is a positive integer that combines the spherical harmonic
index and the space dimension.\footnote{In \cite{Gundlach:2010et}, $p=2l+n$, where $n+1$ is the dimension of space and $l$ is a spherical harmonic index.} This evolution system admits the energy
\begin{align}
  E = \frac{1}{2} \int_0^R\dd r \left(\pi^2 + \psi^2 \right) r^p ,
  \label{eq:flat_wave_energy}
\end{align}
so well-posedness of the associated initial boundary value problem
(with appropriate initial and boundary data) can be shown using this
energy. The SBP operators of~\cite{Gundlach:2010et} are constructed in
a way that they respect a discrete version of this
energy. Consequently, these operators guarantee the stability of the
numerical simulation without the need of artificial
dissipation for the system \eqref{eq:flat_wave_sys}. In~\cite{Gundlach:2010et} the theoretically expected
convergence rate was demonstrated for second- and fourth-order SBP
operators, in a discretized version of the energy
norm~\eqref{eq:flat_wave_energy} as well as pointwisely for the
rescaled variables $r^{p/2} \pi$ and $r^{p/2} \psi$.

The main motivation for choosing these specific SBP operators is that
the system \eqref{eq:flat_wave_sys} matches our evolved quantum
system \eqref{eq:psi_quantum_evol}--\eqref{eq:pi_quantum_evol} for a
flat background, with~$\mu=0$ and $p=2(l+1)$. However, since we are interested in non-flat spacetimes, we extrapolate
the use of these SBP operators. In particular, we combine~$D_r$ and~$\tilde{D}_r(l)$ to replace
all terms~$\sim l g(r)/r$ in the RHS of our evolution equations -- both
classical and quantum -- as
\begin{align*}
  \tilde{D}_r(l) \, g_i - D_r g_i = \frac{l \, g(r_i)}{r_i} + \mathcal{O}(h^2)
  \,.
\end{align*}
For the classical system \eqref{eq:classicalEoMs_matter}--\eqref{eq:classicalEoMs_grav}, we perform the
following replacements:
\begin{equation}
    \begin{aligned}
    \tilde{D}_r(1) \Psi_i - D_r \Psi_i &\simeq \frac{\Psi(r_i)}{r_i},\\
    \tilde{D}_r(1) \lambda_i - D_r \lambda_i &\simeq \frac{\lambda(r_i)}{r_i}, \\
    \tilde{D}_r(1)  D_{\alpha,i} - D_r D_{\alpha,i} &\simeq \frac{D_\alpha (r_i)}{r_i},
    \end{aligned}
    \qquad 
    \begin{aligned}
    \tilde{D}_r(1)  Z_{r,i} - D_r Z_{r,i} &\simeq \frac{Z_r(r_i)}{r_i},\\
    \tilde{D}_r(1) D_{B,i} - D_r D_{B,i} &\simeq \frac{D_B (r_i)}{r_i}, \\
    \tilde{D}_r(1) \tilde{U}_i - D_r \tilde{U}_i &\simeq \frac{\tilde{U}(r_i)}{r_i}.
    \end{aligned}
    \label{eq:classical_SBP_replacements}
\end{equation}
where $\simeq$ denotes equality up to $\mathcal{O}(h^2)$ errors. We made
the choice to use~$\tilde{D}_r(1)$ in all the classical equations,
even though in some cases~$\tilde{D}_r(2)$ would be more appropriate,
because empirically it seemed that otherwise the calculation of the
Hamiltonian constraint~\eqref{eq:our_Ham} was somewhat noisy (although
the noise had a very small amplitude and it did not ruin the stability
of the simulation). We suspect that this is related to the fact that
the weights~$\textrm{w}_i$ have different values for
different~$l$. According to~\cite{Gundlach:2010et},
using~$\tilde{D}_r(1)$ on a centered grid is identical to the Evans
method~\cite{Evans_PhD}, which has been used previously in critical
collapse studies~\cite{SuarezFernandez:2020wqv}.

In the evolution system for the quantum modes~\eqref{eq:quantum_evol_eqs}, we make the replacements
\beq
    \tilde{D}_r(l+1) \psi_{kl;n ,i} - D_r  \psi_{kl;n,i} & \simeq \frac{(l+1)\psi_{kl;n,i}(r_i)}{r_i},
\eeq
and
\begin{equation}
    \begin{aligned}
    \tilde{D}_r(l) D_{A,i} - D_r D_{A,i} &\simeq \frac{l D_A(r_i)}{r_i}, \\
    \tilde{D}_r(l) D_{B,i} - D_r D_{B,i} &\simeq \frac{l D_B (r_i)}{r_i}, 
    \end{aligned}
    \qquad 
    \begin{aligned}
    \tilde{D}_r(l) D_{\alpha, i} - D_r D_{\alpha, i} &\simeq \frac{l D_\alpha (r_i)}{r_i},\\
    \tilde{D}_r(l) \lambda_i - D_r \lambda_i & \simeq \frac{l \lambda(r_i)}{r_i},
    \end{aligned}
\end{equation}
where we first
calculate~$D_A = \tilde{U} + 2 D_B + \frac{4 B \lambda}{A}$ from the
evolved variables. If we had chosen to replace all the above RHSs with
$l \big( \tilde{D}_r(1) - D_r \big)$ instead, the simulation
becomes unstable as~$l$ increases. This is expected since the stiff
term~$\sim l/r$ is the main contribution of the RHS and it is not accurately resolved.

Finally, we need to calculate the Hamiltonian and momentum
constraints~\eqref{eq:our_Ham} and \eqref{eq:our_mom} since they are
needed for the constraint damping evolution equations of~$\Theta$
and~$Z_r$, respectively. The
replacements in \eqref{eq:classical_SBP_replacements} are sufficient to
calculate the Hamiltonian constraint, but not the momentum constraint.  We do not
have any auxiliary variable similar to~$\lambda$ to regularize
the term $(K - 3K_B)/r$ in the momentum
constraint~\eqref{eq:our_mom}. Nevertheless, in principle $K - 3K_B$ should vanish at~$r=0$ due to the fact that the space should remain
locally flat there (see e.g. chapter 10
of~\cite{alcubierre2008introduction}). Even though~$K - 3K_B$ is an
even function of~$r$, we still
use~$( \tilde{D}_r(1) - D_r) (K-3K_B)_i$ to
approximate~$(K(r_i) - 3K_B(r_i) )/r_i$. We have also tried to simply
replace its value at~$r=0$ using the l'Hôspital rule, but this
approach caused instabilities to develop.

\subsection{Technical details and features}
\label{subsec:miscellanious_code}

Here we outline various details of the structure of \texttt{SpheriCo.jl} relevant to the simulations.

\paragraph{Time integration.} The time evolution is performed using the method of lines \cite{alcubierre2008introduction} with the third-order Adams-Bashford (AB3) integrator. Since this method requires data from three distinct
timesteps, we use the explicit fourth-order Runge-Kutta (RK4)
integrator for the first and second timesteps. While AB3 is faster than RK4,
it is less accurate and typically requires a smaller timestep to be
stable  \cite{GusKreOli95}. Nevertheless, given that we need to evolve a large system with as many quantum modes as possible, we chose AB3 for
speed.

\paragraph{Artificial dissipation.} It is common for finite difference numerical schemes to include
artificial dissipation to control high frequency perturbations that
are not resolved by the grid spacing. Even though we use SBP
operators that in principle do not require artificial dissipation 
for their original first-order linear system,
we still add this option since we
heavily extrapolate their use past their original purpose. We use the
standard fourth-order Kreiss-Oliger
dissipation~\cite{kreiss1973methods} by modifying the RHS of our
evolution equations via
\begin{align}
  \textrm{RHS} \left( \p_t f_i \right) \rightarrow
  \textrm{RHS} \left( \p_t f_i \right)
  - \sigma \frac{f_{i-2} - 4 f_{i-1} + 6 f_i - 4 f_{i+1} + f_{i+2}}{16 h}
  \,,
  \label{eq:KO_diss_bulk}
\end{align}
with~$0 < \sigma < 1$, for the grid
points~$\{0, h, \dots, (N_r - 5)\, h \}$. Here, it becomes apparent why
we include two and not just one ghost point on the left of~$r=0$. For
the last two grid points~$\{ (N_r-4)\, h,  (N_r-3)\, h \}$ we apply
\begin{align}
  \textrm{RHS} \left( \p_t f_i \right)
  \rightarrow
  \textrm{RHS} \left( \p_t f_i \right)
  - \sigma \frac{f_{i-4} - 4 f_{i-3} + 6 f_{i-2} - 4 f_{i-1} + f_{i}}{16 h}
  \,.
  \label{eq:KO_diss_boundary}
\end{align}

\paragraph{Initial data generation.} When we generate initial data we need to integrate equation \eqref{eq:ACond_initialHamiltionan} to
obtain~$A(r_i)$, after providing~$A(r_0)$. The RHS
of this equation involves~$\Psi_i$ (the discretized version
of~$\Psi(0,r)$), which is provided only on the grid points. We want to
integrate using RK4, such that we provide initial data with
error smaller than~$\mathcal{O}(h^2)$, which is the dominant error introduced
during the evolution by the finite difference operators. Since RK4
involves steps between the grid points~$r_i$ and~$r_{i+1}$, we need
the value of~$\Psi(0,r)$ in this interval. We obtain this using a
third-order spline interpolation, provided by the~\texttt{Julia}
package~\texttt{Interpolations.jl}~\cite{Interpolations.jl}. Due to
the third-order accuracy of the interpolation method, we expect the
error in the integrated initial data to be~$O(h^3)$.

\paragraph{Discretized quantum modes and asymptotic flatness.} In the discretized version of the quantum field operator $\hat{\Phi}_n$ of \eqref{eq:extended_mode_expansion}, the integral over momentum $k$ is replaced with a discrete and finite sum, ranging from $0$ to $k_\max$. Likewise, the quantum number $l$ ranges from $0$ to $l_\max$, with the representation of the quantum field operator becoming more accurate as $k_\max$ and $l_\max$ increase. The quantum numbers $k_\max$ and $l_\max$ play different roles in achieving a higher degree of accuracy. As discussed already in \cite{Berczi:2021hdh}, 
it is found empirically that, broadly speaking, 
$l_\max$ parametrizes the spatial region over which the expectation value $\av{\chi|\hat{T}_{ab}|\chi}$ is well-represented, while $k_\max$ determines how well $\av{\chi|\hat{T}_{ab}|\chi}$ is represented within that spatial region. In order to find an appropriate choice for $l_\max$ and $k_\max$, we note that the vacuum field configuration should correspond to a Minkowski spacetime. By using Minkowski mode functions and setting the classical part to zero ($\phi=0$), we found empirically that the ratio $l_\max/k_\max\simeq 3$ leads to a sufficiently large $r$-domain in which $\av{\chi|\hat{T}_{ab}|\chi}$ vanishes to sufficient accuracy. In order to further guarantee asymptotic flatness in our simulations, we introduce a filter function $F(r)$ that suppress field sources at large $r$. It is configured as a smooth step function with a transition around a chosen radial point $r_\cut$. We define the outermost point of the radial numerical domain that is causally disconnected from $r_{\max}$ as $r_{\causal} \deff r_{\max} - t_{\max} - \dd t$. We typically choose $r_{\causal} < r_{\cut} < r_{\max}$. The filter function is set as
\beq
    F(r) = \frac{1}{2}\left[1+\tanh(r_\cut^p-r^p)\right], \quad p\in\mathbb{R},
    \label{eq:backreact_filter}
\eeq
which transitions from one to zero around $r=r_\cut$, approaching one as $r<r_\cut$ and zero as $r>r_\cut$, with the steepness controlled by the parameter 
$p$. We set $p=1$, unless stated otherwise explicitly. The filter function is then applied by convolution to the stress-energy tensor quantities in \eqref{eq:stress_tensor_projection_quantum} as well as the cosmological constant $\Lambda$, such that e.g. $\rho_\qu(t,r)\rightarrow F(r)\rho_\qu(t,r)$ and so on. Empirically we find that our simulations are more robust when the filter is also applied first in the individual quantities of \eqref{eq:stress_tensor_projection_quantum} by setting $r \rightarrow F(r) r$. This filter is different from artificial dissipation and is used to protect the simulations from numerical instabilities developing at $r_{max}$.

\paragraph{Infalling outer boundary.} The code provides the option for an evolution setup where the outer
boundary~$r_{\max}$ is infalling at the speed of light. With this
option, the resolution near the origin of the radial domain increases
with time, a technique that has been very useful in the study of
critical phenomena~\cite{Garfinkle:1994jb,
  Rinne:2020asi}. In~\cite{Garfinkle:1994jb} the setup uses a double
null foliation, whereas in~\cite{Rinne:2020asi} it is a spacelike
formulation which includes a dynamical shift condition, that allows
for the outer boundary to drop without regridding. In our formulation,
the shift vanishes, so we perform the drop of~$r_{\max}$ by a
regridding after every timestep by
setting~$r_{\max} \rightarrow r_{\max} - t$, with~$t$ the simulation
time and generating a new uniform grid with the same number of
points~$N_r$, but smaller~$h$. After each regridding, we need to
project all our evolved variables on the new grid, a process that
requires interpolation. Again, we use the third-order spline
interpolation from~\texttt{Interpolations.jl}. This numerical
algorithm is computationally expensive and it is probably more efficient
to include a dynamical shift in the formulation, but we leave this for
future development.

\paragraph{Parallelization.} Since the evolution equations for the quantum
modes~\eqref{eq:quantum_evol_eqs} are independent for each
mode~$u_{kl;n}$, we can easily parallelize them using
Julia's~\texttt{Threads.@threads} command. We also parallelize the
calculation of the stress-energy tensor expectation
values~\eqref{eq:stress_tensor_projection_quantum} in the
same way, where we approximate~$\int \dd k$ with~$\sum_{k} \dd k$. The
current version of the code can only run on a CPU. Given how much the
independent evolution equations grow with increasing the~$k_{\max}$
and~$l_{\max}$ of the quantum modes, it seems that adapting the code to
also run in parallel on a GPU would be beneficial in terms of
speed. This however might include major changes in the architecture of
the code, so we leave such an exploration for the future.

\paragraph{Constraint damping.} The code provides the choice for damping or no damping of the
Hamiltonian and momentum constraints via the variables~$\Theta$ and
$Z_r$. In principle, the classical evolution
system \eqref{eq:classicalEoMs_matter}--\eqref{eq:classicalEoMs_grav} damps the Hamiltonian and
momentum constraints during the evolution, where the amount of damping
is controlled by the parameters~$\kappa_1$ and $\kappa_2$. However, in our numerical tests we noticed instabilities related to the
coupling of~$\Theta$ and $Z_r$ to \eqref{eq:K_eom},
\eqref{eq:KB_eom}, \eqref{eq:alpha_eom} and \eqref{eq:Dalpha_eom} for subcritical data. These instabilities seem to
originate from~$r=0$ and so we believe they are related to the way we
approximate numerically the Hamiltonian and momentum
constraints~\eqref{eq:our_Ham} and \eqref{eq:our_mom}. In addition, as we 
show in subsection~\ref{Subsec:constraint_violation_classical}, even for
supercritical data including constraint damping does not improve the
violation of~\eqref{eq:our_Ham} and \eqref{eq:our_mom}, which defeats their purpose of evolving $\Theta$ and $Z_r$. However, we should mention
here that our formulation and evolved variables are different than
those in~\cite{Bernuzzi:2009ex}, where a conformal splitting is also
used (conformal factor in front of the spatial metric). Furthermore,
we do not include any damping of reduction constraints, which might
also make an important difference \cite{Lindblom:2005qh, Cors:2023ncc}.
To provide the choice of turning off the damping, we
perform $\Theta \rightarrow \textrm{damping} \times\Theta$ and
 $Z_r \rightarrow \textrm{damping}\times Z_r$ (and their derivatives) in
the evolution equations~\eqref{eq:K_eom}, \eqref{eq:KB_eom},
\eqref{eq:alpha_eom} and \eqref{eq:Dalpha_eom}. If one
sets~$\textrm{damping}=0$, then there is no coupling between $\Theta$ and $ Z_r$ 
to the rest of the equations, even though $\Theta$ and $ Z_r$ are
still evolving and provide information about the amount of constraint violation during the evolution.

\section{Code validation: classical setup}
\label{Sec:classical_tests}

We perform a number of tests to check the validity of our code. In
this section we focus on the classical setup and perform different
convergence tests, as well as physical tests that attempt to reproduce
results from the literature.

\subsection{Convergence}
\label{Subsec:convergence_classical}

The dominant error of our numerical scheme comes from the second-order
accurate FD operators (both standard and SBP). For any function~$f$ we
expect~$f = f_{h} + \mathcal{O}(h^2)$, with~$f$ the exact solution and~$f_{h}$
the numerical approximation. For the convergence tests, we solve the same system of evolution equations with the same given data,
with coarse, medium and fine resolutions and inspect the rate at
which the numerical error of the solution decreases. The value of the
theoretically expected convergence rate~$Q$ is dictated by the
resolutions and the order of accuracy~$q$ via 
\begin{align}
  Q = \frac{h_\text{\tiny c}^q - h_\text{\tiny m}^q}{h_\text{\tiny m}^q - h_\text{\tiny f}^q}
  \,,
  \label{eq:self_convergence_rate}
\end{align}
where~$h_\text{\tiny c}$,~$h_\text{\tiny m}$ and~$h_\text{\tiny f}$ denote the grid spacing for the coarse,
medium and fine resolutions, respectively. We define these grid spacings to correspond
to~$N_r= \{1027,\, 2051,\, 4099\}$, such that we half the grid spacing every time
we increase the resolution. These choices of resolutions, combined with the second-order
accuracy, yield~$Q=4$ for our tests. If the solution exhibits perfect
second-order convergence, then we expect to see
$f_\text{\tiny c} - f_\text{\tiny m} = Q (f_\text{\tiny m} - f_\text{\tiny f})$, which is what we monitor in the
simulations. For each resolution we set~$\dd t = h/8$, which satisfies the Courant–Friedrichs–Lewy (CFL) condition \cite{alcubierre2008introduction}. For these tests, our computational domain is defined by~$t_{\max}=15$ and $r_{\max}=30$. We present the data only in the part of the radial domain that is
causally disconnected from $r_{\max}$. In principle, we could consider the whole radial domain since we have implemented outgoing boundary conditions that should control numerical reflections at $r_{\max}$.
However, the implemented outgoing boundary conditions for the quantum modes do not preserve their standing wave behavior in Minkowski spacetime, which is a validation test we perform in subsection \ref{Subsec:vacuum_no_backreact}, but are useful for numerical stability when backreaction is included. Therefore, in the semiclassical case, we are forced to consider only $r \in [0,r_{\causal}]$ when we interpret our results, and for consistency we choose to do the same in the classical setup as well. We emphasize that these boundary conditions are not constraint-preserving, even for the classical setup. Therefore, if we were to consider the whole domain in this case, we would expect additional constraint-violation arising from the boundary conditions.

\begin{figure}[t]
  \includegraphics[width=0.95\textwidth]{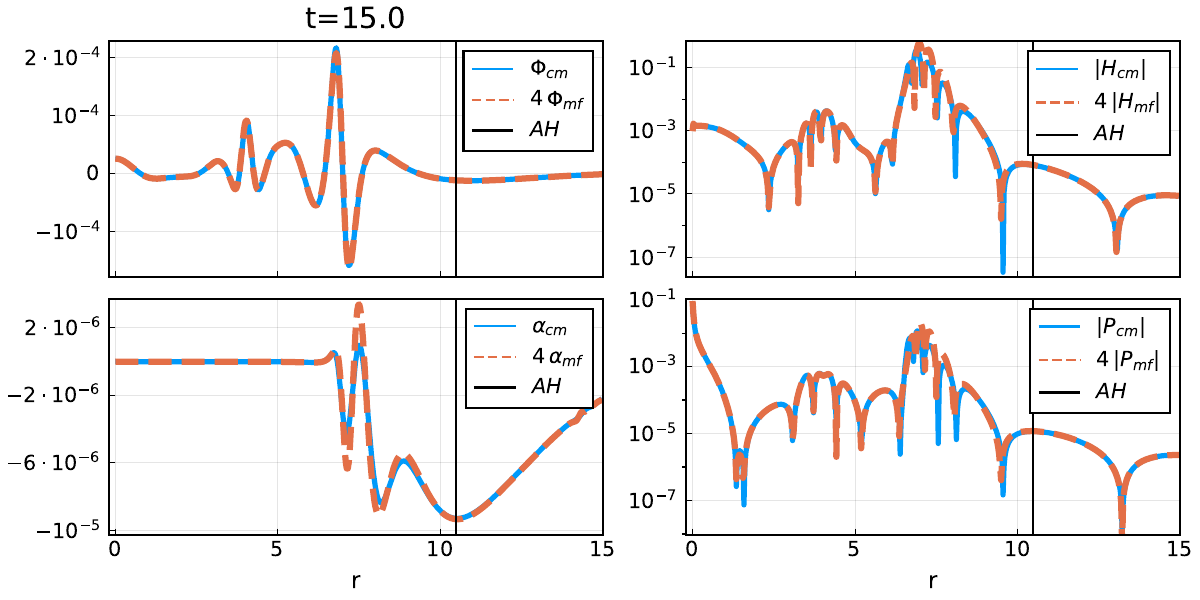}
  \caption{Supercritical data: pointwise convergence for
    $\Phi, \alpha, |H|, |P|$ at~$t_{\max}$. The overlap of~$\Phi_\text{\tiny cm}$
    with~$4 \, \Phi_\text{\tiny mf}$ (similarly for the other gridfunctions)
    indicates good second-order convergence.}
  \label{Fig:strong_data_cmf_conv}
\end{figure}

\begin{figure}[t]
  \includegraphics[width=0.95\textwidth]{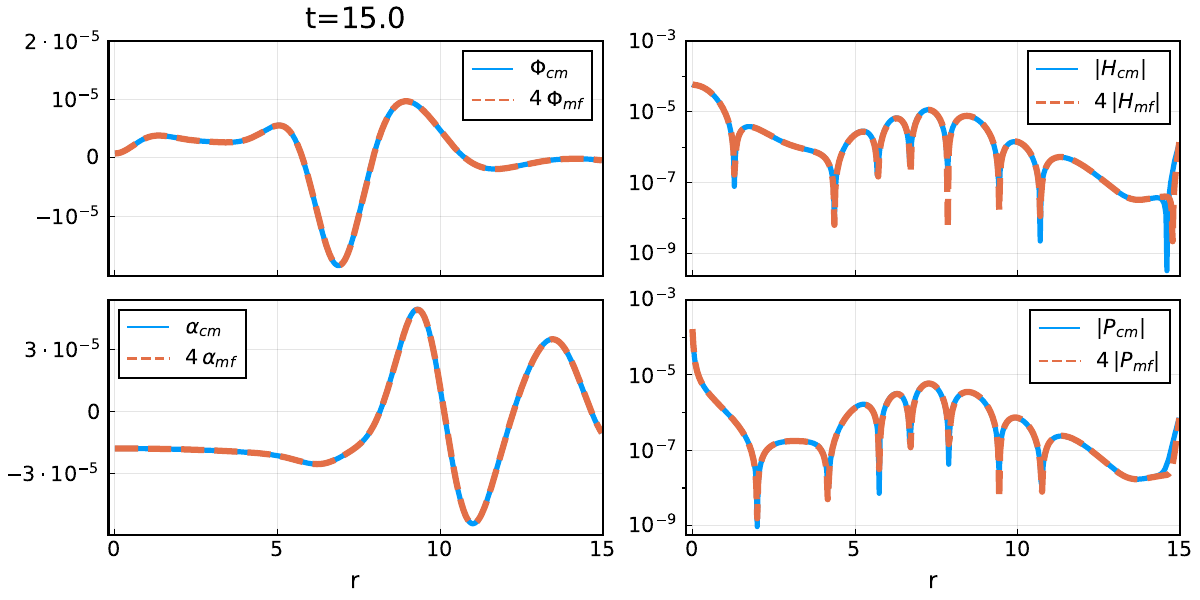}
  \caption{Subcritical data: pointwise convergence for
    $\Phi, \alpha, |H|, |P|$ at~$t_{\max}$. We see again good second
    order convergence, indicated by the overlap of the rescaled
    grid functions.}
  \label{Fig:weak_data_cmf_conv}
\end{figure}

The scalar field initial data are given by \eqref{eq:Phi_ID}
with~$b=5$,~$c=2$ and~$a=0.5$ for supercritical data or~$a=0.05$ for subcritical
data. In this test there is no artificial dissipation, i.e. $\sigma=0$
in~\eqref{eq:KO_diss_bulk}--\eqref{eq:KO_diss_boundary} and we
turn off the constraint damping.
We calculate the error between the resolutions by
\begin{align}
  f_\text{\tiny cm} \deff f_\text{\tiny c} - \perp^\text{\tiny m}_\text{\tiny c} f_\text{\tiny m},
  \qquad
  f_\text{\tiny mf} \deff \perp^\text{\tiny m}_\text{\tiny c} f_\text{\tiny m} - \perp^\text{\tiny f}_\text{\tiny c} f_\text{\tiny f},
  \label{eq:error_diffs_def}
\end{align}
where~$\perp^\text{\tiny m}_\text{\tiny c}$ denotes projection of the medium to the coarse
resolution, i.e., we consider only the common points and similarly
for~$\perp^\text{\tiny f}_\text{\tiny c}$. In figure~\ref{Fig:strong_data_cmf_conv} and
figure~\ref{Fig:weak_data_cmf_conv} we see good pointwise convergence of
the scalar field~$\Phi$, lapse~$\alpha$, absolute value of the
Hamiltonian~$|H|$ and momentum~$|P|$ constraints, for supercritical and subcritical data, respectively. To set the location of the apparent horizon we find the two grid points where  $\theta_{\textrm{exp}}$ changes sign, and take their mean value.

\begin{figure}[t]
  \includegraphics[width=0.999\textwidth]{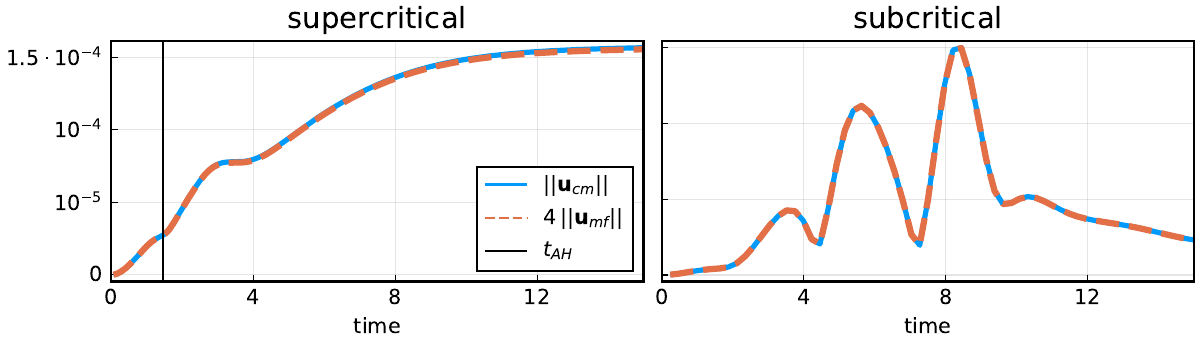}
  \caption{Norm convergence for supercritical (left) and subcritical (right) data,
    with smooth initial data. Both setups exhibit good second-order
    convergence, indicated by the good overlap of the rescaled norms.}
  \label{Fig:L2_smooth_strong_weak}
\end{figure}

In appendix \ref{app:classical_hyp} we show that the classical
evolution system \eqref{eq:classicalEoMs_matter}--\eqref{eq:classicalEoMs_grav} is strongly hyperbolic
and so the initial value problem for its linearized and constant coefficient approximations is well-posed in the~$L^2$-norm~\eqref{eq:L2_classical_sys}. In addition to the pointwise convergence tests, we also perform norm
convergence tests in a discretized version
of~\eqref{eq:L2_classical_sys}, given
by~\eqref{eq:L2_classical_sys_discrete}. We obtain the errors as
defined in~\eqref{eq:error_diffs_def} for each gridfunction of the
state vector~$\mathbf{u}$ and calculate 
\begin{align}
  || \mathbf{u}_\text{\tiny cm}||_{h_\text{\tiny c}}^2
  \deff
  \sum_{r=0}^{r_{\causal}} \sum_j \mathbf{u}_{\text{\tiny cm}, j}^2 \, h_\text{\tiny c},
  \label{eq:L2_classical_sys_discrete}
\end{align}
and similarly for $|| \mathbf{u}_\text{\tiny mf}||_{h_\text{\tiny c}}$. Here $j$ labels the different elements in the state vector $\mathbf{u}$. In
figure~\ref{Fig:L2_smooth_strong_weak} we demonstrate good second-order
convergence in the discretized~$L^2$-norm~\eqref{eq:L2_classical_sys_discrete}, for our supercritical and subcritical data with~$a=\{0.5, 0.05\}$, $b=5$ and~$c=2$.
In appendix~\ref{app:code_tests} we repeat these tests including
artificial dissipation with~$\sigma=0.02$, which also exhibit good
second-order convergence.

\subsection{Robust stability}
\label{Subsec:robust_stability_classical}

We also perform convergence tests with noisy data, typically
called~\textit{robust stability tests}. These often form standard
testbeds for numerical relativity codes~\cite{Alcubierre:2003pc} and
have been widely performed in the literature \cite{Calabrese:2005ft,
  PhysRevD.83.104041, Cao:2011fu, Babiuc:2007vr,
  Giannakopoulos:2023nrb}. Briefly, numerical stability can be
understood as the discrete analog of well-posedness, where the
numerical solution is controlled by the given data, in some
discretized norm, in which the initial data are controlled. We will
not expand more on numerical stability, but the interested reader can
find more information e.g in \cite{Hinder:2005rts}.

For these tests, the initial data are the same as in the previous
subsection, with the addition of random noise on top, with
amplitude~$A_h$. We add random noise to provide a more demanding test
for our code, since often instabilities can develop due to noisy data
that are initially small (e.g. due to round-off errors) and increase
in time. Hence, the random noise simulates numerical error, in an
exaggerated form. To obtain initial data controlled in the~$L^2$-norm~\eqref{eq:L2_classical_sys_discrete}, we need to scale the
amplitude~$A_h$ appropriately for the different resolutions. Since we
halve the grid spacing every time we increase resolution, we need to
drop the amplitude by a factor of four to have initial data that converge to
second-order in the~$L^2$-norm, i.e.,
\begin{align}
  \frac{|| \mathbf{u}_h(0,r_i) ||}{ || \mathbf{u}_{h/2}(0,r_i) ||} =
  \frac{A_h}{A_{h/2}} = 4.
\end{align}
Here we have assumed that the smooth part of all the initial data vanishes, for
simplicity. While this is not the case in our initial data, it helps to
provide an understanding of our chosen scaling for~$A_h$.

\begin{figure}[t]
  \includegraphics[width=0.999\textwidth]{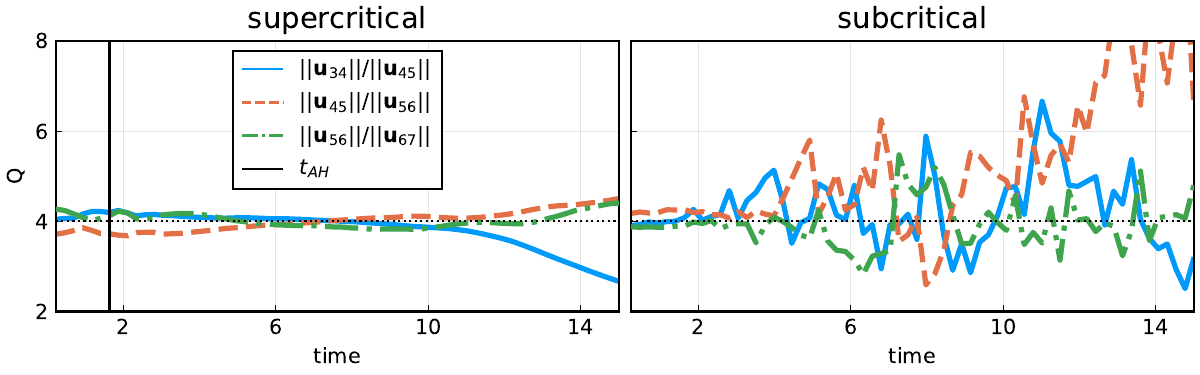}
  \caption{Robust stability tests for supercritical (left) and subcritical (right)
    data with~$b=5$, $c=2$ with $a=0.5$ and $a=0.05$, respectively,
    for~$\Phi(0,r)$. The amplitude of the random noise is high enough
    ($A_{h_D} = 10^{-3}/4^D$) to probe both the linear and non-linear
    terms of the classical setup \eqref{eq:classicalEoMs_matter}--\eqref{eq:classicalEoMs_grav}. We observe that as the resolution increases, the numerically computed convergence rate approaches the theoretical value of four, and the convergence continues for a longer duration. We interpret this as numerical evidence for
    stability and convergence of our code. Note that in this setup
    there is no artificial dissipation.}
  \label{Fig:robust_stability_sigma0_rand_0.001}
\end{figure}

For these tests we choose the resolutions
$N_r = \{1027, 2051, 4099, 8195, 16387\}$, which we label by
$D=\{3,4,5,6,7\}$ respectively, since we follow
$N_r = 128 \cdot 2^D + 3$. We tune the random noise amplitude at
resolution~$D$ via~$A_{h_D} = 10^{-3}/4^D$. With this
choice,~$A^2_{h_D}$ is above the round-off error~$\mathcal{O}(10^{-16})$ for all
resolutions apart from~$D=7$, where it is the same order of magnitude as the round-off error. This allows
us to test also the non-linear part of the equations, as
in~\cite{Calabrese:2005ft}. By calculating the discretized~$L^2$-norm~\eqref{eq:L2_classical_sys_discrete} and forming ratios
like~$||\mathbf{u}_{34}||/||\mathbf{u}_{45}||$ etc., we expect to
obtain the convergence rate~$Q$, where here~$3,4,5$ denote the
different resolutions. In practice, we consider the test as passed
when the numerically computed convergence rate gets closer to the
theoretical value of four with increasing resolution and for
longer. This is exactly the behavior we see in
figure~\ref{Fig:robust_stability_sigma0_rand_0.001} and is qualitatively compatible with the findings
of other passed robust stability tests in the
literature~\cite{PhysRevD.83.104041, Cao:2011fu,
  Babiuc:2007vr}. Furthermore, were an instability triggered in these
tests, we would expect to see a rather different picture (the
numerical convergence rate would be driven away from the theoretical
value faster with increasing resolution, see e.g. the robust stability
tests for weakly hyperbolic systems
in~\cite{Giannakopoulos:2023nrb}). The amount of artificial
dissipation, the CFL, and the noise amplitude $A_h$ can affect the
picture. In appendix \ref{app:code_tests} we provide more
robust stability tests.

\subsection{Constraint violation}
\label{Subsec:constraint_violation_classical}

The classical initial data described in
section~\ref{Subsec:classical_ID_BC} exactly satisfy the momentum
constraint \eqref{eq:our_mom} but introduce
some violation of the Hamiltonian constraint~\eqref{eq:our_Ham} due
to numerical error. During the evolution, the violations of the
Hamiltonian and momentum constraints increase due to accumulation of
numerical errors, but as shown in
section \ref{Subsec:convergence_classical}, this violation
decreases with increasing resolution, following second-order
convergence. Here, we wish to test the performance of the constraint
damping scheme that we have implemented by including the
variables~$\Theta$ and $Z_r$ with evolution equations~\eqref{eq:Theta_eom}
and \eqref{eq:Zr_eom}, respectively.

Unfortunately, our implementation of the constraint damping is
unstable for both supercritical and subcritical data.  The instability grows much
faster for subcritical data near the origin, whereas for supercritical data the
collapse of the lapse function $\alpha$ inside the apparent horizon slows it down and
allows us to perform some diagnostic tests. We choose the supercritical
initial data of subsection~\ref{Subsec:convergence_classical},
namely~$a=0.5$, $b=5$ and $c=2$ in \eqref{eq:Phi_ID}, discretized
by $N_r=4099$, $\dd t=h/8$, with $r_{\max}=30$,~$t_{\max}=15$ and no
artificial dissipation. To investigate the performance of the
implemented constraint damping, we perform a simulation with damping
turned off (such that $\Theta$ and $Z_r$ do not couple to the rest of
the evolution equations) and compare it against simulations with
damping turned on, for different values of the parameters $\kappa_1$
and $\kappa_2$. We refer to the former as \textit{no-damping} and the
latter as \textit{with-damping}. We report with-damping explorations
for
$(\kappa_1, \kappa_2) = \{(0.002, 0),\, (0, 0.002),\, (0.02, 0),\, (0,
0.02)\}$, which we label by~$1-4$. The no-damping case is labeled
with~$0$.

\begin{figure}[t]
  \includegraphics[width=0.999\textwidth]{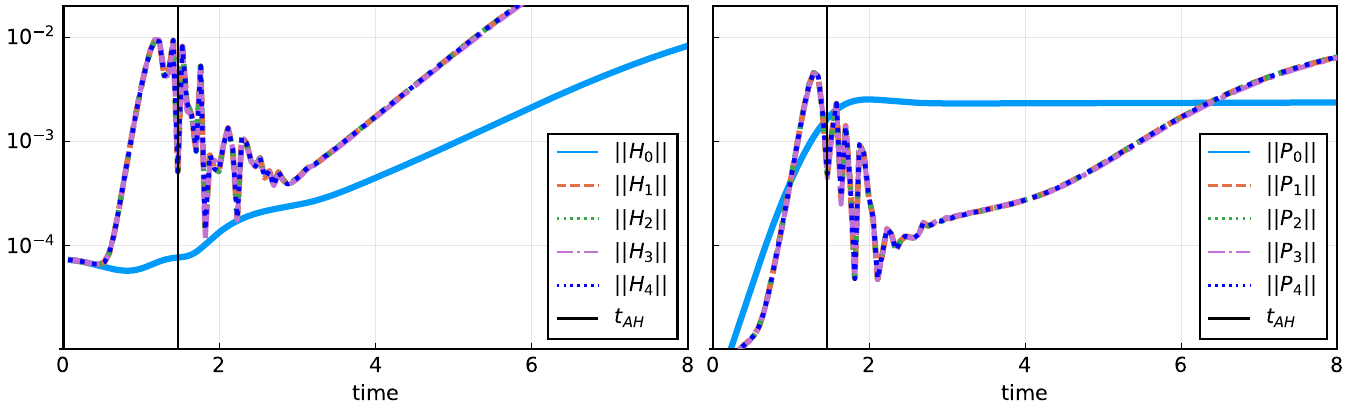}
  \caption{$L^2$-norm of the Hamiltonian (left) and momentum (right)
    constraint violation, in time. We label as $0$ the no-damping and
    as $1-4$ the with-damping for
    $(\kappa_1, \kappa_2) = \{(0.002, 0),\, (0, 0.002),\, (0.02, 0),\,
    (0, 0.02)\}$. The Hamiltonian constraint violation is greater for
    all the with-damping cases, at all times. The momentum constraint
    violation is successfully damped for a brief period, for all
    with-damping simulations. As described in the text, the violation increases in the whole domain, but more significantly in the region $r=(6,8)$, that is inside the AH.
    }
  \label{Fig:l2_t_Ham_Mom}
\end{figure}

\begin{figure}[t]
  \includegraphics[width=0.999\textwidth]{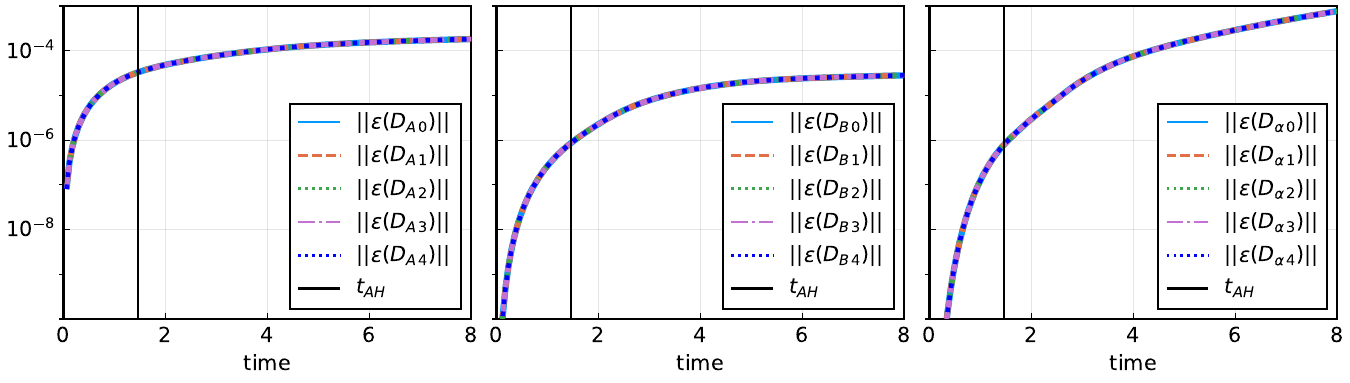}
  \caption{$L^2$-norm of the violation for the reduction constraints
    of $D_A$ (left), $D_B$ (middle) and $D_\alpha$ (right), in
    time. We label as $0$ the no-damping and as $1-4$ the with-damping
    for
    $(\kappa_1, \kappa_2) = \{(0.002, 0),\, (0, 0.002),\, (0.02, 0),\,
    (0, 0.02)\}$. The violation is similar for all cases, both no- and
    with-damping.}
  \label{Fig:l2_t_reduction_constraints}
\end{figure}

\begin{figure}[t]
  \includegraphics[width=0.999\textwidth]{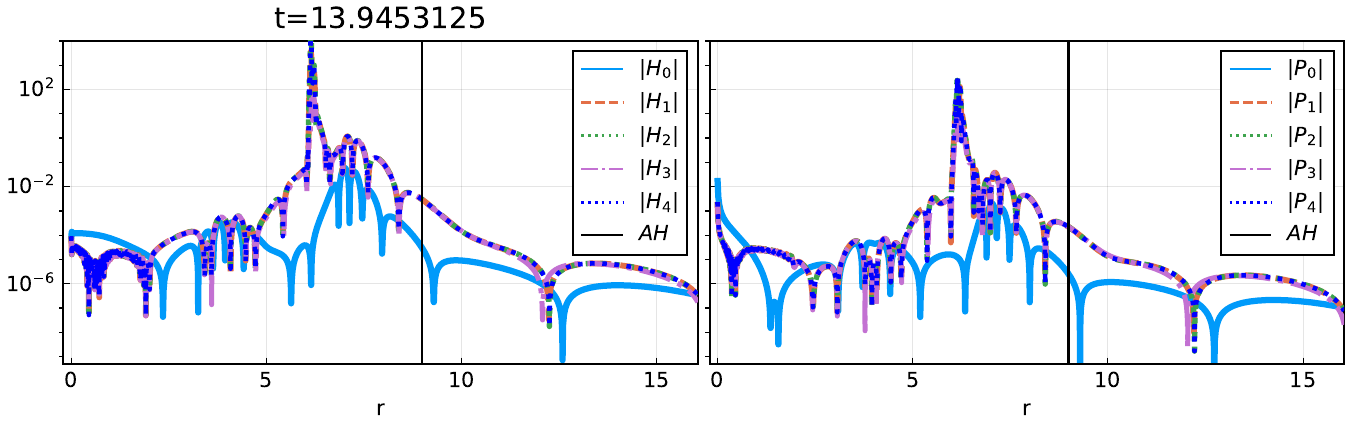}
  \caption{The absolute value of the Hamiltonian (left) and momentum
    (right) constraint violation, in the radial domain that is
    causally disconnected from $r_{\max}$, at $t=8.02734375$. The subscript $0$ labels the no-damping case and the subscripts $1-4$ the with-damping case for
    $(\kappa_1, \kappa_2) = \{(0.002, 0),\, (0, 0.002),\, (0.02, 0),\,
    (0, 0.02)\}$. For both constraints, the violation is greater in
    the with-damping case, especially in the region $r=(6,8)$, except
    from $r=0$ for the momentum constraint.}
  \label{Fig:tf_Ham_Mom}
\end{figure}

\begin{figure}[t]
  \includegraphics[width=0.999\textwidth]{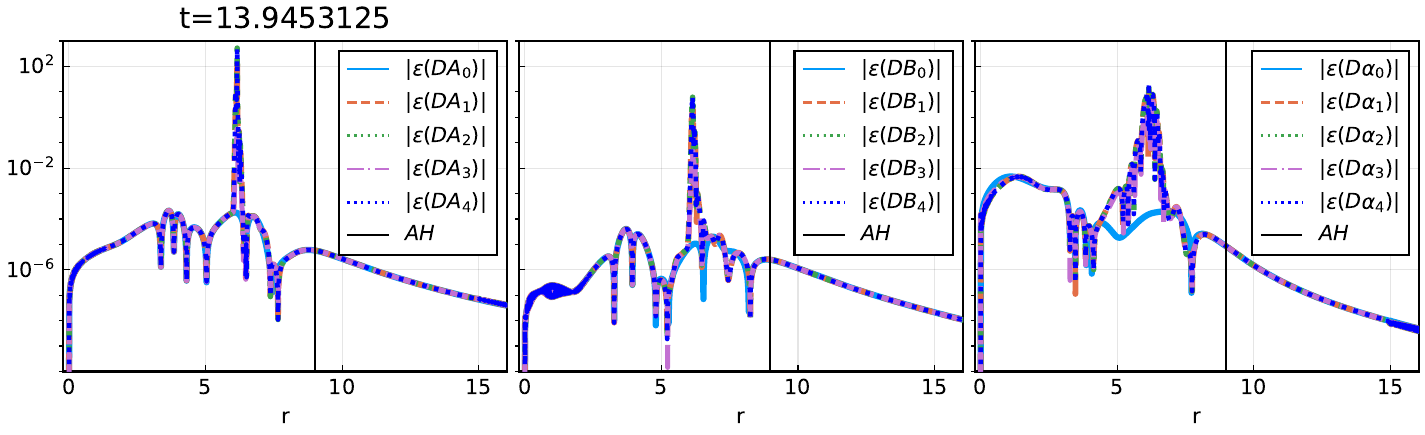}
  \caption{The absolute value of the violation for the reduction
    constraints of $D_A$ (left), $D_B$ (middle) and $D_\alpha$
    (right), in the radial domain that is causally disconnected from
    $r_{\max}$, at $t=8.02734375$. We label as $0$ the no-damping and
    as $1-4$ the with-damping for
    $(\kappa_1, \kappa_2) = \{(0.002, 0),\, (0, 0.002),\, (0.02, 0),\,
    (0, 0.02)\}$. The violation is similar for all cases.}
  \label{Fig:tf_reduction_constraints}
\end{figure}

In all the with-damping simulations the code crashes around
$t = 13.9$. In figures \ref{Fig:l2_t_Ham_Mom},
\ref{Fig:l2_t_reduction_constraints}, \ref{Fig:tf_Ham_Mom},
\ref{Fig:tf_reduction_constraints}, we present the results until
around $t = 8$, because later the violation grows too big to
meaningfully compared with the no-damping case. In figure \ref{Fig:l2_t_Ham_Mom} we show the Hamiltonian- ($H$) and momentum constraint ($P$) violation in time, by evaluating the
expressions \eqref{eq:our_Ham}, \eqref{eq:our_mom}, respectively, and
compute their discretized $L^2$-norm
\begin{align}
  || f_h||_{h}^2
  \deff
  \sum_{r=0}^{r_{\causal}} f_h^2  h,
  \label{eq:L2_function_discrete}
\end{align}
where in this case the grid function is $f_h=\{H,P\}$. In figure \ref{Fig:l2_t_Ham_Mom} see that the
Hamiltonian constraint violation is always greater for the
with-damping cases, irrespective of the choice of $\kappa_1$ and
$\kappa_2$. In contrast, the momentum constraint violation, after
initially growing larger for the with-damping cases, is successfully
damped below that of the no-damping one, up until around $t=6$. After
this, it again grows larger than the no-damping case, which has
reached a plateau.

We emphasize that even though the constraint damping we implemented is
based on~\cite{Bernuzzi:2009ex}, we evolve different variables and we have
not implemented damping of the reduction constraints, which could be
important~\cite{Lindblom:2005qh, Cors:2023ncc}. Hence, we also examine the reduction constraint violation in time, by computing their $L^2$-norm. For our system, we define the reduction
constraints violation as
\begin{subequations}
  \begin{align}
    \epsilon(D_A)
    & \deff
      \frac{A'}{A} - D_A
      = \frac{A'}{A} -
      \tilde{U} - 2D_B - \frac{4 B \lambda}{A}
      \,, \label{eq:DA_reduction_constr}
    \\
    \epsilon(D_B)
    & \deff
      \frac{B'}{B} - D_B
      \,, \label{eq:DB_reduction_constr}
    \\
    \epsilon(D_\alpha)
    & \deff
      \frac{\alpha'}{\alpha} - D_\alpha
      \,, \label{eq:Dalpha_reduction_constr}  
  \end{align}
  \label{eq:reduction_constr}%
\end{subequations}
which we evaluate in post-processing from our evolved variables, using
the standard second-order accurate FD
operator~\eqref{eq:standard_FD_bulk}, and \eqref{eq:standard_FD_rmax} at $r_\max$. In
figure \ref{Fig:l2_t_reduction_constraints} we see that the violation of
the reduction constraints grows during the evolution in the same way
for all cases, both no- and with-damping. This is expected, since we
have not implemented any scheme for the damping of reduction
constraints at this stage. In figures \ref{Fig:tf_Ham_Mom} and
\ref{Fig:tf_reduction_constraints} we show the violation of the
Hamiltonian and momentum as well as reduction constraints, respectively, at
$t \simeq 8$. For the Hamiltonian and momentum constraints, the violation is greater for the
with-damping cases, especially in the region $r = (6,8)$, which
eventually grows significantly, leading to the crashing of the code
(the relevant evolved variables $\Theta$ and $Z_r$ become very steep in
this region). The reduction constraints violation is
very similar for all cases, both no- and with-damping, which is
expected. We do notice a significant violation in the region $r=(6,8)$
as well, especially for~$D_A, D_B$. It is possible that this has an
important contribution to the growth of the Hamiltonian and momentum
constraint violation and eventually the crashing of the code, but it
is not clear to us at this stage how much. We have also experimented
with the values
$(\kappa_1, \kappa_2) = \{(0.002, 0.002),\, (0.02, 0.02),\, (0.2,
0),\, (0, 0.2),\, (0.2,0.2)\}$, but their results are very similar. We
leave further investigation and development on this topic for future
work.

\subsection{Critical phenomena}
\label{Subsec:critical_phenomena_classical}

Critical phenomena in gravitational collapse were first discovered
numerically by Choptuik \cite{Choptuik:1992jv} and have since then
been studied extensively; see \cite{Gundlach:2007gc} for a review and
related references. Choptuik's prototypical example is a massless
scalar field, minimally coupled to gravity, the same as our classical
setup. 

Consider a one-parameter family of initial data, say the classical initial conditions
described by~\eqref{eq:Phi_ID} and \eqref{eq:Psi_Pi_ID}, for fixed
position and width and varying amplitude $a$ of the classical scalar field profile. By tuning the amplitude,
the evolved data can be supercritical, leading to black hole
formation, or subcritical, such that no black hole forms and the
spacetime tends to Minkowski spacetime as the energy disperses to
infinity. The gravitational critical phenomena occur at the threshold
of black hole formation, in between super- and subcritical data and exhibit
interesting structure such as echoing (specific oscillations) of the
scalar field, power-law scaling of the black hole mass and
universality of solutions (that is same behavior of the scalar field
irrespective of the family of initial data considered, for near critical
solutions).

We explore the performance of \texttt{SpheriCo.jl} in the regime of
critical phenomena, and focus on the echoing behavior and the universality of solutions. We leave the study of the power-law scaling with our code for future work. To the best of our knowledge, the specific combination of numerical methods we employ -- namely, the use of standard FD and SBP operators as described in Section \ref{Sec:numerical_implementation} -- has not been tested before. Therefore, we aim to assess whether the numerical solution for the classical setup can accurately reproduce well-known physical results. We consider this as a non-trivial test of the code,
and we recognize that the formulation and numerical methods we use are
not well-adapted to this problem. Techniques better suited to study
criticality may include refinement near the origin via an adaptive
mesh as in Choptuik's original work~\cite{Choptuik:1992jv}, having an infalling outer boundary (either in characteristic
\cite{Garfinkle:1994jb} or standard spacelike \cite{Rinne:2020asi}
formulations), using a non-uniform radial grid 
\cite{Baumgarte:2018fev}, or using co-ordinates appropriate
for self-similar spacetimes \cite{Cors:2023ncc}. To explore
criticality more economically in terms of computational time, we run
the simulations with a numerically infalling outer boundary 
$r_{\max}$, as described in
subsection \ref{subsec:miscellanious_code}, since our
metric ansatz \eqref{eq:lineelement} has a vanishing shift.

The one-parameter family of initial data we choose here is given by
\eqref{eq:Phi_ID} and \eqref{eq:Psi_Pi_ID}, with $b=0$, $c=1$ and $a$
determining whether the initial data is sub- or supercritical. This is the same family used in
\cite{Baumgarte:2018fev} (in the spherical case), though here $a$ is
half the amplitude of the initial Gaussian when it is centered at
$r=0$, due to the definition of our initial data. We refer to this as \textit{initial data family 1}.
In appendix~\ref{app:code_tests} we repeat the
analysis for initial data with $b=5$, $c=1$ and $a$ tunable, which we refer to
as \textit{initial data family 2}. Close to criticality and at $r=0$, the scalar
field is expected to oscillate fast (in terms of the simulation time)
around the values $\pm 0.6$ \cite{Baumgarte:2018fev}, irrespective of the initial data family used (universaliy of the solution). To have a
clearer picture of this oscillation, the similarity time $T$ is often
used, defined as
\begin{align}
  T \deff - \ln \left( \tau_* - \tau \right), \label{eq:similarity_time}
\end{align}
where the proper time $\tau$ is given by
\begin{align}
  \tau(t_\text{\tiny f},r) = \int_{0}^{t_\text{\tiny f}} \dd t\, \alpha(t,r), \label{eq:proper_time}
\end{align}
where $t_\text{\tiny f}$ is some final time and $\tau_*$ the accumulation time. The latter can
be calculated by considering a pair of subsequent zero-crossings
$\tau_n$ and $\tau_{n+1}$ together with another zero-crossing pair $\tau_m$ and
$\tau_{m+1}$, via
\begin{align}
  \tau_* \deff \frac{\tau_n \tau_{m+1} - \tau_{n+1} \tau_m}{\tau_n - \tau_{n+1} - \tau_m + \tau_{m+1}},
  \label{eq:accumulation_time}
\end{align}
and we refer to e.g. \cite{Baumgarte:2018fev} for more details on this.

To have increasing resolution near the origin with time we need to tune
$t_{\max}$ and $r_{\max}$ appropriately, such that we have the highest
resolution near the accumulation time. Since we do not have prior
knowledge of $\tau_*$ before the simulation, we make an educated guess
based on a few trial simulations. For initial data family 1 we obtain our best
results by setting $t_{\max}=6$ and $r_{\max}=6.01$ (different values
might be even better). We discretize the radial grid with
$N_r=128 \cdot 2^D +3$ and $D=\{5,6,7\}$ and use artificial
dissipation with $\sigma=0.02$. Without artificial dissipation the
classification of supercritical and subcritical data is sometimes false near the end
of the simulation because of noise near the origin, which possibly
arises due to the decreasing grid spacing. We are able to tune the
critical amplitude $a_*$ up to five decimals for all three
resolutions, that means supercritical and subcritical data are the same in the first
five decimal points of $a_*$. Empirically, it is expected that the
critical parameter tuned in numerical explorations depends on the
overall accuracy of the simulation, which in our case
translates into $a_*$ being a bit different for different values of
$D$, making convergence tests near criticality hard. The critical
amplitude found here is $a_* \in (0.16801640, 0.16801718)$, which
differs in the second decimal to that found in
\cite{Baumgarte:2018fev} (the comparison is between $2 a$ in our setup
and $\eta$ of \cite{Baumgarte:2018fev}). This difference is expected, as our code is not specifically designed for critical collapse investigations, and the critical amplitude is highly sensitive to both the input data and numerical resolution.
\begin{figure}[t]
  \includegraphics[width=0.999\textwidth]{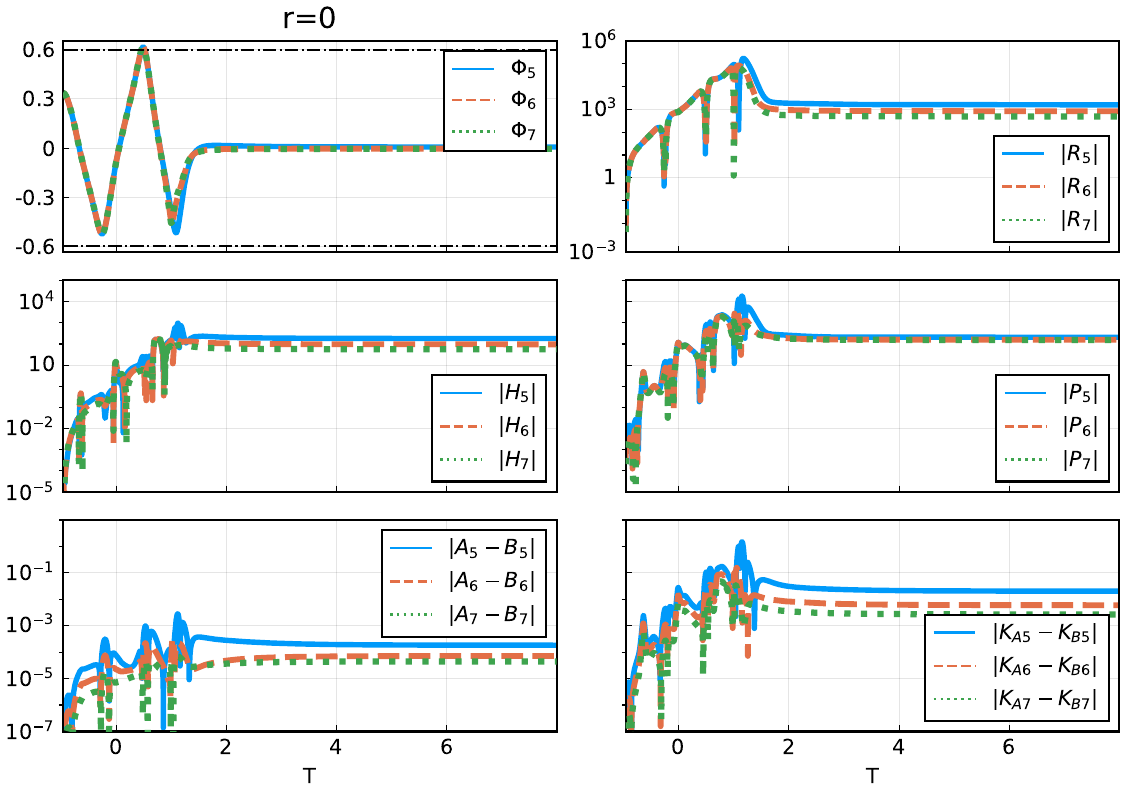}
  \caption{The different quantities monitored against the similarity
    time $T$ as defined in \eqref{eq:similarity_time}, at $r=0$, for
    three different resolutions labelled by $D=\{5,6,7\}$. Top left:
    the scalar field, exhibiting a complete oscillation (echo) between
    $\pm 0.6$, as expected for near critical solutions. Top right: the
    absolute value of the Ricci scalar $R$, the maximum value of which
    differs at late times between the lowest resolution $D=5$ and the
    other two, indicating a possible lack of accuracy. Middle: the
    absolute value of the Hamiltonian $H$ and momentum $P$ constraints
    on the left and right, respectively. The violation decreases as we
    increase the number of points in the radial grid. Bottom: the
    absolute value of the difference between the metric function $A,B$
    (left) and the extrinsic curvature components $K_A, K_B$
    (right). The latter is greater than the former by a few orders of
    magnitude, which could related to the greater violation of $P$ in
    comparison to $H$ and the fact that the formulation involves the
    auxiliary variable $\lambda$ related to $A,B$, but nothing similar
    for $K_A, K_B$.}
  \label{Fig:critical_ID_family_1_plots}
\end{figure}

\begin{figure}[t]
  \includegraphics[width=0.999\textwidth]{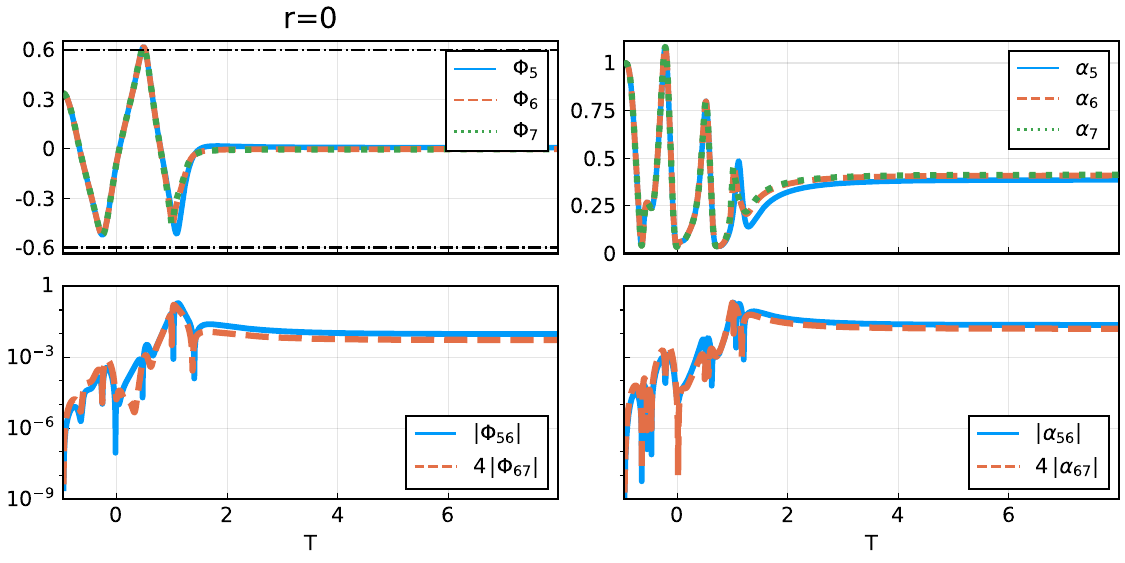}
  \caption{Pointwise convergence at $r=0$ for the classical scalar
    field $\Phi$ (left) and the lapse function $\alpha$ (right), as a
    function of the similarity time $T$. The different resolutions
    $D=\{5,6,7\}$ are denoted with the respective $D$ in the
    subscript. The numerical error (bottom) is computed by taking the
    difference between the numerical solutions of two different
    resolutions, denoted by both their $D$ labels in the
    subscript. The good overlap of the rescaled errors in the bottom,
    indicates good second-order convergence. The rescaling factor
    comes from the theoretically expected convergence rate computed in
    \eqref{eq:self_convergence_rate}.}
  \label{Fig:critical_ID_family_1_conv}
\end{figure}

In figures~\ref{Fig:critical_ID_family_1_plots} and
\ref{Fig:critical_ID_family_1_conv} we show our analysis for subcritical data
with $a = 0.16801640$, for all three resolutions. We focus on the
behavior of the different quantities at
$r=0$. Figure \ref{Fig:critical_ID_family_1_plots} shows about one
echoing period of the scalar field, oscillating around the values
$\pm 0.6$ as expected by the universality of near critical solutions,
against the similarity time. The same oscillating behavior is also
apparent in initial data family 2, shown in
appendix~\ref{app:code_tests}. Additionally, figure \ref{Fig:critical_ID_family_1_plots} shows the Ricci scalar $R$, as well as the Hamiltonian and momentum constraint
violations $H$ and $P$. We also monitor the difference between the metric
functions $A$ and $B$ as well as the components of the extrinsic curvature
$K_A$ and $K_B$. We do this to assess if the introduction of the auxiliary
variable $\lambda$ defined in \eqref{eq:FirstOrderGravVariables}, which
drives $A = B$ at $r=0$, has an effect on the violation of the
Hamiltonian constraint $H$ \eqref{eq:our_Ham}. In contrast, there is
no such auxiliary variable for $K_A, K_B$ (the evolved variable here
is $K = K_A + 2 K_B$) but the term $K_A - K_B$ affects the violation
of momentum constraint $P$, as seen from the definition
\eqref{eq:our_mom}.

Interestingly, in figure~\ref{Fig:critical_ID_family_1_plots} we see
that the difference between $K_A$ and $K_B$ at $r=0$ is a few orders of
magnitude larger than that of $A$ and $B$, and similarly for $P$ in
comparison to $H$. This result suggests that introducing another
auxiliary variable similar to $\lambda$ but for $K_A$ and $K_B$ might lead to a
smaller momentum constraint violation, allowing for a better
simulation of near critical phenomena. In the current version of the
code, we see that the maximum momentum constraint violation for the
smallest resolution is only an order of magnitude lower than the Ricci
scalar at $r=0$, which is often interpreted as having insufficient accuracy to
model critical phenomena \cite{Marouda:2024epb}. The introduction of an auxiliary
variable for $K_A, K_B$ similar to $\lambda$ has already been
suggested and successfully used \cite{Jimenez-Vazquez:2022fix}. In
figure \ref{Fig:critical_ID_family_1_conv} we see good second-order
convergence for the scalar field and lapse at $r=0$, against the
similarity time.

\section{Code validation: semiclassical setup}
\label{Sec:semiclassical_tests}

We also perform tests to examine the validity of our code in the
semiclassical setup. Although our primary interest lies in black hole spacetimes or the regime near criticality, we first focus on Minkowski spacetime, where the behavior of the quantum modes is well-known and well-understood.

\subsection{Vacuum without backreaction}
\label{Subsec:vacuum_no_backreact}

\begin{figure}[t]
  \includegraphics[width=0.999\textwidth]{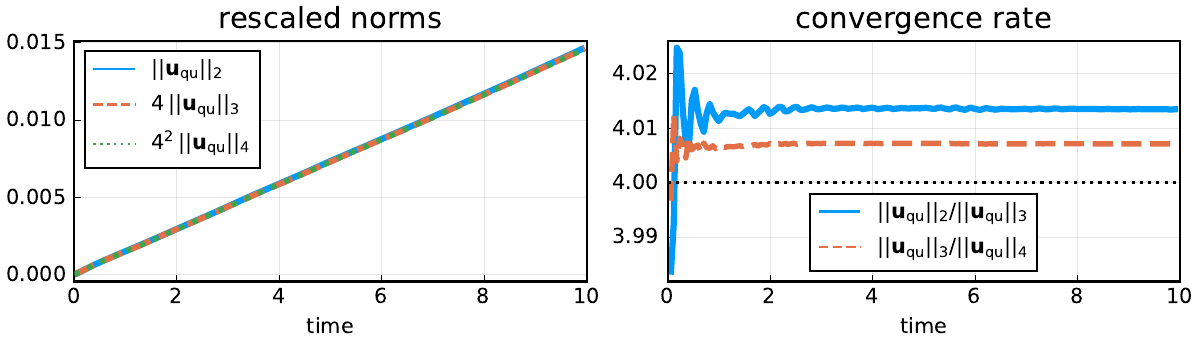}
  \caption{The rescaled
    norm~\eqref{eq:u_q_L2_norm_all_exact_conv_test} (left) and exact
    convergence rate (right), for a semiclassical simulation with
    vanishing classical scalar field, $k_{\max}=10$, $l_{\max}=60$ and
    $M_{\pv}=1$. The simulation exhibits good second-order convergence
    for the whole simulation time, when all the quantum modes are
    included.}
  \label{Fig:standing_waves_norm_all}
\end{figure}

\begin{figure}[t]
  \includegraphics[width=0.999\textwidth]{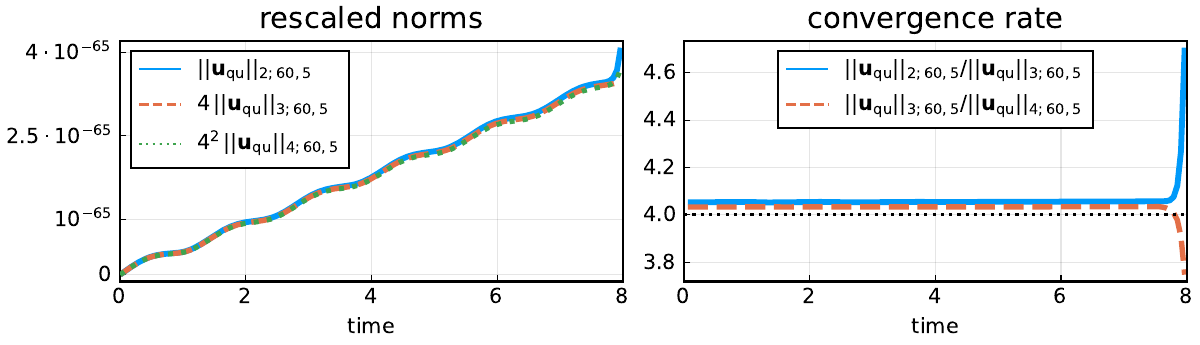}
  \caption{The rescaled
    norm~\eqref{eq:u_q_L2_norm_604_exact_conv_test} that includes only
    the fastest oscillating quantum modes (left) and the respective
    exact convergence rate (right), for a semiclassical simulation
    with vanishing classical scalar field, $k_{\max}=10$, $l_{\max}=60$
    and $M_{\pv}=1$. These modes exhibit good second-order convergence
    for a shorter part of the simulations, but this effect is not seen
    in the norm \eqref{eq:u_q_L2_norm_all_exact_conv_test}, due to
    their small relative amplitude. }
  \label{Fig:standing_waves_norm_604}
\end{figure}

We consider the setup where the classical scalar field is vanishing,
so that the spacetime is Minkowski. We perform an evolution of the quantum modes in Minkowski spacetime, neglecting backreaction. The classical
geometry is obtained for every timestep of the evolution by running
\texttt{SpheriCo.jl} with $a=0$ for the classical
initial data~\eqref{eq:Phi_ID}. For the quantum modes we choose~$k_{\max}=10$,
$l_{\max}=60$, $\dd k = \pi/30$ and $M_{\pv}=1$. The ratio $l_{\max}/k_{\max}$ here is 6 instead of 3, which is our choice in general, just to make less expensive the calculation of norm \eqref{eq:u_q_L2_norm_all_exact_conv_test}. This ratio is not important here as we are not considering backreaction. As mentioned in
subsection \ref{Subsec:quantum_ID_BC}, the analytical solution in vacuum
for the quantum modes is
\begin{subequations}
\begin{align}
  u_{kl;n}^{\text{\tiny Mink}}(t,r)
  & = e^{-i \omega_n t}  \frac{k}{\sqrt{\pi \omega_n}} \frac{j_l(k r)}{r^l}
    \,,
    \label{eq:u_klmu_mink_sol}
  \\
  \psi_{kl;n}^{\text{\tiny Mink}}(t,r)
  & = e^{-i \omega_n t}  \frac{k}{\sqrt{\pi \omega_n}}
    \left[ \frac{\p_r j_l(kr)}{r^l} - \frac{l j_l(kr)}{r^{l+1}} \right]
    \,,
    \label{eq:psi_klmu_mink_sol}
  \\
  \pi_{kl;n}^{\text{\tiny Mink}}(t,r)
  & = - i \omega_n e^{-i \omega_n t}  \frac{k}{\sqrt{\pi \omega_n}} \frac{j_l(kr)}{r^l}
    \,,
    \label{eq:pi_klmu_mink_sol}
\end{align}
\label{eq:quantum_mink_sol}%
\end{subequations}
with~$\omega_n = \sqrt{k^2 + \mu_n^2}$. We simulate this setup with
$r_{\max}=30$ fixed, $t_{\max} = 10$, $N_r = 128 \cdot 2^D +3$, for
$D=\{2,3,4\}$ and $\dd t_D = h_D/16$ in all resolutions, without any
artificial dissipation. In this test we use the convention $1/M_\pl^2 = 1$ 
which was also used in~\cite{Berczi:2021hdh}. We want to compare the
numerical against the analytical solution and examine the rate at
which the former converges to the latter, with increasing
resolution. For this we calculate the following norm 
\begin{align}
  ||\mathbf{u}_\text{\tiny qu}||_D^2(t)
    \deff
    \sum_{k l n} \sum_{r=0}^{r_{\causal}}
    &h_D \left[
    \left(u_{kl;n,D}(t,r) - u_{kl;n}^{\text{\tiny Mink}}(t,r) \right)
    \left(u_{kl;n,D}(t,r) - u_{kl;n}^{\text{\tiny Mink}}(t,r) \right)^*
    \right.
    \nonumber
  \\
  & +
    \left.
    \left(\psi_{kl;n,D}(t,r) - \psi_{kl;n}^{\text{\tiny Mink}}(t,r) \right)
    \left(\psi_{kl;n,D}(t,r) - \psi_{kl;n}^{\text{\tiny Mink}}(t,r) \right)^*
    \right.
    \label{eq:u_q_L2_norm_all_exact_conv_test}
  \\
  & +
    \left.
    \left(\pi_{kl;n,D}(t,r) - \pi_{kl;n}^{\text{\tiny Mink}}(t,r) \right)
    \left(\pi_{kl;n,D}(t,r) - \pi_{kl;n}^{\text{\tiny Mink}}(t,r) \right)^*
    \right]
    \,,
    \nonumber
\end{align}
where $\mathbf{u}_\text{\tiny qu}$ is the part of the state vector that contains all the quantum field modes, the subscript $D$ denotes the numerical solution at resolution
$D$, the superscript $*$ the complex conjugate and $t,r$ are to be
understood as the evolution timestep and radial grid point where the data is saved at each resolution. Similarly to
subsection \ref{Subsec:convergence_classical}, the expected convergence
rate for this test is
\begin{align}
  Q = \frac{h_D^2}{h_{D+1}^2} = \frac{h_D^2}{h_D^2/4}=4
  \,,
\end{align}
which we call \textit{exact convergence rate} since the comparison is
against an exact (analytic) solution and not a numerical one.
Figure~\ref{Fig:standing_waves_norm_all} illustrates the second-order convergence of the numerical solution
for~\eqref{eq:quantum_mink_sol} in the norm
\eqref{eq:u_q_L2_norm_all_exact_conv_test}, including all the evolved
modes. However, we know that the modes with high $l$ and $\mu_5$
oscillate faster and so are harder to simulate
accurately. Furthermore, we know that the amplitude of the modes with
high $l$ is smaller than those with lower and hence a possible lack
of convergence for high $l$ might not be seen in the norm
\eqref{eq:u_q_L2_norm_all_exact_conv_test}. For this reason we also
examine convergence in the norm 
\begin{align}
  ||\mathbf{u}_{\text{\tiny qu}}||_{D;60,5}^2
    =
    \sum_{k} \sum_{r=0}^{r_{\causal}}
    &h_D \left[
    \left(u_{k 60 ;5,D} - u_{k 60 ;5}^{\text{\tiny Mink}} \right)
    \left(u_{k 60 ;5,D} - u_{k 60 ;5}^{\text{\tiny Mink}} \right)^*
    \right.
    \nonumber
  \\
  & +
    \left.
    \left(\psi_{k 60 ;5,D} - \psi_{k 60 ;5}^{\text{\tiny Mink}} \right)
    \left(\psi_{k 60 ;5,D} - \psi_{k 60 ;5}^{\text{\tiny Mink}} \right)^*
    \right.
    \nonumber
  \\
  & +
    \left.
    \left(\pi_{k 60 ;5,D} - \pi_{k 60 ;5}^{\text{\tiny Mink}} \right)
    \left(\pi_{k 60 ;5,D} - \pi_{k 60 ;5}^{\text{\tiny Mink}} \right)^*
    \right],
    \label{eq:u_q_L2_norm_604_exact_conv_test}
\end{align}
which is the part of the state vector that includes all the quantum modes with $l=60$ and $n=5$ and we have suppressed the argument $(t,r)$.  Figure \ref{Fig:standing_waves_norm_604} shows the convergence of this part of the numerical solution in the norm
\eqref{eq:u_q_L2_norm_604_exact_conv_test}. Up until $t \simeq 7.9$
the solution exhibits good second-order convergence, which improves
with increasing resolution. However, around $t \simeq 7.9$, there is a
fast loss of convergence, which is not captured by the
norm~\eqref{eq:u_q_L2_norm_all_exact_conv_test}, possibly because the
slower modes that continue to converge well dominate the norm, as they have a
much greater amplitude. This loss of convergence does not lead to a code instability within the evolution time.

\subsection{Vacuum with backreaction}
\label{Subsec:vacuum_backreact}

We also test the evolution of vacuum initial data with
backreaction. In this scenario, we wish to test if the numerical
solution remains closer to Minkowski with an increasing number of
quantum modes backreacting on the geometry. We set $a=0$ for the
classical scalar initial data~\eqref{eq:Phi_ID}, $r_{\max}=30$, $t_{\max}=10$,
$\dd k=\pi/30$, $\dd t=h/16$, $N_r = 128 \cdot 2^D +3$ with $D=3$ and no
artificial dissipation. In the filter function \eqref{eq:backreact_filter}, we set $r_{\cut}=20$. When we increase the number of quantum modes we use
$(k_{\max},l_{\max})=\{ (5,15), (10,30), (20,60) \}$. This choice keeps
the ratio $l_{\max}/k_{\max}=3$ fixed, which we found empirically 
allows us to model accurately the backreaction in the region up to
$r \simeq \pi/\dd k$. By this we mean that for a vanishing classical scalar field, the stress-energy tensor components
at initial time have a flat profile in this region, as
described in \cite{Berczi:2021hdh}. 

\begin{figure}[t]
  \includegraphics[width=0.999\textwidth]{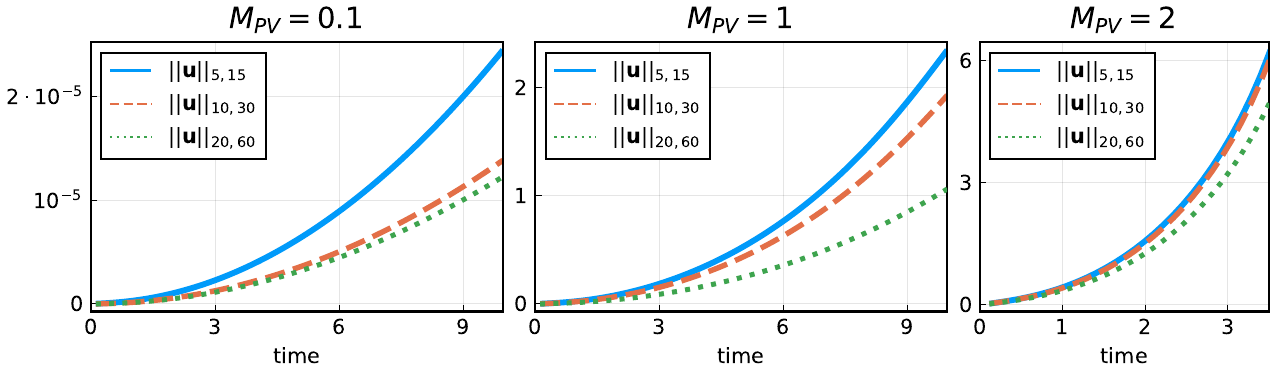}
  \caption{The deviation of the numerical solution with backreaction
    from Minkowski, as measured by the norm
    $||\mathbf{u}||_{k_{\max}, l_{\max}}$
    \eqref{eq:u_geom_Mink_norm}. The deviation decreases with
    increasing number of quantum modes backreacting and with smaller
    values of $M_{\pv}$. }
  \label{Fig:mink_backreaction_norms}
\end{figure}

To measure the deviation of the numerical solution to the Minkowski spacetime solution we
calculate the following norm 
\beq
  || \mathbf{u}||^2_{k_{\max}, l_{\max}
}(t)
  \deff
    \sum_{r=0}^{r_\text{\tiny f}}& h
    \left[
    \left( A(t,r) - A^{\text{\tiny Mink}}(t,r) \right)^2
    + \left( B(t,r) - B^{\text{\tiny Mink}}(t,r) \right)^2
    \right.
  \\
  & \left.
    + \left( D_{B}(t,r) - D_{B}^{\text{\tiny Mink}}(t,r) \right)^2
    + \left( \tilde{U}(t,r) - \tilde{U}^{\text{\tiny Mink}}(t,r) \right)^2
    \right.
  \\
  &
    \left.
    + \left( K(t,r) - K^{\text{\tiny Mink}}(t,r) \right)^2
    + \left( K_B(t,r) - K_B^{\text{\tiny Mink}}(t,r) \right)^2
    \right.
  \\
  &
    \left.
    + \left( \lambda(t,r) - \lambda^{\text{\tiny Mink}}(t,r) \right)^2
    + \left( \alpha(t,r) - \alpha^{\text{\tiny Mink}}(t,r) \right)^2
    \right.
  \\
  & \left.
    + \left( D_{\alpha}(t,r) - D_{\alpha}^{\text{\tiny Mink}}(t,r) \right)^2
    \right]
    \,,
    \label{eq:u_geom_Mink_norm}
\eeq
where $r_\text{\tiny f}$ is some final radial grid point. We perform this test for
$M_{\pv} = \{0.1, 1, 2\}$ and we choose $r_\text{\tiny f} = 10$ for the upper limit in \eqref{eq:u_geom_Mink_norm}, so that we are safely
within the region where backreaction is well-modeled  and causally disconnected from
the outer boundary $r_{\max}$ and $r_{\cut}$ of the
filter function \eqref{eq:backreact_filter}. In figure \ref{Fig:mink_backreaction_norms} we show the results of this test. We see that with increasing number of quantum modes backreacting to the classical geometry, the numerical solution is closer to that of Minkowski spacetime during the evolution, i.e., the size of the deviation from Minkowski as measured with the norm~\eqref{eq:u_geom_Mink_norm} decreases. We also see that this deviation is significantly smaller for low values of $M_{\pv}$ and even detect instabilities that lead to the simulation crashing for $M_{\pv}=2$.

\begin{figure}[t]
  \includegraphics[width=0.999\textwidth]{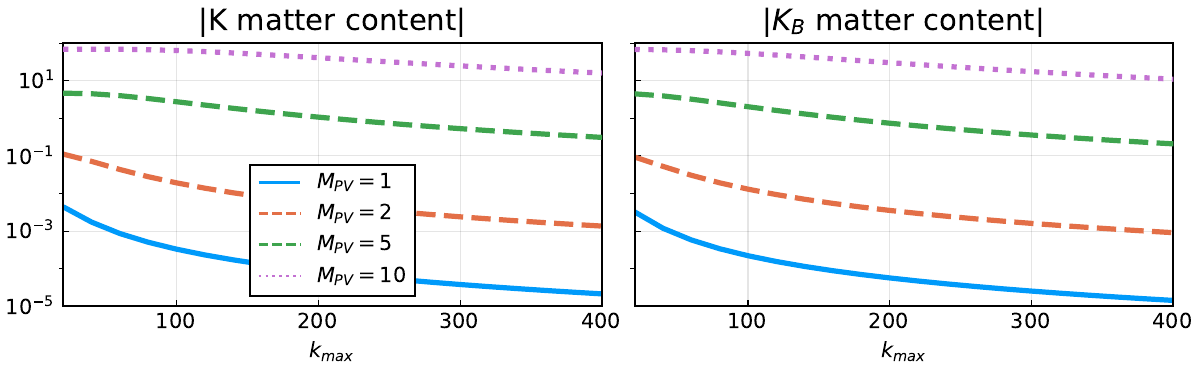}
  \caption{The matter content for the RHS of $K$ (left) and $K_B$ (right) for Minkowski classical initial data at $(t,r)=(0,0)$ for increasing $k_{\max}$, with $l_{\max}=10$ and $M_{\pv}=\{1,2,5,10\}$.}
  \label{Fig:mink_backreaction_ID_kmax_vs_mPV}
\end{figure}

To better understand the interaction between the values of $M_{\pv}$, $k_{\max}$ and $l_{\max}$ in backreaction, we also inspect the quality of the initial data, for different values of $M_{\pv}$, fixed $l_{\max}$ and increasing $k_{\max}$. Based on the results of \cite{Berczi:2021hdh}, empirically we know that keeping $l_{\max}$ fixed and increasing $k_{\max}$ provides a better approximation of the quantum operator, but for a smaller region of $r$, close to $r=0$. If instead $k_{\max}$ is fixed and $l_{\max}$ increases, we get a poorer approximation of the quantum operator but for a larger $r$ region. This result also depends on the value of $\dd k$. We want to understand how fast the matter content of the evolution equation \eqref{eq:K_eom} for $K$ and \eqref{eq:KB_eom} for $K_B$ tend to zero at $t=r=0$, with increasing $k_{\max}$ and fixed $l_{\max}$.
These are given by
\begin{align}
   \frac{1}{M_\pl^2}
      \left( \frac{S_A}{2} + S_B  - \Lambda + \frac{\rho}{2}
      \right)
      \,,
      \quad
    \frac{1}{2 M_\pl^2} \left( S_A - 2 \Lambda - \rho \right)
    \,,
    \label{eq:K_KB_matter_content}
\end{align}
for  \eqref{eq:K_eom} and \eqref{eq:KB_eom}, respectively. This behavior can inform us which are the possible realistic Pauli-Villars masses that we can use in backreaction. If the deviation of these matter contents at the level of initial data and for $r=0$ is significantly far from zero for a given $M_{\pv}$, then we cannot hope to have a good time evolution that would pass our earlier Minkowski consistency test, let alone trust it in more interesting scenarios. We explore this behavior for $l_{\max}=10$, $k_{\max}=[20,400]$, $M_{\pv}=\{1,2,5,10\}$, $\dd k=\pi/r_{\max}$ and $r_{\max}=30$. In figure \ref{Fig:mink_backreaction_ID_kmax_vs_mPV} we see that the $K$ and $K_B$ matter contents \eqref{eq:K_KB_matter_content} converge slower to zero with increasing values of $M_{\pv}$. Therefore, to study quantum correlators we use $M_{\pv}=1$, since larger values of $M_{\pv}$ provide less accurate approximations of the quantum operator, for realistic values of $k_{\max}$ and $l_{\max}$. In our physical explorations of section \ref{Sec:quantum_correlators}, we set $k_{\max}$ at most to 30.

\section{Quantum correlators in black hole formation} \label{Sec:quantum_correlators}

\begin{figure}[t]
    \centering
    \begin{minipage}[b]{0.49\textwidth}
        \centering
        \includegraphics[width=0.85\textwidth]{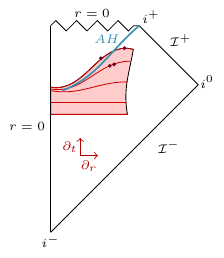}
    \end{minipage}
    \begin{minipage}[b]{0.49\textwidth}
        \centering
        \includegraphics[width=\textwidth]{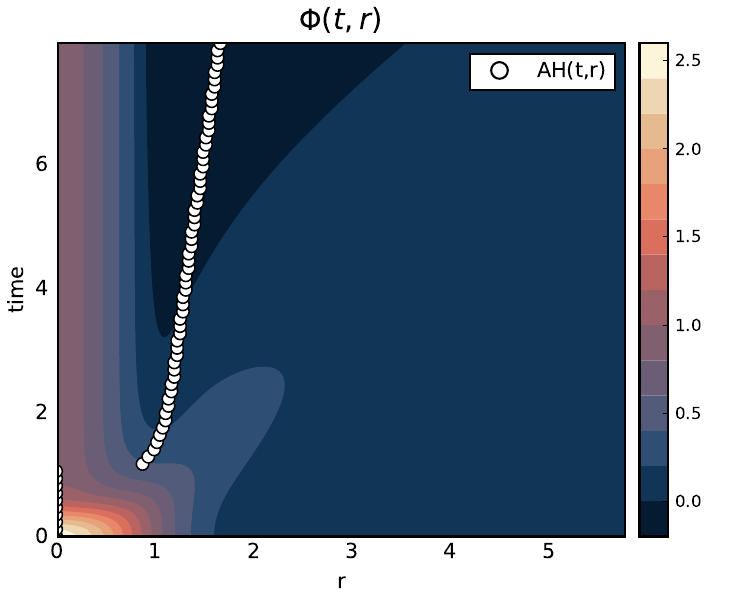}
    \end{minipage}
     \caption{Left: part of the conformal diagram of an asymptotically flat spacetime, where an apparent horizon dynamically forms. The red region depicts our numerical domain and the dots are two arbitrary points on a spacelike hypersurface (solid red lines) for which we calculate the equal time quantum correlators. Right: the evolution of a scalar field in the $t-r$ plane, with $a=1.25$, $b=0$ and $c=1$, which leads to a dynamical formation of an apparent horizon (white circles).}
     \label{Fig:quantum_correlators_and_domain}
\end{figure}

Hawking radiation is a phenomenon that requires quantum fields in the environment of a black hole. Using \texttt{SpheriCo.jl} we can explore such scenarios, where the apparent horizon of a black hole dynamically forms, as for instance seen in the right of figure~\ref{Fig:quantum_correlators_and_domain}. In this scenario, we can examine if there are any non-trivial correlations of the scalar field operator at different spacetime events across the horizon of the black hole. Such a correlation of the scalar field may be interpreted as correlating pairs of Hawking quanta. Videos from simulations of this section can be found in \cite{animations}.

\subsection{Without backreaction}

\begin{figure}[h]
\begin{center}
  \includegraphics[width=0.87\textwidth]{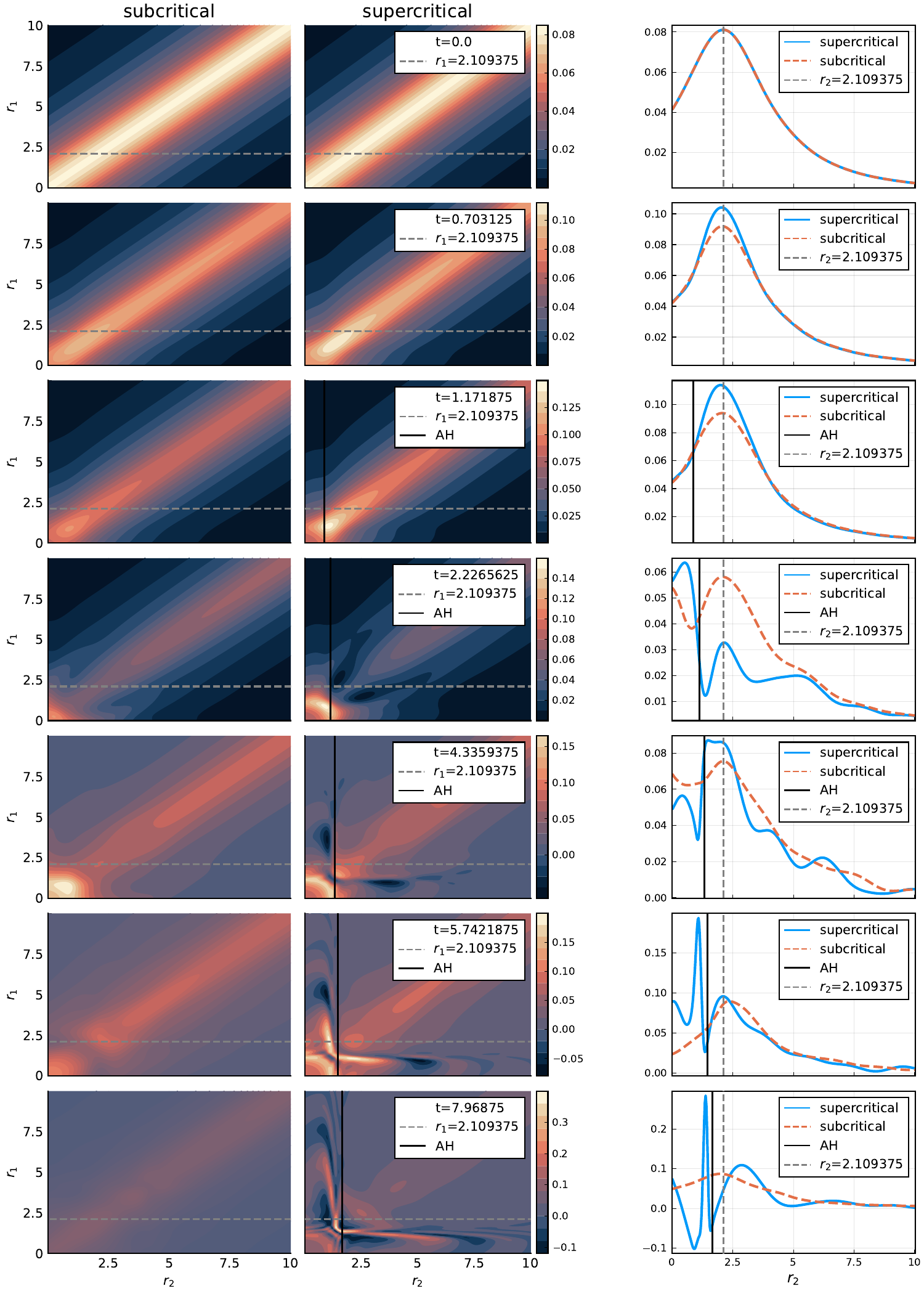}
  \caption{The real part of the equal time correlator \eqref{eq:rr_correlator} for $\Phi(0,r)$ with $b=0$, $c=1$, $k_{\max}=30$, $l_{\max}=90$ and $M_{\pv}=1$ without backreaction. The subcritical case (left) has $a=0.75$ and the supercritical (middle) $a=1.25$. The correlation of point $r_1=2.109375$ with the rest of the domain, at different timesteps, for both sub- and supercritical setups, is shown on the right. For $r_1=r_2$ we see the correlation of the chosen point to itself.}
\label{Fig:rr_correlators_a1.25_vs_a0.75_backreact_false_kmax30_lmax90_mPV1}
\end{center}
\end{figure}

From the two-point correlation function \eqref{eq:twopointcorrelator} of the quantum field, we define the correlation function $C$ of the fluctuations around the field expectation value $\phi$, i.e., 
\beq
  C(t_1,t_2;r_1,r_2) &\deff \bra{\chi}\hat{\Phi}(t_1,r_1)\hat{\Phi}(t,r_2)\ket{\chi}
  - \phi(t_1,r_1)\phi(t_2,r_2)\\
  &=  \frac{\hbar c^2}{4\pi}\int \dd k 
    \sum_{n=0}^5
    \sum_{l=0}^{l_{\max}} 
    (-1)^n (2l+1)
    \left[
    \Tilde{u}_{kl;n}(t_1,r_1) \Tilde{u}^*_{kl;n}(t_2,r_2) 
    \right].
    \label{eq:rr_correlator}
\eeq
In the following we are interested in equal-time correlations, and so we will evaluate $C(t_1,t_2;r_1,r_2)=C(t;r_1,r_2)$ at time $t_1=t_2=t$.
In \texttt{SpheriCo.jl}, one can also calculate analogous correlation functions as \eqref{eq:rr_correlator} for $\p_t \tilde{u}_{kl;n}$, $\p_r \tilde{u}_{kl;n}$, $\pi_{kl;n}$ and $\psi_{kl;n}$, by replacing $\tilde{u}_{kl;n}$ with the appropriate variable.

We perform simulations for both a sub- and supercritical setup, without backreaction and calculate the correlator \eqref{eq:rr_correlator} at different times. For these simulations, the scalar pulse is initially set with $b=0$, $c=1$ and $a=0.75$ for the subcritical and $a=1.25$ for the supercritical case, respectively. We also set $k_{\max}=30$, $l_{\max}=90$, $\dd k=\pi/r_{\max}$, with $r_{\max}=30$ and $M_{\pv}=1$. The radial grid consists of $N_r=1027$ points, the artificial dissipation is $\sigma = 0.02$ and the timestep is set to $\dd t = h/16$. In figure \ref{Fig:rr_correlators_a1.25_vs_a0.75_backreact_false_kmax30_lmax90_mPV1} we present our findings. We shall call the \textit{main diagonal} the one that connects the bottom left corner to the top right corner in the heatmaps of figure \ref{Fig:rr_correlators_a1.25_vs_a0.75_backreact_false_kmax30_lmax90_mPV1}. At $t=0$ the correlation is the same in both cases and it is the same as in Minkowski, which is more prominent on the main diagonal, indicating maximum correlation of a point to itself. For the supercritical case, as time progresses, we see that the correlator exhibits a non-trivial behavior away from the main diagonal, that correlates a point inside the apparent horizon with one outside. We use the terminology of \cite{Carusotto:2008ep} and refer to these as \textit{tongues}. These tongues of correlation grow in size as time progresses, also acquiring negative values. In contrast, in the subcritical case, after an initial increase in size of the correlator around the origin of the heatmap, and a mild oscillatory behavior, the correlator relaxes back to a profile similar to its original one.

Hawking radiation has been studied and observed in analogue black holes \cite{Carusotto:2008ep, Steinhauer:2015saa, Jacquet:2022vak}. There are both similarities and differences between the correlators we compute in the supercritical case, and those from analogue systems with a horizon. In both cases there is significant non-trivial behavior of the correlator away from the main diagonal, which has an oscillatory profile and takes both positive and negative values. However, in analogue systems the correlator maintains a dominant profile in the main diagonal at all times, which is not the case in our simulations. In addition, the non-trivial correlation away from the main diagonal appears in greater angles in the analogue system compared to ours. In our case we expect the details of the correlation tongues to depend on the choice of lapse function, but the qualitative structure of correlations outside and within the horizon to remain. It is possible that our picture is greatly affected by the collapse of the lapse inside the apparent horizon, which effectively freezes the simulation near the origin of our computational domain, leading to that the tongues appear at smaller angles. However, in analogue systems there is no singularity, which might be responsible for the quantitative difference in angle. It might be useful to repeat our calculation with a different gauge choice (e.g. harmonic), and examine if a different qualitative picture arises. We note that a different gauge does not change the value of the scalar field at a given spacetime point, but rather the gauge choice determines the foliation, and hence the choice of points that are used in the correlator. It is also possible that calculating the two-point function at equal proper time and using proper radius instead of $r$ could provide a picture closer to that of the analogue systems. We leave this for future work.

We should emphasize that with this setup we were not able to respect the condition $M_\pv\gg2\pi/\tilde{\lambda}$, given in section~\ref{Sec:numerical_implementation}, due to numerical instability. Here we have $2 \pi/ \tilde{\lambda} \simeq 3.6$, which we chose for practical reasons. To respect $M_\pv\gg2\pi/\tilde{\lambda}$, a choice of $M_{\pv} =100$ could be more appropriate, but as explained in section~\ref{Subsec:vacuum_backreact} and seen in figure \ref{Fig:mink_backreaction_ID_kmax_vs_mPV}, we would not trust this simulation with backreaction and for the number of modes we can practically include. Therefore, we also do not consider it in the setup without backreaction, since it might provide misleading results. Another reason is that Hawking radiation is expected to be more intense for smaller black holes, which is what we try to form dynamically. In principle, we could perform simulations with a wider initial scalar profile, such that we respect the conditions outlined in section~\ref{Sec:numerical_implementation}. However, such a setup either leads to a bigger black hole, or becomes unstable if the amplitude is tuned very low, in an attempt to form a smaller black hole. Maybe there is a region of the parameter space -- in terms of initial data -- that better respects these conditions, that more thorough exploration could reveal.

\subsection{With backreaction}

\begin{figure}[t]
\begin{center}
  \includegraphics[width=0.999\textwidth]{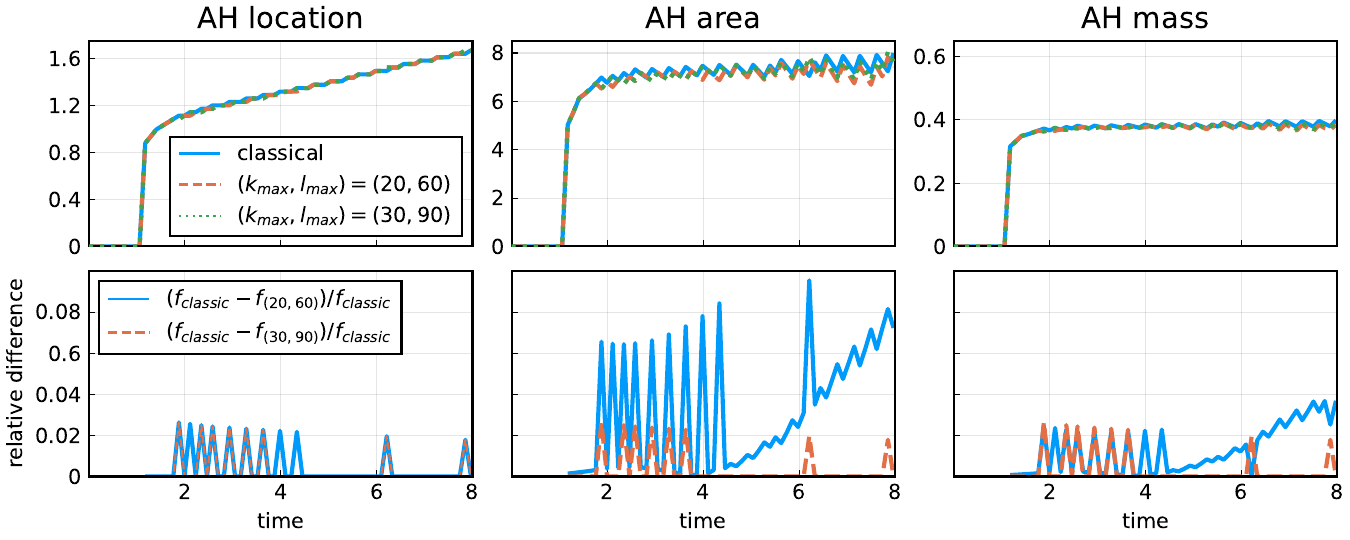}
  \caption{We compare a classical supercritical setup, against its semiclassical version with backreaction, for different values of $(k_{\max},l_{\max})$. For the initial data we consider the scalar pulse \eqref{eq:Phi_ID} with $a=1.25, \, b=0$ and $c=1$. From left to right we present in time the location, area, and mass of the apparent horizon. The bottom row illustrates the relative difference between the semiclassical and classical setups, using the classical one as the benchmark. For all quantities we see that the relative difference between classical and semiclassical is below $10\%$, and decreases with increasing values of $(k_{\max},l_{\max})$.}
\label{Fig:bh_classical_vs_backreact}
\end{center}
\end{figure}

We repeat the supercritical simulations from the previous subsections, including backreaction. To preserve the asymptotics of the spacetime near $r_{\max}$ we use the filter function \eqref{eq:backreact_filter} with $r_{\cut} = 20$. We explore two backreacting setups with $(k_{\max},l_{\max})=\{(20,60),(30,90)\}$ and analyze the effect they have on the location of the apparent horizon, as well as the area and the mass of the black hole. As mentioned in subsection \ref{Subsec:classical_ID_BC}, we track the location of the apparent horizon $r_{\AH}$ by finding the outermost radius at which the expansion of null rays $\theta_\text{\tiny exp}$ vanishes. The  area of the apparent horizon is defined as
\begin{align}
    A_{\AH} \deff 4 \pi B \,r_{\AH}^2,
    \label{eq:AH_area}
\end{align}
and the apparent horizon mass \cite{alcubierre2008introduction} as
\begin{align}
    M_{\AH} \deff \frac{r_{\AH}}{2} \sqrt{B|_{r_{\AH}}}.
    \label{eq:AH_mass}
\end{align}
In figure \ref{Fig:bh_classical_vs_backreact} we see that all these three quantities are very similar in the classical and backreaction cases. More specifically, their relative difference is always below $10\%$ and decrease with increasing values of $k_{\max}$ and $l_{\max}$. We expect the semiclassical solution to be more reliable with increasing number of quantum modes. We also expect the deviation of the semiclassical picture to be bigger for smaller black holes. With the current status of our code we cannot simulate very small black holes, and so we do not expect a significant deviation from the classical geometry. Even though it is not clear what the semiclassical solution would converge to in the limit of infinite modes, the trend we see in figure \ref{Fig:bh_classical_vs_backreact} is not unexpected.

\begin{figure}[t]
\begin{center}
  \includegraphics[width=0.999\textwidth]{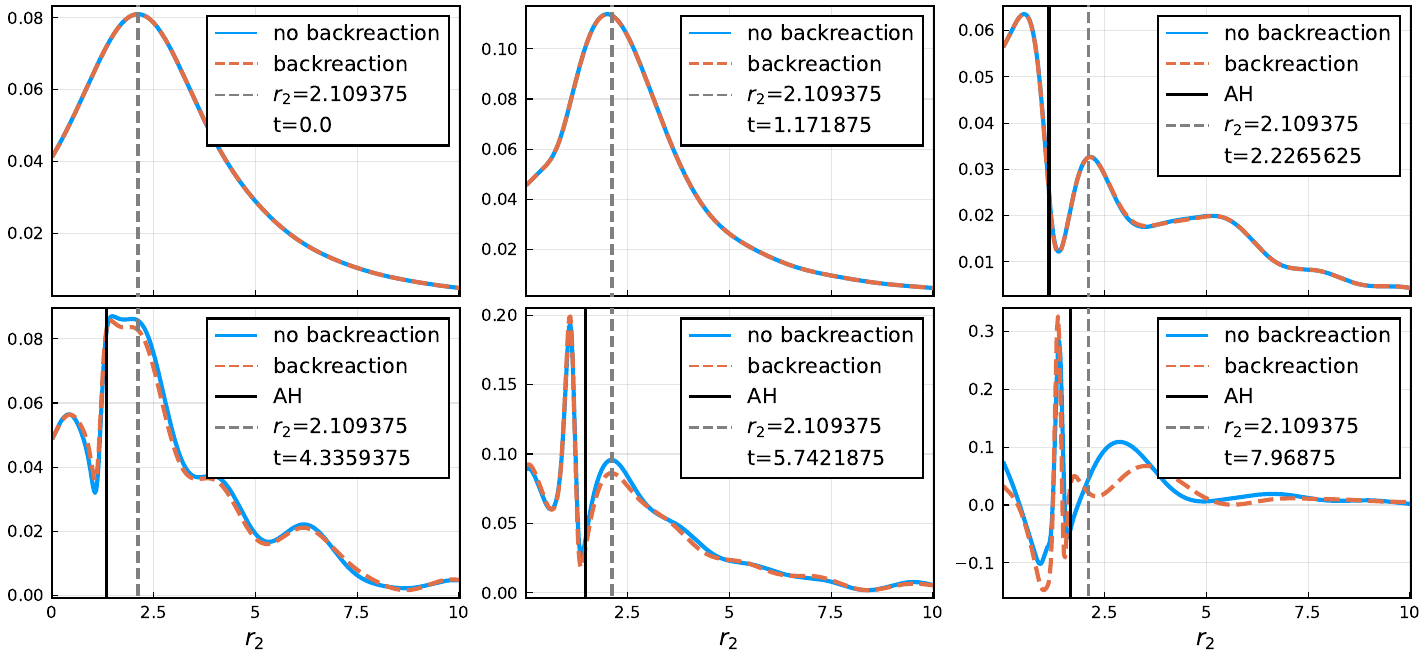}
  \caption{The real part of the correlation function \eqref{eq:rr_correlator} at $r_1=2.109375$ for a semiclassical supercritical setup with $M_\pv=1$ and $(k_{\max},l_{\max})=(30,90)$, with and without backreaction. The initial scalar profile is given by  \eqref{eq:Phi_ID} with $a=1.25,\, b=0,\, c=1$. The behavior is very similar in both cases, with the backreacting one presenting slightly enhanced oscillations at late times around the location of the apparent horizon.}
\label{Fig:bh_classical_vs_backreact_crlts}
\end{center}
\end{figure}

In addition, we compare the correlation \eqref{eq:rr_correlator} of the case with $(k_{\max},l_{\max})=(30,90)$ with backreaction, against the supercritical case without backreaction and with the same quantum modes, from the previous subsection. Qualitatively the behavior appears to be very similar. In figure \ref{Fig:bh_classical_vs_backreact_crlts} we demonstrate this by presenting the equal-time correlation function for $r_1=2.109375$, with and without backreaction. In the backreacting case we see that the oscillatory behavior around $r_{\AH}$ is more intense and that the negative value of the correlator inside the apparent horizon increases in absolute magnitude. In the neighborhood of the parameters we have explored, we expect this feature to be similar. However, whether it is robust across the wider parameter space is unclear at the moment and requires further investigation. We leave this for future work.

\section{Outlook} \label{Sec:outlook}

In this paper we have introduced \texttt{SpheriCo.jl} to simulate the spherically symmetric collapse of a scalar field into a black hole, in classical and semiclassical gravity. \texttt{SpheriCo.jl} is modular, open-access and written in the \texttt{Julia} programming language. A challenge present in many studies of gravitational collapse is the approximation of the evolved variables near $r=0$, since terms of the form $1/r$ appear in their equations of motion. This challenge becomes increasingly difficult in the semiclassical setup, where these terms can become very large and render the equations stiff. This seems to have been the main reason for driving earlier semiclassical simulations on the topic only stable for a short time. By combining the summation-by-parts operators of \cite{Gundlach:2010et} with the convenient evolved variables of \cite{Alcubierre:2004gn} we manage to perform semiclassical simulations for longer than it was previously possible, and explore quantum correlations of events across an apparent horizon.

Constraint violations can have a big impact on numerical simulations by driving the numerical solution significantly far from the physical one, or even making the simulation unstable. We attempt to control the numerical violation of the classical Hamiltonian and momentum constraints by combining the $Z_4$ constraint damping scheme of \cite{Bernuzzi:2009ex} with the system of \cite{Alcubierre:2004gn}. In the latter, formally singular terms for $r=0$ of the form $1/r$ appear, but are practically regular due to the specific reduction variables used. We approximate these terms with the second-order accurate summation-by-parts operators of \cite{Gundlach:2010et}. We perform multiple tests on the classical module of our code. We recover the expected second-order convergence for both smooth and noisy data, either sub- or supercritical. We also test the performance of our implemented constraint damping scheme, which seems to be unsuccessful. More specifically, the violation of the Hamiltonian and momentum constraints with 
 the damping scheme is equal or greater than without, or even leads to unstable simulations for certain subcritical data. We should highlight that we do not include damping of reduction constraints in our system and that we use different evolved variables than \cite{Bernuzzi:2009ex}, features that might be responsible for this failure. One possible future improvement would be to explore the constraint damping scheme further and possibly add the damping of reduction constraints.

 In terms of physical tests for the classical module, we also recover the expected near critical oscillating behavior of a scalar field, as well as the universality of the solution. This is a non-trivial result, especially given that to our knowledge the specific summation-by-parts operators have not been used again in this case, and are originally constructed for a different setup. We should also highlight that to obtain this result we had to allow our code to run with an infalling outer boundary. This is achieved numerically, since our formulation has a vanishing shift. We expect that it is numerically more economic to have a dynamic shift and perform the infalling outer boundary technique analytically. Adding the option of a dynamical shift condition could be another possible future improvement. It could make the code more appropriate for near critical simulations, especially in the semiclassical setup, where the numerical infalling outer boundary technique seems to be unstable.

In the semiclassical setup the geometry is still classical, but the scalar field is promoted to a quantum field. To model this we utilize an expansion in spherical harmonics and radial mode functions labeled by the quantum numbers $k$ and $l$, for which we solve the resulting equations of motion. The numerical approximation of the quantum field is improved with increasing number of modes included in the system. In \texttt{SpheriCo.jl} the semiclassical setup can be simulated with or without backreaction to the classical geometry. Including  backreaction means the expectation value of the quantum stress-energy tensor replaces that of the classical one. To obtain a non-divergent expectation value, we use the Pauli-Villars regularization scheme, where massive ghost fields are included in the system. In theory, one would like to take the number of quantum modes, and the Pauli-Villars mass, to infinity. In practice we are limited by computational resources and the performance of our code, and truncate them to certain values. For the setup without backreaction we test our code against an exact solution for the quantum modes and recover good second-order convergence, even for high values of $l$, higher than those tested in \cite{Gundlach:2010et}. To test the backreaction implementation, we examine whether we recover the Minkowski solution with smaller error, as we include more modes in the simulation. The code passes this test, but its performance depends on the value of the Pauli-Villars mass, with better performance for smaller values. 

Finally, we explore the correlation of Hawking pairs around black holes, by calculating quantum two-point correlation function of the scalar in geometries with a dynamically forming apparent horizon. Our results suggest a non-trivial correlation of events inside and outside the apparent horizon. Regarding the number of quantum modes simulated, the value of the Pauli-Villars mass, and the size of the apparent horizon, we limit our simulations to setups where we trust the performance of our code, based on tests we conducted earlier. A more extended study that scans a wider region of the parameter space  would be necessary to understand the phenomenon deeper, but we leave this for future work. We also highlight certain similarities and differences of our findings to those of analogue black hole systems.

\begin{acknowledgments}
\label{Sec:acknowledgements}

We would like to thank Miguel Bezares, Carsten Gundlach, David
Hilditch, Krinio Marouda and Isabel Su\'arez Fern\'andez for helpful
discussions, Carsten Gundlach for sharing with us his code related to
the summation-by-parts operators and David Hilditch for sharing with
us his notebook on the $Z_4$ constraint damping formulation. This work received partial support from the Science and Technology Facilities Council (STFC) [Grant
Nos. ST/T000732/1, ST/X000672/1 and ST/V005596/1].
\end{acknowledgments}

\section*{Data Access Statement}

Code may be found in \cite{SpheriCo.jl}. Data may be found in \cite{zenodo}. All the data were generated using \texttt{SpheriCo.jl} on a laptop with a 13th Gen Intel Core i7-1370P 14-core processor and 64 GB RAM. The longest simulation took around 15 clock-time hours to complete.

\bibliographystyle{JHEP}
\bibliography{refs}{}

\begin{appendix}

\section{Hyperbolicity of the classical system}
\label{app:classical_hyp}
The hyperbolicity of a partial differential equation (PDE) system is connected to the well-posedness of the corresponding PDE problem, which is a necessary condition for trustable numerical simulations. It can also be used to understand what are appropriate boundary conditions for a specific problem. To study the hyperbolicity of the evolved system \eqref{eq:classicalEoMs_matter}--\eqref{eq:classicalEoMs_grav}, we linearise it around an arbitrary background 
\beq
    \mathbf{u}_0 = \left( \Phi_0, \Pi_0, \Psi_0, A_0, B_0, D_{B 0}, \tilde{U}_0,
  K_0, K_{B0}, \lambda_0, \alpha_0, D_{\alpha 0}, \Theta_0, Z_{r0}\right)^T\,,
  \label{eq:u0}
\eeq
and work in the constant coefficient approximation \cite{kreiss1973methods}. Since the Einstein equations are quasi-linear (i.e. linear in highest derivatives), the degree of hyperbolicity of the system is the same as its corresponding linear system \eqref{eq:u0}. The linearised Einstein equations can then be written as
\beq
    \dot{\mathbf{u}} = \mathbf{A}^r\left( \mathbf{u}_0 \right) \, \mathbf{u}'
  + \mathbf{S}\left( r, \mathbf{u} , \mathbf{u}_0, \mathbf{u}_0'\right),
  \label{eq:app:lin_sys}
\eeq
where $\mathbf{A}^r$ is the principal part matrix containing the coefficients of the first-order (radial) derivative of $\mathbf{u}$, and $\mathbf{S}$ only containing lower-order (in derivatives) terms. 
The linearised system can be brought into the form
\beq
  \dot{\mathbf{v}} \simeq \mathbf{J}^r\left( \mathbf{u}_0 \right)\mathbf{v}',
\eeq
where $\simeq$ denotes equality in the principal terms and 
\begin{align}
  \mathbf{J}^r  \left( \mathbf{u}_0 \right)
  &
    \deff
  \mathbf{T}^{-1} \left( \mathbf{u}_0 \right)
  \mathbf{A}^r \left( \mathbf{u}_0 \right)
  \mathbf{T} \left( \mathbf{u}_0 \right)
    \label{eq:Jordan_form_Ar}
  \\
  & =
    \textrm{diag} \left(0, 0, 0, 0, 0, 0,
    -\sqrt{\frac{2 \alpha_0}{A_0}}, \sqrt{\frac{2 \alpha_0}{A_0}},
    - \frac{\alpha_0}{\sqrt{A_0}}, - \frac{\alpha_0}{\sqrt{A_0}}, - \frac{\alpha_0}{\sqrt{A_0}},
    \frac{\alpha_0}{\sqrt{A_0}}, \frac{\alpha_0}{\sqrt{A_0}}, \frac{\alpha_0}{\sqrt{A_0}}
    \right)
    \,,
    \nonumber
\end{align}
is the Jordan normal form of $\mathbf{A}^r$, $\mathbf{T}$ is the transformation
matrix and $\mathbf{v} \deff \mathbf{T}^{-1}\left(\mathbf{u}_0\right) \mathbf{u}$ is a vector containing the characteristic variables of the system. If $\mathbf{J}^r$ is real and diagonal, then the system \eqref{eq:app:lin_sys} is called strongly hyperbolic and each characteristic variable satisfies an advection equation up to lower order terms.

The eigenvalues $\mathcal{E}_i$ of $\mathbf{A}^r$ (entries of $\mathbf{J}^r$) determine the radial propagation speeds of the characteristic variables (entries of $\mathbf{v}$) and are referred to as the characteristic speeds of the system. In a spherically symmetric spacetime, stationary modes ($\mathcal{E}_i=0$) do not propagate radially. Outgoing modes ($\mathcal{E}_i<0$) propagate outward towards increasing $r$ and ingoing modes ($\mathcal{E}_i>0$) propagate inward towards decreasing $r$. The static, outgoing and ingoing modes corresponding to \eqref{eq:Jordan_form_Ar} are
\beq
    \mathbf{v}_{\text{\tiny stat}} =
  \left(
  - \alpha_0 D_\alpha - 4 D_B + \tilde{U} + 4 Z_r ,
  \frac{A_0 D_B}{B_0} + \lambda ,
   \alpha , B , A , \Phi
  \right)^T, 
\eeq
and 
\beq
     \mathbf{v}_{\text{\tiny out}} = \left(v_{\text{\tiny out,1}},v_{\textrm{\tiny out,2}},v_{\textrm{\tiny out,3}},v_{\textrm{\tiny out,4}}\right)^T,\qquad
     \mathbf{v}_{\text{\tiny in}}  = \left(v_{\text{\tiny in,1}}, v_{\textrm{\tiny in,2}}, v_{\textrm{\tiny in,3}}, v_{\textrm{\tiny in,4}}\right)^T,
\eeq
with 
\beq
    v_{\textrm{\tiny out,1}} &= \frac{1}{2 \alpha_0 - 4 }\bigg\{\sqrt{2 A_0 \alpha_0}\left[K \left( \alpha_0 -                        2 \right) - 4 K_B \alpha_0 + 2 \left( \alpha_0 + 2 \right) \Theta \right]\\
    &\quad + \alpha_0 \left[ D_\alpha\left(\alpha_0 - 2 \right)+ 4 D_B -8 Z_r \right] \bigg\},\\
    v_{\text{\tiny out,2}} &= \frac{1}{2}\left[ \sqrt{A_0} \left( 2 K_B - \Theta \right) + Z_r \right],\\
    v_{\text{\tiny out,3}} &= \frac{\sqrt{A_0} \Theta + D_B - Z_r}{2 \sqrt{A_0}},\\
    v_{\text{\tiny out,4}} &= \frac{1}{2} \left(\Pi - B_0 \Psi \right),
\eeq
and 
\beq
    v_{\text{\tiny in,1}} &= \frac{1}{2 \alpha_0 - 4 }\bigg\{-\sqrt{2 A_0 \alpha_0}\left[K \left( \alpha_0 -                        2 \right) - 4 K_B \alpha_0 + 2 \left( \alpha_0 + 2 \right) \Theta \right]\\
    &\quad + \alpha_0 \left[ D_\alpha\left(\alpha_0 - 2 \right)+ 4 D_B -8 Z_r \right] \bigg\},\\
    v_{\text{\tiny in,2}} &= \frac{1}{2}\left[ \sqrt{A_0} \left(\Theta - 2 K_B \right) + Z_r \right],\\
    v_{\text{\tiny in,3}} &= \frac{\sqrt{A_0} \Theta - D_B + Z_r}{2 \sqrt{A_0}},\\
    v_{\text{\tiny in,4}} &= \frac{1}{2} \left(\Pi + B_0 \Psi \right).
\eeq

In order to obtain boundary conditions using the above analysis, the background values \eqref{eq:u0} are replaced with the evolved variables. We expect these values to be close to the Minkowski solution for the spatial region near $r_{\max}$, provided that $r_\max$ is placed far enough from the centre of the scalar Gaussian profile.

When the classical evolution system \eqref{eq:classicalEoMs_matter}--\eqref{eq:classicalEoMs_grav} is strongly hyperbolic, its initial value problem is well-posed in the $L^2$-norm \cite{GusKreOli95}, i.e.,
\begin{align}
  ||\mathbf{u}(t,r)||  \leq c_1 \, e^{c_2 t} || \mathbf{u}(0,r)||,
  \label{eq:well-posendess_ineq}
\end{align}
for some constants $c_1\geq 1$ and $c_2\in \mathbb{R}$, with
\begin{align}
  || \mathbf{u}(t,r)||^2
  &
    = \int_0^{r}\dd r'
    \sum_i u_i^2  (t,r')  ,
    \label{eq:L2_classical_sys}
\end{align}
where $i$ denotes the elements of $\mathbf{u}$. In subsections \ref{Subsec:convergence_classical} and
\ref{Subsec:robust_stability_classical} we utilise the
norm \eqref{eq:L2_classical_sys} to perform convergence tests, where
we take $r = r_\text{\tiny causal}$, which is the maximum value of $r$ that is
causally disconnected from $r_{\max}$.

\section{Double null co-ordinates for postprocessing}
\label{app:double_null}

While the line element (\ref{eq:lineelement}) is useful for the simulation, double-null co-ordinates are useful for understanding the global structure of the geometry. This also gives another natural set of co-ordinates in which to look at $\langle\hat\Phi\hat\Phi\rangle$ correlators, namely correlators along a null surface. For this reason, we added in \texttt{SpheriCo.jl} the option for
the following post-processing calculation. We start by writing the line element as
\begin{align}
    \ud s^2&=-\left[\alpha(t,r)\ud t-\sqrt{A(t,r)}\ud r\right]\left[\alpha(t,r)\ud t+\sqrt{A(t,r)}\ud r\right]+r^2B(t,r)\ud\Omega
    ,
\end{align}
and introduce new co-ordinates $U(t,r)$, $V(t,r)$ defined by
\begin{align}\label{eq:U_def}
    \frac{1}{f(t,r)}\ud U&=\alpha(t,r)\ud t-\sqrt{A(t,r)}\ud r,\\\label{eq:V_def}
    \frac{1}{g(t,r)}\ud V&=\alpha(t,r)\ud t+\sqrt{A(t,r)}\ud r,
\end{align}
for some $f$ and $g$ to be determined. This leads to
\begin{align}
    \ud s^2&=-\frac{\ud U\;\ud V}{fg}+r^2B(t,r)\ud\Omega .
\end{align}
Now we take the defining relations (\ref{eq:U_def}), (\ref{eq:V_def}) to find
\begin{align}\label{eq:U_eom}
    \alpha f&=\frac{\del U}{\del t},\qquad-f\sqrt{A}=\frac{\del U}{\del r},\\\label{eq:V_eom}
    \alpha g&=\frac{\del V}{\del t},\qquad g\sqrt{A}=\frac{\del V}{\del r},
\end{align}
noting that integrability requires
\begin{align}
    \frac{\del}{\del r}\frac{\del U}{\del t}&=\frac{\del}{\del t}\frac{\del U}{\del r},\qquad\frac{\del}{\del r}\frac{\del V}{\del t}=\frac{\del}{\del t}\frac{\del V}{\del r},
\end{align}
which leads to
\begin{align}
    \frac{\del}{\del t}f&=-\frac{\del_r(\alpha f)+f\del_t\sqrt{A}}{\sqrt{A}},
    \label{eq:app:f_evol}
    \\
    \frac{\del}{\del t}g&=-\frac{\del_r(\alpha g)-g\del_t\sqrt{A}}{\sqrt{A}}.
    \label{eq:app:g_evol}
\end{align}
These differential equations for $f(t,r)$ and $g(t,r)$ may be solved once inital profiles for them have been set. A natural choice is $f(t=0,r)=g(t=0,r)=1$ and we furthermore choose $U(t=0,r=0)=V(t=0,r=0)=0$. Now we have initial data for $U(t,r)$ and $V(t,r)$, we may evolve to find $U(t,r)$, $V(t,r)$ using the time evolution equations in (\ref{eq:U_eom}) and (\ref{eq:V_eom}). 
For convenience, we solve the evolution equations \eqref{eq:U_eom}, \eqref{eq:V_eom}, \eqref{eq:app:f_evol} and \eqref{eq:app:g_evol} together with our classical system \eqref{eq:classicalEoMs_grav}, and store the respective variables.

\section{Pauli-Villars counterterms}
\label{app:counterterms}

The UV divergences contained in the expectation values of the stress-energy tensor can be cancelled using metric and curvature tensors, as expressed in \eqref{eq:Treg}. The coefficients of these counterterms are therefore independent of the background metric they are computed in. However, since some metrics have vanishing curvature tensors, not all metrics are appropriate to determine all possible counterterms. For instance, in Minkowski spacetime, only the cosmological constant $\Lambda$ in \eqref{eq:Treg} can be computed. Interesting discussions on regularization in curved spacetimes may be found in \cite{Pla:2022spt, Baacke:1997rs, Baacke:1999gc, Markkanen:2013nwa}.

A relatively simple metric choice with non-vanishing curvature tensors is the FRW metric, 
\beq
    \dd s^2=-\dd t^2+a^2(t)\dd\mathbf{x}^2,
\eeq
where $a(t)$ is the scale factor. In this case, the regularized stress-tensor can be written 
\beq
    T_{ab}^\text{reg} = \av{\chi|\hat{T}_{ab}|\chi} + \Lambda g_{ab} + \delta M_\pl^2G_{ab}+\delta\epsilon_1 H^{(1)}_{ab}+\delta\epsilon_2 H^{(2)}_{ab}+\delta\epsilon_3 H_{ab}, 
\eeq
where the $\mathcal{O}(R^2)$ gravitational terms $H^{(1)}_{ab}, H^{(2)}_{ab}$ and $H^{}_{ab}$ can be found in \cite{Paz:1988mt}. On dimensional grounds we expect that the counterterms depend on the Pauli-Villars mass as $\Lambda\propto M_\pv^4$, $\delta M_\pl^2\propto M_\pv^2$ and $\delta\epsilon_{1,2,3}\propto \ln M_\pv$. As in section \ref{Subsec:RegSEtensor}, we set the renormalisation condition that $\delta\epsilon_{1,2,3}=0$ for the range of Pauli-Villars masses considered.

To compute the counterterms $\Lambda$ and $\delta M_\pl^2$ we consider a free massive scalar field $\hat{\phi}$, with mode expansion 
\beq
    \hat{\phi}(t,\mathbf{x}) = \int \dd^3\mathbf{k}\left[\hat{a}_ku_k(t)e^{i\mathbf{k}\cdot\mathbf{x}} + \hat{a}_k^\dagger u_k^*(t)e^{-i\mathbf{k}\cdot\mathbf{x}}\right],
\eeq
where the modes satisfy the equation of motion
\beq
    \ddot{u}_k+3H\dot{u}_k+\mu^2u_k^2=0,
    \label{eq:modeEoM_PVApp}
\eeq
where $H\deff \dot{a}/a$ is the Hubble constant. The conjugate momentum is $\hat{\pi}=a^3\hat{\phi}$ and fulfils the commutation relations $[\hat{\phi}(t,\mathbf{x})\hat{\pi}(t',\mathbf{x'})] = \frac{i}{a^3(t)}\delta^{(3)}(\mathbf{x}-\mathbf{x'})$, where the normalisation has been chosen according to $[\hat{a}_k,\hat{a}_{k'}^\dagger]=\delta^{(3)}(\mathbf{k}-\mathbf{k'})$.

In terms of the modes $u_k$, the expectation values of the stress-energy tensor components are 
\begin{subequations}
\begin{align}
    \av{\chi|\hat{T}_{00}|\chi} &= \int\dd^3k\, \frac{1}{2}\left[ |\dot{u}_k|^2 + \frac{k^2}{a^2}|u_k|^2 + \mu^2|u_k|^2\right], \label{eq:T_00FRWgeneral}\\
    \av{\chi|\hat{T}^i{}_i|\chi} &= \int\dd^3k\, \frac{1}{2}\left[ 3|\dot{u}_k|^2 - \frac{k^2}{a^2}|u_k|^2 - 3\mu^2|u_k|^2\right],
    \label{eq:T_iiFRWgeneral}
\end{align}
\label{eq:T_FRWgeneral}
\end{subequations}
Assuming the background to be slowly varying, we may use the WKB approximation for the mode functions, 
\beq
    u_k(t) = \frac{1}{\sqrt{2(2\pi)^3a^3(t)W(t)}}e^{-i\int^t\dd t'W(t')},
\eeq
where the equation of motion \eqref{eq:modeEoM_PVApp} leads to an equation for $W(t)$,
\beq
    W^2 = \omega_k^2-\frac{3}{2}\frac{\ddot{a}}{a}-\frac{3}{4}H^2+\frac{3\dot{W}^2}{4W^2}-\frac{\ddot{W}}{2W},
    \label{eq:WadiabaticExp}
\eeq
with $\omega_k^2\deff k^2/a^2+\mu^2$. This can be solved iteratively, where at zeroth adiabatic order $W_{(0)}^2=\omega_k^2$. Substituting $W_{(0)}^2$ into the right-hand side of \eqref{eq:WadiabaticExp}, we find to the next order
\beq
    W_{(0)}^2+W^2_{(2)} &= \omega_k^2-\frac{3}{2}\frac{\ddot{a}}{a}-\frac{3}{4}H^2+\frac{3}{4}\frac{\dot{\omega}_k^2}{\omega_k^2}-\frac{\ddot{\omega}_k}{2\omega_k}\\
    &= \omega_k^2-\frac{2\omega_k^2+\mu^2}{2\omega_k^2}\frac{\ddot{a}}{a}-\frac{4\omega_k^4+4\mu^2\omega_k^2-5\mu^4}{4\omega_k^4}H^2,
    \label{eq:approxW}
\eeq
where we have expressed $\dot{\omega}_k$ and $\ddot{\omega}_k$ in terms of $a$ and $\ddot{a}$ using the dispersion relation. In principle one may continue to higher WKB order, but it will be necessary to only compute the counterterms $\Lambda$ and $\delta M_\pl^2$. We can then set \eqref{eq:approxW} as our approximation for $W^2$ and use it to compute the stress-energy components in \eqref{eq:T_FRWgeneral}. Here we note that, at this order in the adiabatic expansion, 
\beq
    \frac{\dot{W}}{W} = -\frac{\omega_k^2-\mu^2}{\omega_k^2}H, \quad 
    \frac{1}{W} = \frac{1}{\omega_k}\left[1-\frac{W^2_{(2)}}{2\omega_k^2}\right].
\eeq
For the temporal component of the stress-energy tensor \eqref{eq:T_00FRWgeneral}, we have 
\beq
    \av{\chi|\hat{T}_{00}|\chi} = \int\frac{\dd k\, k^2 }{(2\pi)^2a^3}\left[ \sqrt{k^2/a^2+\mu^2} + \frac{H^2}{8}\frac{(2k^2/a^2+3\mu^2)^2}{(k^2/a^2+\mu^2)^{5/2}}\right].
    \label{eq:T_00FRW1}
\eeq
The above expression applies to a single scalar field. For our system with a physical massless scalar and five additional Pauli-Villars ghost fields with masses as in \eqref{eq:PVmasses}, the first integral in \eqref{eq:T_00FRW1} becomes proportional to
\beq
    \int_0^\infty \dd \xi \, \xi^2\left(\xi-2\sqrt{\xi^2+1}+2\sqrt{\xi^2+3}-\sqrt{\xi^2+4}\right)
    = \frac{1}{8}\ln\left(\frac{3^9}{2^{16}}\right),
\eeq
with $\xi\deff k/(a M_\pv)$.
Similarly, for the second integral,
\beq
    \int\dd\xi\, \xi^2\left( \frac{4}{\xi} - 2\frac{(2\xi^2+3)^2}{(\xi^2+1)^{5/2}} 
    + 2\frac{(2\xi^2+9)^2}{(\xi^2+3)^{5/2}} - \frac{(2\xi^2+12)^2}{(\xi^2+4)^{5/2}}\right) 
    = \ln\left(\frac{2^8}{3^6}\right).
\eeq
Collecting these terms, we find 
\beq
    \av{\chi|\hat{T}_{00}|\chi} &= \frac{M_\pv^4}{8(2\pi)^2}\ln\left(\frac{3^9}{2^{16}}\right) + \frac{M_\pv^2}{24(2\pi)^2}\ln\left(\frac{2^8}{3^6}\right)3H^2\\
    &= -\frac{M_\pv^4}{8(2\pi)^2}\ln\left(\frac{3^9}{2^{16}}\right)g_{00} + \frac{M_\pv^2}{24(2\pi)^2}\ln\left(\frac{2^8}{3^6}\right)G_{00},
\eeq
where on the second row we have expressed the result in terms of the metric and Einstein tensor component in FRW spacetime, $G_{00}=3H^2$.

For the spatial part of the stress-energy tensor \eqref{eq:T_iiFRWgeneral}, we have
\beq
    \av{\chi|\hat{T}^i{}_i|\chi} &= \int\frac{\dd^3 \mathbf{k} }{(2\pi)^3}\frac{1}{a^3\omega_k}\left[ \frac{\omega_k^2-\mu^2}{2} + \frac{H^2}{16}\frac{(2\omega_k^2+\mu^2)(2\omega_k^4-m^2\omega_k^2+5\mu^4)}{\omega_k^{6}}-\frac{(2\omega_k^2+\mu^2)^2}{8\omega_k^{4}}\frac{\ddot{a}}{a}\right]\\
    &=\int\frac{\dd^3\mathbf{k}  }{(2\pi)^3}\frac{1}{a^3\omega_k}\left[  \frac{1}{2} \tilde{k}^2+\frac{H^2}{16}\frac{(2\tilde{k}^2+3\mu^2)(2\tilde{k}^4+3\mu^2\tilde{k}^2+6\mu^4)}{(\tilde{k}^2+\mu^2)^3}-\frac{(2\tilde{k}^2+3\mu^2)^2}{8(\tilde{k}^2+\mu^2)^{2}}\frac{\ddot a}{a}\right],
    \label{eq:T_iiFRW1}
\eeq
where $\tilde{k}\deff k/a$. Including the Pauli-Villars fields, we may compute \eqref{eq:T_iiFRW1} analogously to \eqref{eq:T_00FRW1}, to obtain
\beq
    \av{\chi|\hat{T}^i{}_i|\chi} 
    &= -3\frac{M_{\pv}^4}{8(2\pi)^2}\ln\left( \frac{3^9}{2^{16}}\right)
    -\frac{M_{\pv}^2}{24(2\pi)^2}\ln\left( \frac{2^8}{3^6}\right)3H^2
    -\frac{M_{\pv}^2}{24(2\pi)^2}\ln\left( \frac{2^8}{3^6}\right)\frac{ 6\ddot a}{a}\\
    &=-\frac{M_{\pv}^4}{8(2\pi)^2}\ln\left( \frac{3^9}{2^{16}}\right)\eta^i{}_i+\frac{M_{\pv}^2}{24(2\pi)^2}\ln\left( \frac{2^8}{3^6}\right)G^i{}_i.
\eeq
where we have used $G^i{}_i=-3\left[H^2+2\frac{\ddot{a}}{a}\right]$. Hence, we recover the result \eqref{eq:TregExplicit} for the regularized stress-energy tensor, which sets the counterterms according to \eqref{eq:CountertermsEq}.

\section{Additional code testing}
\label{app:code_tests}

\begin{figure}[t]
  \includegraphics[width=0.9\textwidth]{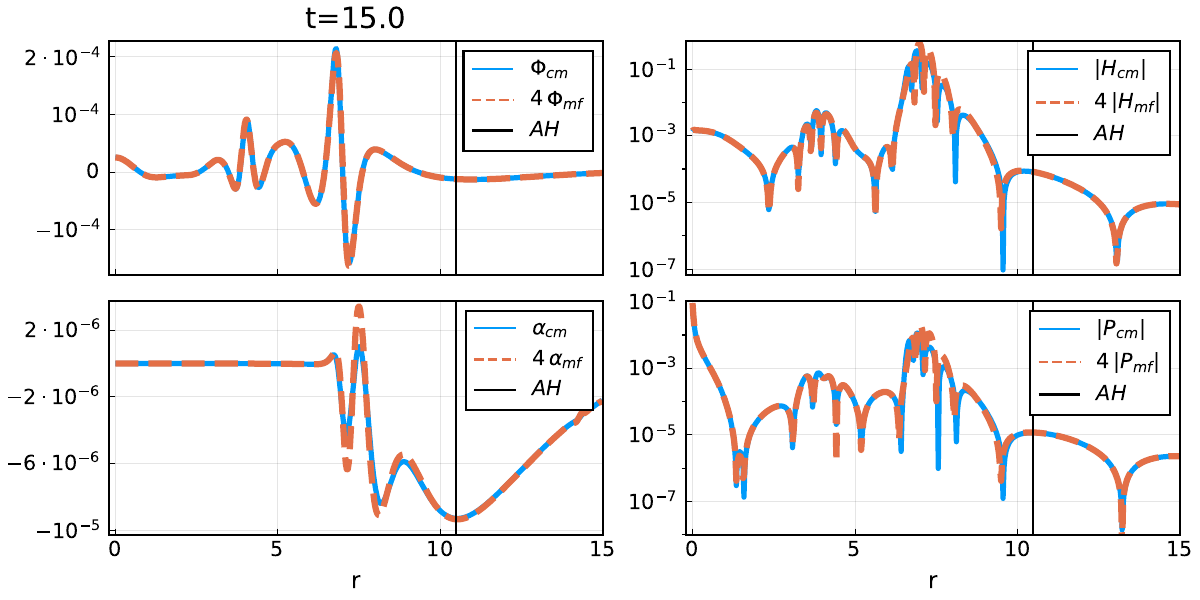}
  \caption{Supercritical data: pointwise converngence for
    $\Phi, \alpha, |H|, |P|$ at~$t_{\max}$, for evolution with
    artificial dissipation~$\sigma=0.02$. We interpret the overlap of
    the rescaled grid functions as evidence for good second-order
    convergence, as in the case with no artificial dissipation.}
\end{figure}
\begin{figure}[t]
  \includegraphics[width=0.9\textwidth]{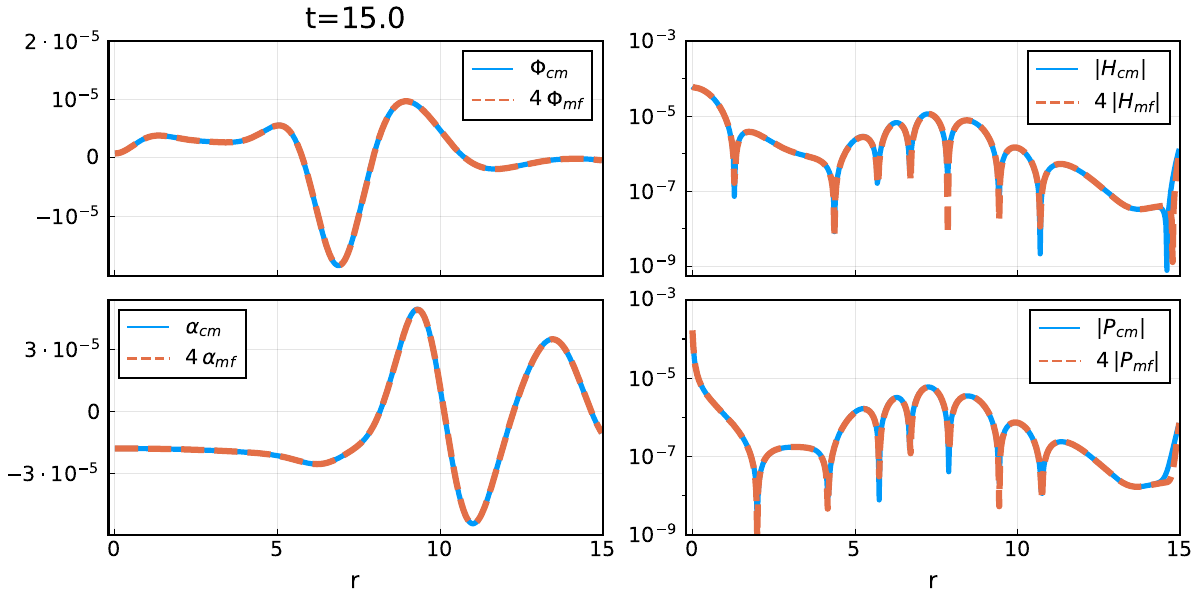}
  \caption{Subcritical data: pointwise converngence for
    $\Phi, \alpha, |H|, |P|$ at~$t_{\max}$, for an evolution with
    artificial dissipation~$\sigma=0.02$. The code exhibits second
    order convergence.}
\end{figure}
\begin{figure}[t]
  \includegraphics[width=0.9\textwidth]{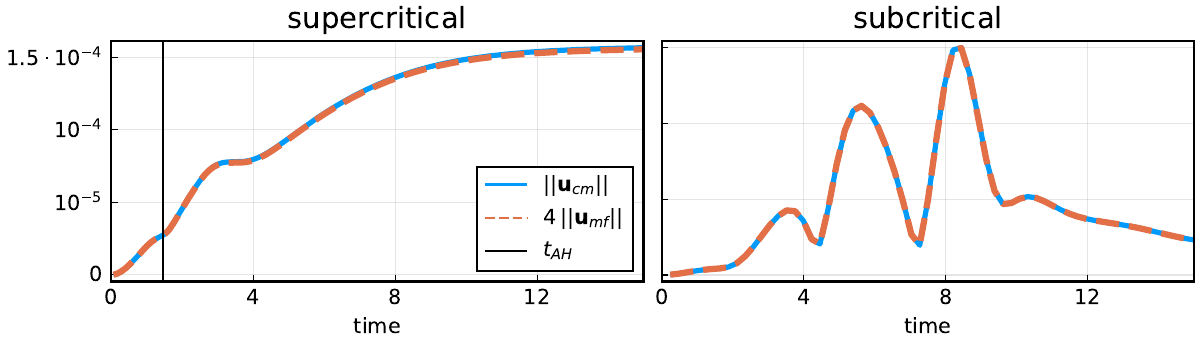}
  \caption{Norm convergence for supercritical (left) and subcritical (right) data,
    with smooth initial data, for an evolution with artificial
    dissipation~$\sigma=0.02$. Both setups exhibit good second-order
    convergence, indicated by the good overlap of the rescaled norms.}
\end{figure}

\begin{figure}[t]
  \includegraphics[width=0.9\textwidth]{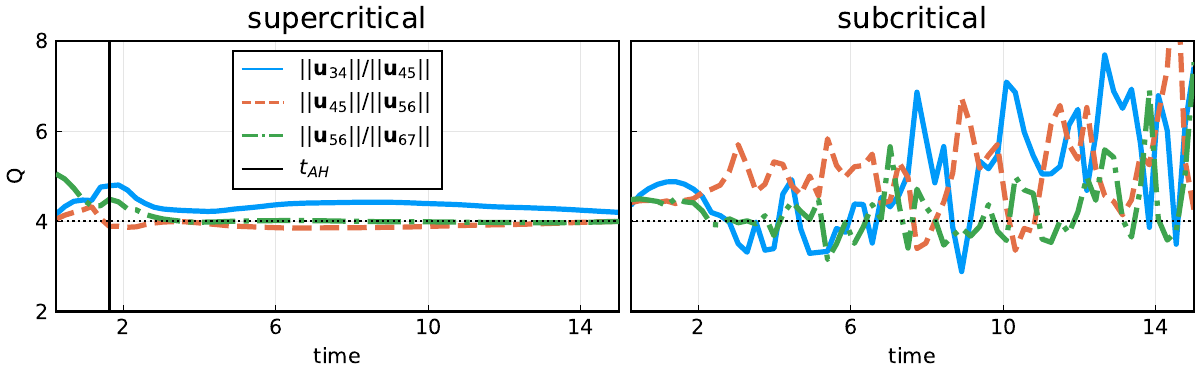}
  \caption{Robust stability tests for supercritical (left) and supcritical (right)
    data with $A_{h_D} = 10^{-3}/4^D$ and artificial dissipation
    $\sigma=0.02$. }
  \label{Fig:robust_stability_sigma002_rand_0.001}
\end{figure}

\begin{figure}[t]
  \includegraphics[width=0.9\textwidth]{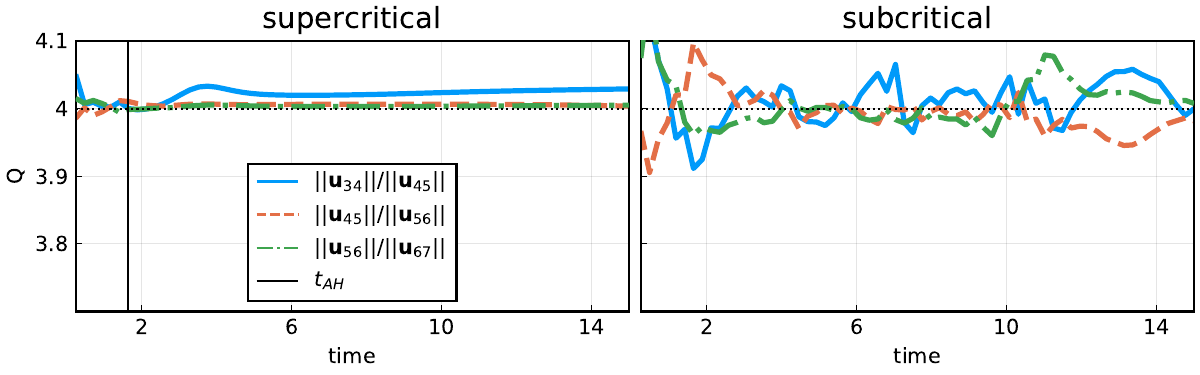}
  \caption{Robust stability tests for supercritical (left) and subcritical (right)
    data with $A_{h_D} = 10^{-5}/4^D$ and no artificial dissipation.}
  \label{Fig:robust_stability_sigma0_rand_1.0e-5}
\end{figure}

\paragraph{Convergence and robust stability.} We repeat the convergence tests of
subsection~\ref{Subsec:convergence_classical} for the classical
setup. The given data are the same, the difference is that there is
artificial dissipation with~$\sigma=0.02$. We see good second-order
convergence for both super- and subcritical data, pointwisely and in
the~$L^2$-norm of the evolved system. We also repeat the robust stability tests of
subsection~\ref{Subsec:robust_stability_classical}, with artificial
dissipation~$\sigma=0.02$ and random noise
amplitude~$A_{h_D} = 10^{-3}/4^D$ in
figure~\ref{Fig:robust_stability_sigma002_rand_0.001}, as well as no
artificial dissipation and~$A_{h_D} = 10^{-5}/4^D$ in
figure~\ref{Fig:robust_stability_sigma0_rand_1.0e-5}. In both cases the
convergence rate is improved in comparison to the results of
figure~\ref{Fig:robust_stability_sigma0_rand_0.001}. This is expected
for both instances: for the former, artificial dissipation is designed
to reduce noise and for the latter, the lower amplitude of the random
noise means that all resolutions apart from~$D=3$ have an
amplitude~$A_{h_D}$ which is below round-off error when
squared. Consequently, only the linear part of the evolution system is
probed with amplitude.


\begin{figure}[h]
  \includegraphics[width=0.88\textwidth]{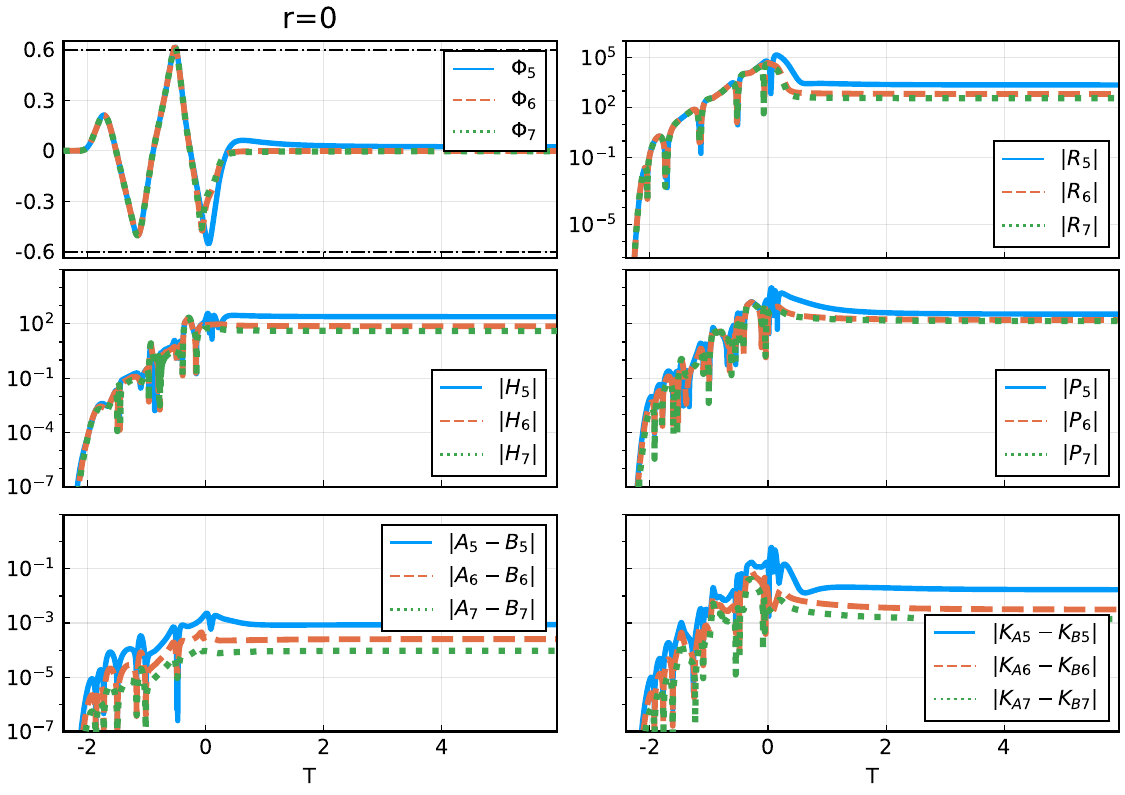}
  \caption{The different quantities monitored against the similarity
    time $T$ as defined in \eqref{eq:similarity_time}, at $r=0$, for
    three different resolutions labelled by $D=\{5,6,7\}$. Top left:
    the scalar field, top right: the absolute value of the Ricci
    scalar $R$. Middle: the absolute value of the Hamiltonian $H$ and
    momentum $P$ constraints on the left and right,
    respectively. Bottom: the absolute value of the difference between
    the metric function $A,B$ (left) and the extrinsic curvature
    components $K_A, K_B$ (right).}
  \label{Fig:critical_ID_family_2_plots}
\end{figure}

\begin{figure}[h]
  \includegraphics[width=0.88\textwidth]{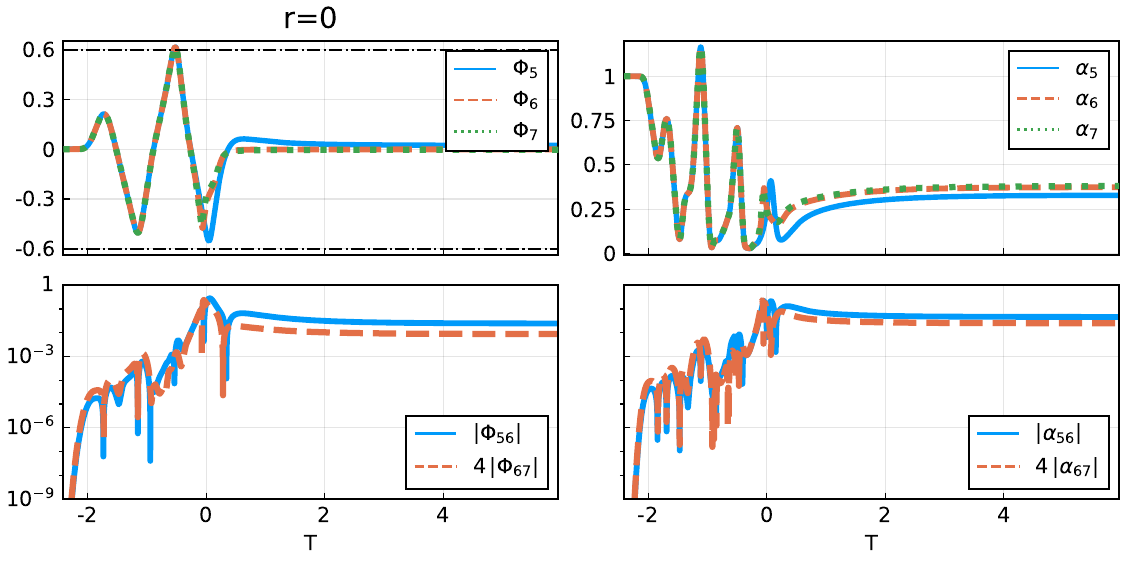}
  \caption{Pointwise convergence at $r=0$ for the classical scalar
    field $\Phi$ (left) and the lapse function $\alpha$ (right), as a
    function of the similarity time $T$. The different resolutions
    $D=\{5,6,7\}$ are denoted with the respective $D$ in the
    subscript. The numerical error (bottom) is computed by taking the
    difference between the numerical solutions of two different
    resolutions, denoted by both their $D$ labels in the
    subscript.}
  \label{Fig:critical_ID_family_2_conv}
\end{figure}

\paragraph{Critical phenomena.} We report the results of ID family 2 with $b=5$ and $c=1$ in the
initial profile of $\Phi$ in \eqref{eq:Phi_ID}. For these runs we
choose $r_{\max}=11.76$, $t_{\max}=11.75$ and dissipation with
$\sigma=0.02$. As in
subsection~\ref{Subsec:critical_phenomena_classical}, the radial grid has
$N_r = 128 \cdot 2^D + 3$ points, with $D=\{5,6,7\}$. We find that the critical
amplitude is $a_* \in (0.044900, 0.044901)$ and in
figure \ref{Fig:critical_ID_family_2_plots} we show the value of the
classical scalar field $\Phi$ at $r=0$ against the similarity time $T$
defined in \eqref{eq:similarity_time}, as well as the Ricci scalar
$R$, Hamiltonian and momentum constraint violations and differences
$A-B$ and $K_A - K_B$. The results are qualitatively the same as for
ID family 1, with the scalar field oscillating again between
$\pm 0.6$, as expected from the universality of the near critical
spherically symmetric
solutions. In figure \ref{Fig:critical_ID_family_2_conv} we demonstrate the
second-order convergence for $\Phi$ and $\alpha$ at $r=0$.

\end{appendix}

\end{document}